\documentclass[12pt]{article}
\input{xy}
\xyoption{all}
\xyoption{poly}
\usepackage[all]{xy}
\usepackage{amsthm,stmaryrd,fancyhdr,mathtools}
\usepackage{color}
\usepackage{graphicx}
\usepackage{float}
\usepackage{amsfonts}
\usepackage{bbm}
\usepackage{bm}
\usepackage{amsmath}
\usepackage{amscd}

\usepackage{amssymb,lscape}
\usepackage{epsfig}
\usepackage{tikz}

\usetikzlibrary{shapes,snakes,shadows,arrows}
\def\ellip{(0,0) ellipse (12 and 6)}
\def\ellipp{(2,0) ellipse (6 and 3)}
\tikzstyle{P_3} = [draw,very thick]
\tikzstyle{P_4} = [draw,fill=blue!20,very thick]

\allowdisplaybreaks

\newcommand{\bmat}{\left(\begin{array}}
\newcommand{\emat}{\end{array}\right)}

\def\gtrsim{\mathrel{\raise.3ex\hbox{$>$\kern-.75em\lower1ex\hbox{$\sim$}}
}
}

\def\-{\hphantom{-}}

\def\s2{\frac{1}{\sqrt2}}

\def\beq{\begin{equation}}
\def\eeq{\end{equation}}
\def\beqa{\begin{eqnarray}}
\def\eeqa{\end{eqnarray}}

\def\mg{m_{3/2}}
\def\mg2{m^2_{3/2}}

\def\Dsl{\,\raise.15ex\hbox{/}\mkern-13.5mu D} 

\def\be{\begin{equation}}
\def\ee{\end{equation}}
\def\bea{\begin{eqnarray}}
\def\eea{\end{eqnarray}}

\newcommand{\nn}{\nonumber}

\makeatletter
\@addtoreset{equation}{section}
\makeatother

\begin{document}
\pagestyle{plain}
\begin{titlepage}
\begin{center}
~

\LARGE{\bf Double Field Theory: \\ A Pedagogical Review\\[10mm]}
\large{\bf  Gerardo  Aldazabal${}^{a,b}$,  Diego
Marqu\'es${}^c$ and Carmen N\'u\~nez$^{c,d}{}$
 \\[4mm]}
\small{
${}^a${\em Centro At\'omico Bariloche,} ${}^b${\em Instituto Balseiro
(CNEA-UNC) and CONICET.} \\[-0.3em]
{\em 8400 S.C. de Bariloche, Argentina.}\\
[0.3cm]

${}^c${\em Instituto de Astronom\'ia y F\'isica del Espacio
(CONICET-UBA)} \\
C.C. 67 - Suc. 28, 1428 Buenos Aires, Argentina.  \\
[0.3cm]

${}^d${\em Departamento de F\' isica, FCEN, Universidad de Buenos Aires}\\
[0.7cm]

{\verb"aldazaba@cab.cnea.gov.ar , diegomarques@iafe.uba.ar , carmen@iafe.uba.ar"} \\[1cm]

\small{\bf Abstract} \\[0.5cm]

}\end{center}

Double Field Theory (DFT) is a proposal to incorporate  T-duality, a distinctive
symmetry of string theory, as a symmetry of a field
theory defined on a double configuration space.
The aim of this review is to provide a pedagogical presentation of DFT and its
applications.
We first introduce some basic ideas on T-duality and
supergravity in order to proceed to the construction of generalized
diffeomorphisms
and an invariant action on the double space.
Steps towards the construction
of a geometry
on the
double space are discussed. We then address
generalized  Scherk-Schwarz compactifications of DFT
and their  connection to gauged supergravity and flux
compactifications. We also discuss U-duality extensions, and  present a brief
parcours on world-sheet approaches to DFT. Finally, we provide a summary of
other developments and applications that are not discussed in
detail in the review.

\vfill

\today

\end{titlepage}


\begin{small}
\tableofcontents
\end{small}

\newpage

\section{Introduction}
\label{seci}

Double Field Theory (DFT) \cite{Siegel:1993th,Hull:2009mi}
 is a proposal to incorporate T-duality, a distinctive symmetry of string
(or M-)theory,
 as a symmetry of a field theory. At first sight, such attempt could appear to lead  to a {\it blind
alley} since the very presence of T-duality requires
extended objects like strings which, unlike field theory particles, are
able
to wrap non-contractible cycles. It is the very existence of winding modes
(associated to these wrappings)  and momentum modes that underlies
T-duality, which manifests itself by connecting the physics of
strings defined on geometrically
very different backgrounds.
Then, a T-duality symmetric field theory must  take  information about windings into account.

A  way
to incorporate such information  is suggested by compactification of
strings on a torus. In string toroidal compactifications, there are
compact momentum modes, dual to compact coordinates $y^m$, $m = 1,\dots , n$,
as
well as
string
winding modes. Therefore, it appears that  a new set
of coordinates $\tilde y_m$,  dual to  windings, should be
considered for the compactified sector in the   field theory description.
It is in this sense that DFT is a ``doubled'' theory: it doubles the coordinates of the compact space.
Formally, the non-compact directions $x^\mu$, $\mu = 1,\dots,d$
are also assigned duals $\tilde x_\mu$ for completion, although this is merely aesthetical since  nothing really
depends on them. The DFT  proposal is that, for a $D$-dimensional space with
$d$ non-compact space-time
dimensions and $n$ compact dimensions, i.e. $D=n+d$, the fields
depend on coordinates $X^M=(\tilde x_\mu, \tilde y_m, x^\mu, y^m)$,
where
$x^{\mu}$ are space-time coordinates, $\tilde x_\mu$ are there simply for
decoration,   and
 $\mathbb{Y}^A=(\tilde y_m,\, y^m)$
are $2n$ compact
coordinates, with ${
A}=1,\dots, 2n$.

When the compactification scale is   much bigger than the string size,
it is hard for strings to wrap cycles and winding modes are ineffective at low energies.  In the DFT
framework, this corresponds to the usual situation where there is no dependence
on dual coordinates. Oppositely, in the T-dual description, if the
compactification scale is small, then the momentum (winding) modes are heavy
(light), and DFT only depends on dual coordinates. Either way, these
(de)compactification limits
typically amount on the DFT side to constrain the theory to depend only on a
subset of coordinates. In particular, when all the
coordinates are non-compact, one finds
complete  correspondence with
supergravity in $D=10$ dimensions.

The T-duality group associated to  string
toroidal compactifications on $T^n$ is $O(n,n)$.
The doubled internal coordinates ${\mathbb Y}^A$ mix (span a vector representation) under the action
of this group. However, it proves  useful to formulate the theory in a double
space with full duality group
$O(D,D)$, where all $D$ coordinates are doubled, mimicking a string theory
where all dimensions are compact.\footnote{$\bf$ Time is
treated here at the same level of other space  coordinates
for simplicity, but it can be restored by a standard Wick rotation.}

The next step in the construction of DFT is to choose the
defining fields. In the simplest formulation of DFT, the field content involves the  $D$-dimensional metric
$g_{ij}$, a two-form field $b_{ij}$ and a scalar dilaton
field $\phi$. From a string perspective, they  correspond to the
universal  gravitational massless bosonic sector, present in the
bosonic, heterotic and Type II
string theories as well as in the closed sector of
Type I strings, in which case $b_{ij}$ would be a Ramond-Ramond (R-R)
field.  However, since we are looking for an  $O(D,D)$
invariant theory, the fundamental fields should be  $O(D,D)$ tensors
with $2D$ dimensional indices. In fact, in DFT the
 $g_{ij}$ and $b_{ij}$ fields are {\it unified} in a single object: a generalized  $O(D,D)$ symmetric  metric ${\cal H}_{MN}$, with $M, N=1,\dots,2D$, defined in the double space. Then, based on {\it symmetries}, DFT {\it unifies} through {\it geometrization}, since it incorporates the two-form into a generalized geometric picture.
There is also a field
$d$, which is a T-scalar
combining  the dilaton $\phi$ and the determinant of the metric $g$.
The first  part of this  review  will
be dedicated to discuss the
consistent construction of a DFT action as a functional of  these generalized fields on a
doubled configuration space.

In the decompactification limit (taking for example $D = 10$ so as to make
contact with string theory), when the dual coordinates are
projected
out,  the DFT action reproduces the action of the universal massless bosonic sector of
supergravity
\begin{eqnarray}
S &=&\int d x \sqrt{g}e^{-2\phi} \left( R ~+ ~
4 ~(\partial \phi)^2 -
\frac{1}{12}  H_{ijk}H^{ijk}\right)\, ,\nn
\end{eqnarray}
where $H_{ijk} = 3 \partial_{[i}b_{jk]}$ is the  field strength of the two-form. This limit action is invariant under the usual diffeomorphisms of General
Relativity and gauge transformations of the two-form. Following with the unification route, we then expect to combine these transformations into
``generalized  diffeomorphisms" under which the DFT action should be invariant.
They  should then reduce to standard general coordinate and gauge
transformations in the decompactification
limit.
In Section \ref{sec: DFT} we will define
these transformations and discuss constraint
equations  required  by gauge consistency of the generalized diffeomorphisms.
 Generically, these constraints restrict the space of configurations for which
DFT is a consistent theory, i.e. DFT is a {\it restricted theory}.

The constraints of the theory are solved in particular when a {\it section condition} or
{\it strong constraint}, is imposed. This restriction was proposed in the original
formulations
of DFT, inspired by string field theory constraints.
It implies that the fields of the theory only depend on a slice of the double space parameterized by half of
the coordinates, such that there always exists a frame in which, locally,
the configurations do not depend on the dual coordinates. Since the strong constraint
is covariant under the global symmetries, the theory can
still be covariantly formulated, but it is  actually   not truly doubled after
it is solved.

Nevertheless, one can also find other solutions to the constraints that violate the strong constraint. In particular,
Scherk-Schwarz (SS) dimensional
reductions  of DFT, where
the space-time fields are twisted by functions of the internal
coordinates,  have proven to be interesting scenarios where consistent strong-constraint violating configurations are allowed.
Interestingly enough, the SS reduction of (bosonic) DFT  on
the doubled space leads to an action that can be identified with (part
of) the  action of the bosonic sector  of four-dimensional half-maximal gauged
supergravities.
 Recall that gauged supergravities are deformations of ordinary abelian
supergravity theories, in which the deformation parameters (gaugings) are encoded in the embedding tensor.  DFT  provides a higher dimensional interpretation of these
gaugings  in terms of SS double  T-duality twists.
Moreover, the quadratic constraints
on gaugings are in one to one correspondence with the  closure constraints of the generalized
diffeomorphisms.

Gauged supergravities describe superstring compactifications with fluxes,
where the gaugings correspond to the quantized fluxes.
Therefore it is instructive to look at the connection between
SS reductions of DFT and string flux compactifications. This connection
 is subtle. It is known that orientifold compactifications of $D  = 10$
effective supergravity actions, corresponding to the low energy limit of
string theories,  lead to four dimensional superpotentials in which the
coefficients are the fluxes.
However, by looking at flux
compactifications of string theories, expected to be T-duality related
(for instance, type IIA and type IIB theories) the effective
superpotentials
turn out not to be  T-dual. Namely, these compactifications  are gauged
supergravities  but with different orbits of gaugings turned on,
not connected by T-duality.
By invoking symmetry arguments,  it has been suggested
that new fluxes should be included in order for the full superpotentials
to be T-duals, so as to repair the mismatch. Similarly, more fluxes are required  by
invoking type IIB S-duality, M-theory or heterotic/type I S-duality, etc.
Then,  by imposing duality invariance at the level of the
four
dimensional effective theory,
the full (orientifold truncated) supergravity theory is
obtained
with all allowed gaugings.

Hence,  we can conclude that four dimensional gauged supergravity
incorporates stringy information  that, generically,
is not present in the reduction of a ten dimensional effective supergravity action.
Compactification of DFT contains this stringy information from
the start and provides a {\it geometric} interpretation for fluxes,
even for those that are non-geometric from a supergravity point of view.

There have also been different proposals  to extend  DFT
ideas to incorporate
the full  stringy U-duality symmetry group. Take $E_{7(7)}$ as an example, which
includes
T-dualities and
strong-weak duality. The symmetrization now requires an
Extended Geometry on which one can define an Extended Field
Theory (EFT). Interestingly enough, from a string theory perspective such
formulation  automatically incorporates information on NS-NS
and R-R fields.  While in DFT with $O(n,n)$ symmetry a doubled
$2 n$ compactified space is needed, in EFT coordinates span a mega-space with
more dimensions, where SS compactifications lead to four-dimensional gauged maximal supergravity.

Closely related to DFT (or EFT) is the framework of
 Generalized Geometry (or Exceptional Generalized Geometry), a program that
also incorporates duality as a building block.
In Generalized Geometry (GG), the tangent space,
where the vectors generating diffeomorphisms live, is enlarged to include
the one-forms corresponding to gauge transformations of the two-form. The
internal space is not extended, but the notion of geometry is
still modified. DFT and GG  are related when the  section condition
(which un-doubles the double space) is imposed.

~

To summarize, DFT is
 a T-duality invariant reformulation of supergravity which appears to offer a
way to go
beyond the supergravity limits of string theory by introducing some
stringy features into
particle physics.
DFT
is all about T-duality
{\it symmetries}, {\it unification} and {\it geometry}.
It is a rather young theory, still under development, but it has already produced plenty of new perspectives and results. There are still many things to understand, and the number of applications is increasing. Here we intend to review this beautiful theory and some of its applications, in as much a pedagogical fashion as we can.

\newpage

\section{Some references and a guide to the review}

In this review we intend to provide a self-contained pedagogical introduction to
DFT. We will introduce the basics of the
theory in lecture-like fashion, mostly intended for non-experts that are
willing to know more about this fascinating theory. We will mainly review the recent literature on the formulation and
applications of the theory. The field is undergoing a quick expansion, and many
exciting results are still to appear. Given the huge amount of material in this
active area of
research, we are forced to leave out many developments that are as important and
stimulating as those that we consider here. With the purpose of reducing the
impact of this restriction, we provide an updated
list of references,
were the reader can find more specific information. We apologize if,
unintentionally,  we have omitted important references.

Let us first start with a brief list of books on string theory \cite{books}.
There are already some very good and complete reviews and lectures on this and
related topics, that we strongly suggest.
In   \cite{Giveon:1994fu,Alvarez:1994dn} the reader will find a complete exposition on
T-duality.  Flux compactifications
are nicely reviewed in \cite{Grana:2005jc}. Comprehensive reports on
non-geometric fluxes
and their relation to gauged supergravities are those in \cite{Wecht:2007wu} and
\cite{Samtleben:2008pe}, respectively.
DFT has also been reviewed in \cite{Zwiebach:2011rg}, and Generalized Geometry
in \cite{lecturesGG}. A complete review on duality symmetric string and M-theory can be found in \cite{RevBerman}.

Historically, the idea of implementing T-duality as a manifest symmetry goes back to
M. Duff \cite{duff} and  A. Tseytlin \cite{tseytlin}, where many of the building blocks of DFT were
introduced.
In \cite{duff}, one can identify already the double coordinates, the generalized
metric and frame, the notion of duality symmetric sigma models and the extension of these concepts to U-duality. In  \cite{tseytlin}, the idea of DFT was essentially present: double coordinates were considered, an effective action for the metric in double space presented, and the necessity of consistency constraints noted.
 Soon after, Siegel contributed his pioneer work \cite{Siegel:1993th}, in which a full duality symmetric action  for the low-energy superstring was built in superspace formalism.
More recently, C. Hull and B. Zwiebach combined their expertise on double
geometry \cite{doublegeom} and
string field theory \cite{Kugo:1992md} to build Double Field Theory
\cite{Hull:2009mi}. Later, together with O. Hohm,
they constructed a background independent \cite{Backgroundindependent} and
generalized metric \cite{Generalizedmetric}
formulation of the theory. The relation of their work to
Siegel's was analyzed in
\cite{framelikegeom}. Closely related to DFT is the  Generalized Geometry
introduced by N. Hitchin and M. Gualtieri \cite{Hitchin}
and related to string theory in the works by M. Gra\~na, T. Grimm, J. Louis, L. Martucci, R.
Minasian, M. Petrini,
A. Tomasiello and D. Waldram \cite{GGafter}, among others.

The inclusion of heterotic vector fields in the theory was discussed in
\cite{heteroticHohm} (see also \cite{Andriot:2011iw}). R-R fields and a unification of Type II theories were included in
\cite{TypeIIZwiebach,TypeIIWaldram,TypeIIPark},
while the massive Type II theory was treated in \cite{MassiveTypeII}. The
inclusion of fermions and supersymmetrization was performed in \cite{SDFT,TypeIIWaldram}.
There are many works devoted to explore the geometry of
DFT \cite{GeometryPark,TypeIIWaldram,GeometryZwiebach,GeometryBerman}. A fully
covariant supersymmetric Type II formulation was constructed by I. Jeon, K. Lee
and J. Park in \cite{TypeIIParkII}. The gauge symmetries and equations of motion
were analyzed by S. Kwak \cite{EOMsKwak}, and the gauge algebra and constraints
of the theory were discussed in \cite{Hull:2009zb,GDFT}. The connection with
duality symmetric non-linear sigma models was established by D. Berman, N. Copland and D.
Thompson in \cite{bt,WorldsheetDFT,copland}. Many of these works were inspired by
Siegel's construction \cite{Siegel:1993th}.

Covariant frameworks extending T-duality to the full U-duality group were built
as well.
These include works by C. Hull \cite{HullU}, P. Pacheco and D. Waldram
\cite{PachecoU}, D. Berman and M.
Perry \cite{BermanPerry}, the $E_{11}$ program by P. West et. al.
\cite{E11programme} and \cite{Koepsell:2000xg,Hillmann:2009ci}. More recent DFT-related developments
can be found in
\cite{Uduality1,Uduality2,Udualities3,localsymm,Thompson:2011uw,SL5Uduality,Cederwall,E8}.
Also in this direction, but more related to non-geometry and gauged
supergravities we have
 \cite{Udualfluxes,SSMtheory,Musaev,Extended geometry}.

The ideas introduced in  \cite{Dasgupta:1999ss,Kachru:2002sk,Hellerman:2002ax,Dabholkar:2002sy} led to the development of non-geometry, and
T-dual non-geometric fluxes were named as such in \cite{stw} (see also \cite{Othernongeom}). Later, S-dual fluxes were introduced in
\cite{acfi}, and finally the full U-dual set of fluxes was completed in
\cite{Udualfluxes}. Fluxes were considered from a generalized geometrical point
of view in \cite{GGafter}, and also from a double geometrical point of view in
\cite{doublegeom,doubletorus,HackettJones:2006bp}.  The relation between DFT,
non-geometry and gauged supergravities was explored in
\cite{Andriot:2012an,Aldazabal:2011nj,Geissbuhler:2011mx,Dibitetto:2012rk,
Exploring,Blumenhagen:2013aia,Berman:2013cli,oxidation,Hohm:2013nja,Condeescu:2013yma}. Different  world-sheet perspectives for fluxes were
addressed in \cite{avra,Halmagyi:2008dr,Mylonas:2012pg}.

Some other developments on DFT
and related works can be found in \cite{OtherDFT}. In the final section we
include more references, further developments and applications of DFT.

The present review covers the following topics:
\begin{itemize}

\item {\bf Section \ref{sec: DFT}} provides a general introduction to DFT.
Starting
with
some  basics on  T-duality as a motivation, double space and generalized
fields are
then defined. A generalized Lie derivative encoding usual
diffeomorphisms and two-form gauge transformations is introduced, together with
its consistency constraints. We then present the DFT action, its symmetries
and equations of motion.

\item {\bf Section \ref{sec: Double Geometry}} reviews the  construction of an
underlying double geometry for
DFT.
Generalized connections, torsion, and curvatures are discussed, and their similarities and
differences with ordinary Riemannian geometry are examined.

\item {\bf Section \ref{sec:Dimensional reductions}} is
devoted to a discussion of
dimensional reductions of DFT. After a brief introduction of usual
SS compactifications, the procedure is applied to deal with
generalized SS compactifications of DFT. The notions of geometric and non-geometric fluxes are addressed and the connection with  gauged supergravity is
established.

\item {\bf Section \ref{sec:EFT}} considers the U-duality extension of DFT,
Extended Geometries, Extended Field Theories and their relation to maximal
gauged supergravity.

\item {\bf Section \ref{sec: WSDFT}} reviews the various attempts to construct
$O(D,D)$
  invariant non-linear sigma models, and their relation to  DFT.

\item {\bf Section \ref{sec:Further developments}}
 provides a brief summary of different
developments related to DFT (and guiding references), together with open problems,  that are not discussed in
detail  in the review.
\end{itemize}

\newpage
\section{ Double Field Theory}
 \label{sec: DFT}

Strings feature many amazing properties that particles lack, and this
manifests
in the fact that string theory has many stringy symmetries that are
absent in
field theories like supergravity. Field
theories usually describe the dynamics of particles, which have no
dimension.
Since the string is one-dimensional,
closed
strings can wind around non-contractible cycles if the space is compact.
So clearly, if we aimed at describing the dynamics of strings with a field
theory,
 the particles should be assigned more degrees of freedom,
to
account for their limitations to reproduce stringy dynamics like winding.
Double Field Theory is an attempt to incorporate some stringy features
into a field theory in which the new degrees of freedom are introduced
by doubling the space of coordinates.

DFT
can be thought of as  a T-duality
invariant
formulation of the ``low-energy'' sector of string theory on a compact space.
The reason why {\it low-energy} is quoted here is because, although it is $O(D,D)$ symmetric, DFT keeps the levels
that would be massless in the decompactification limit of the string spectrum. In some sense, DFT can
be thought of
as a T-duality symmetrization of supergravity.
Our route will begin with  the NS-NS
sector,
and later we will see how these ideas can be extended to the other
sectors. As a
starting point, we will briefly introduce the basic notions of T-duality and
supergravity,
mostly in an ``informal'' way, with the only purpose of introducing the
fundamental
concepts that will then be applied and extended for DFT. A better and more
complete exposition of these topics
can be found in  the many books on string theory \cite{books}.

\subsection{T-duality basics}\label{secTduality}

T-duality is a symmetry of string theory that relates winding modes in a
given
compact space with momentum modes in another (dual) compact space.
Here we summarize the basic ingredients of T-duality. For a complete and
comprehensive review see \cite{Giveon:1994fu}.\footnote{For recent progress on non-Abelian T-duality see \cite{nonAbelianT}.}

Consider the mass spectrum of a
closed
string on a circle of radius $R$
\be
M^2 = (N + \tilde N - 2) + p^2\ \frac{l_s^2}{R^2} + \tilde p^2\
\frac{l_s^2}{\tilde R^2} \label{massformulaT}
\ee
where $l_s$ is the string length scale and ${\tilde R} = \frac{l_s^2 }{R
}$, the dual radius.
The first terms contain the infinite mass levels of the string spectrum, and the
last
two
terms are proportional to  their quantized momentum $p$ and winding
$\tilde
p$. The modes are constrained to satisfy the Level Matching Condition (LMC)
\be
N - \tilde N = p \tilde p\, ,
\ee
reflecting the fact that there are no special points in a closed string.

If we take the decompactification limit ${R} \gg {l_s}$, the winding modes
become heavy, and the mass spectrum for the momentum modes becomes a
continuum. On the other hand, if we take the opposite limit ${R}\ll
{l_s}$, the
winding modes become light and the momentum modes heavy. These behaviors
are
very reasonable: if the compact space is large it would demand a lot of
energy
to stretch a closed string around a large circle so that it can wind, but
very
little if the space were small.

Notice that for any level, the mass spectrum is invariant under the
following
exchange
\be
\frac{R}{l_s} \leftrightarrow \frac{\tilde R}{l_s} = \frac{l_s }{R }\ , \
\ \ \
\ p \leftrightarrow \tilde p \, ,\label{Tdualitycircle}
\ee
so if we could only measure masses, we would never be able to distinguish
between a
closed string moving with a given momentum $k$ on a circle of radius $R$,
and a
closed string winding $k$ times on a circle of radius $l_s^2 / R$. This
symmetry
not only holds for the mass spectrum, but it is actually a symmetry of any
observable
one can imagine in the full theory!

DFT currently restricts to the modes of the string that are massless in the
decompactified
limit, i.e. with $N+\tilde N = 2$,  but considers them on a
compact
space
(actually, some or all of these dimensions can be taken to be non-compact). These
modes
correspond to the levels\footnote{For the cases
$(N,\tilde N) = (1,0)$ and $(0,1)$ there is a particular enhancement of
the
massless degrees of freedom at $R  =l_s$, which has not been contemplated in
DFT so far.}  $N = \tilde N = 1$ (notice that the LMC forbids
the
possibilities $(N,\tilde N) = (2,0)$ and $(0,2)$ when $p \tilde p = 0$) corresponding to a symmetric metric $g_{ij}$, an
antisymmetric two-form
$b_{ij}$ and a dilaton $\phi$.

The T-duality symmetry of circle compactifications is generalized to $O(D,D,
{\mathbb Z})$ in toroidal compactifications with constant background metric
and antisymmetric field.
The elements of the infinite discrete group $O(D,D)$ (we will drop the ${\mathbb
Z}$ in this review because it is irrelevant for our purposes of introducing DFT
at the classical level) can be defined as the set of $2D \times 2D$
matrices $h_{MN}$
that preserve the $O(D,D)$ invariant metric $\eta_{MN}$
\be
h_M{}^P\ \eta_{PQ} \ h_N{}^Q = \eta_{MN}
\ee
where
\be
\eta_{MN} = \left(\begin{matrix}0 & \delta^i{}_j \\ \delta_i{}^j & 0
\end{matrix}\right) \ , \ \ \ \ \eta^{MN} = \left(\begin{matrix}0 &
\delta_i{}^j
\\ \delta^i{}_j & 0 \end{matrix}\right) \ , \ \ \ \ \eta^{MP}\eta_{PN} =
\delta^M_N \, ,\label{eta}
\ee
raises and lowers all the $O(D,D)$ indices $M, N = 1, \dots , 2D$.

The momentum and winding modes are now
$D$-dimensional objects $p^i$ and $\tilde p_i$ respectively. They can be
arranged
into a larger object (a generalized momentum)
\be
{\cal P}^M = \left(\begin{matrix}\tilde p_i \\ p^i\end{matrix}\right) \,
,\label{genmomenta}
\ee
in terms of which the mass operator becomes
\be
M^2 = (N+\tilde N-2) + {\cal P}^P {\cal H}_{PQ} {\cal P}^Q \, ,\label{m0m}
\ee
where
\be
 \label{gennmetTduality}
  {\cal H}_{MN} \  \ = \
  \begin{pmatrix}    g^{ij} & -g^{ik}b_{kj}\\[0.5ex]
  b_{ik}g^{kj} & g_{ij}-b_{ik}g^{kl}b_{lj}\end{pmatrix}
 \ee
is called the {\it generalized metric} \cite{Giveon:1988tt,Maharana:1992my}. The LMC now takes the form
\be
N - \tilde N = \frac 1 2 {\cal P}^M {\cal P}_M \, ,\label{LMC}
\ee
and implies that, for the DFT states $N = \tilde N = 1$, the generalized
momenta
must be orthogonal with respect to the $O(D,D)$ metric $p^i \tilde p_i = 0$.

Any element of $O(D,D)$ can be decomposed as successive products of the
following transformations:
\bea
&&{\rm Diffeomorphisms:} \ \ \ h_M{}^N = \left(\begin{matrix}E^i{}_j & 0\\
0 & E_i{}^j \end{matrix}\right)\ , \ \ \ \ \ \ E \in GL(D)\nn\\\nn\\
&&{\rm Shifts:} \ \ \ \ \ \ \  h_M{}^N = \left(\begin{matrix}\delta^i{}_j &
0\\
B_{ij} & \delta_i{}^j \end{matrix}\right) \ , \ \ \ \ \ B_{ij} = - B_{ji}\label{Tdualities}\\\nn\\
 &&\begin{matrix}{\rm Factorized}\\{\rm T-dualities: }\end{matrix}\ \ \ \ h^{(k)}{}_M{}^N =
\left(\begin{matrix}\delta^i{}_j
-t^i{}_j & t^{ij}\\ t_{ij} & \delta_i{}^j - t_i{}^j  \end{matrix}\right) \
, \ \
\ \ t = {\rm diag}(0\dots 0\ 1 \ 0\dots 0) \, ,\nn
\eea
If the antisymmetric $D\times D$ matrix $B_{ij}$ in the shifts were written in the North-East
block, the resulting transformation is usually called $\beta$-transformation, for reasons that will become clear latter.
The diffeomorphisms correspond to basis changes of the lattice underlying the torus, and the factorized T-dualities generalize the $\frac R {l_s} \leftrightarrow \frac {\tilde R}{l_s}$ symmetry discussed above. The $1$ in  the $D\times D$ matrix $t$ is in the $k$-th position. It is
therefore common to find statements about T-duality being performed on a
given
$k$-direction, in which case the resulting transformations for the metric
$g_{ij}$ and two-form $b_{ij}$ are named {\it Buscher rules}
\bea
&& g_{kk} \to \frac 1{g_{kk}} \ , \ \ \ \ g_{k i} \to  \frac {b_{k i
}}{g_{kk}}
\ , \
\ \ \ g_{ij} \to g_{ij} - \frac{g_{ki} g_{kj} - b_{ki}b_{kj}}{g_{kk}}\, ,
\nn\\\nn\\
&& b_{k i} \to   \frac{g_{k i }}{g_{kk}}\ , \ \ \ \ b_{ij} \to b_{ij} -
\frac{g_{ki} b_{kj} - b_{ki}g_{kj}}{g_{kk}} \, .\label{Buscher}
\eea
These transformation rules were first derived by T. Buscher from a
world-sheet
perspective in \cite{buscher1,buscher2}, and they rely on the fact that
the T-duality is performed in an
isometric direction (i.e., a direction in which the fields are constant).
Notice that $g$ and $g^{-1}$ get exchanged in the
$k$-th
direction, just as it happens in the circle with the inversion $R/l_s$ and
$l_s/R$. Also notice that the metric (\ref{eta}) corresponds to a
product of $n$ successive T-dualities, and for this reason this matrix is usually called the inversion metric
(as we will see, it inverts the full generalized metric
(\ref{gennmetTduality})).

Summarizing, the T-duality symmetry of the circle compactification is
generalized in toroidal compactifications to $O(n,n)$ acting as
\be
{\cal H}_{MN}\ \leftrightarrow\ h_M{}^P \ {\cal H}_{PQ}\ h_N{}^Q\ , \ \ \
\ \
{\cal P}^M\ \leftrightarrow\ h^M{}_N \ {\cal P}^N \ , \ \ \ \ h \in O(n,n)\, ,
 \ee
on constant backgrounds. More generally, T-duality in DFT is allowed in
non-isometric directions, as we will see.

Let us now consider the dilaton, on which T-duality acts non-trivially. The
closed string coupling in $D$-dimensions, $g_s^{(D)}=e^{-2\phi}$,  is
related to
the $(D-1)$-dimensional coupling
when one dimension is compactified on a circle as $g_{s}^{(D-1)} =
\sqrt{R/l_s}\
g_s^{(D)}$.
Given that the scattering amplitudes for the dilaton  states
are invariant under T-duality, so must be the $(D-1)$-dimensional coupling.
Therefore, the dilaton of two theories compactified on circles of dual radii
$R$ and  $l_s^2/R$ must be related. When the compact space is
$n$-dimensional,
the T-duality invariant $d$ is given by the following combination
\be
e^{-2 d}= \sqrt {g} e^{-2\phi}\label{gendil}
\ee

This intriguing symmetry of string theory is not inherited
 by the fully decompactified low energy effective theory
(supergravity), because all the winding modes are infinitely heavy and play
no
role in the low energy dynamics. Therefore, decompactified supergravity
describes the ``particle limit'' of the massless modes of the string.
However, it is likely that a fully compactified supergravity in
$D$-dimensions (i.e. where all dimensions are compact, and then $D=n$)
can be rewritten  in a T-duality, or more generally $O(D,D)$
covariant
way, such that the symmetry becomes manifest at the level of the field
theory.
Then, DFT can be thought of as a T-duality invariant formulation of supergravity
with
compact dimensions. Actually, as we will see, DFT is more general than just
a compactification of a
fully decompactified theory (where the winding modes have been integrated
out).
The generalization
relies on
the fact that the winding dynamics is kept from the beginning, and at low-energies
winding modes only decouple
when the corresponding  directions of the fully compactified theory are
decompactified.

 In order to begin with the construction of DFT, it is instructive
to first introduce
supergravity in $D$  decompactified dimensions.

 \subsection{Supergravity basics}
Before trying to assemble the NS-NS sector of supergravity in a T-duality
invariant
formulation, let us briefly review the bosonic sector of the theory that we will
then try to covariantize.
The degrees of freedom are contained in a $D$-dimensional metric (of
course, we
always keep in mind that the relevant dimension is $D = 10$) $g_{ij} =
g_{(ij)}$, with $i,j,\dots = 1,\dots,D$, a $D$-dimensional two-form
$b_{ij} =
b_{[ij]}$ (also known as the $b$-field or the Kalb-Ramond field) and a
dilaton
$\phi$. All these fields depend on the $D$ coordinates of space-time
$x^i$.

There is a pair of local gauge transformations under which the  physics
does not
change:
\begin{itemize}
\item Diffeomorphisms, or change of coordinates, parameterized by
infinitesimal vectors $\lambda^i$
\bea
g_{ij} &\to& g_{ij} + L_\lambda g_{ij} \ , \ \ \ \ \ L_\lambda g_{ij} =
\lambda^k \partial_k g_{ij}  + g_{kj} \partial_i \lambda^k + g_{ik}
\partial_j
\lambda^k \, ,\nn\\
b_{ij} &\to& b_{ij} + L_\lambda b_{ij}\ , \ \ \ \ \ L_\lambda b_{ij} =
\lambda^k
\partial_k b_{ij} + b_{kj} \partial_i \lambda^k + b_{ik} \partial_j
\lambda^k\, ,
\nn\\
\phi &\to& \phi + L_\lambda \phi\ , \ \ \ \ \ \ \ \; L_\lambda \phi =
\lambda^i
\partial_i \phi \, .\label{diffeossugra}
\eea
Here $L_\lambda$ is the Lie derivative, defined as follows for arbitrary vectors $V^i$
\be
L_\lambda V^i = \lambda^j \partial_j V^i - V^j \partial_j \lambda^i = [\lambda, V]^i \label{Lie}
\ee
In the last equality we have defined the Lie Bracket, which is antisymmetric and satisfies the Jacobi identity. It is very important to keep the Lie derivative in mind, because
it will be generalized later, and the resulting generalized Lie derivative is one of the building blocks of DFT. The action of the Lie derivative amounts to diffeomorphic transformations, and the invariance of the action signals the fact that the physics remains unchanged under a change of coordinates.

\item Gauge transformations of the two-form, parameterized by infinitesimal
one-forms $\tilde \lambda_i$
\be
b_{ij} \to b_{ij} + \partial_i \tilde \lambda_j - \partial_j \tilde
\lambda_i\, .
\label{twoformgaugesugra}
\ee
\end{itemize}

The supergravity action takes the following form
\be\label{sugraaction}
   S \ = \
   \int d^Dx \sqrt{g} e^{-2 \phi}
   \bigg[ R + 4(\partial \phi)^2
   - \frac{1}{12} {H}^{ijk} {H}_{ijk}\bigg]\, ,
 \ee
 where we have defined the following three-form with corresponding Bianchi
identity (BI)
 \be
 H_{ijk} = 3 \partial_{[i} b_{jk]} \ , \ \ \ \ \
\partial_{[i} H_{jkl]} = 0 \, ,\label{threeform}
 \ee
and $R$ is the Ricci scalar constructed from $g_{ij}$ in the usual
Riemannian
sense. It is an instructive warm-up exercise to show that this action is
invariant
under diffeomorphisms (\ref{diffeossugra}) and the two-form gauge
transformations
(\ref{twoformgaugesugra}).

The equations of motion derived from the supergravity action take the form
\bea \label{eom} R_{ij}  - \frac{1}{4} {H_{i}}^{ p q} H_{j p q}  + 2 \nabla_i
\nabla_j \phi &=& 0  \label{eomg}\\[1.5ex] \frac{1}{2} \nabla^p H_{p i j} -
H_{p i j} \nabla^p \phi &=& 0 \, , \\ [1.5ex] R+4\left (\nabla^i \nabla_i \phi
-
(\partial \phi)^2\right ) - \frac{1}{12}H^2 &=&0 \label{eomd}\eea
From the string theory point of view, they imply the Weyl invariance of
the
theory at the one loop
quantum level.

We have described in this section  the bosonic NS-NS sector of
supergravity. This sector  is interesting on its own because it determines the
moduli
space of the theory. Given that the fermions are charged with respect to
the
Lorentz group, for any given configuration the vacuum expectation value
(VEV) of
a fermion would break Lorentz invariance. In order to preserve this
celebrated
symmetry, one considers vacua in which the fermions have vanishing VEV.
For this reason, and also for simplicity, in this review we will restrict to
bosonic degrees of freedom.

\subsection{Double space and generalized fields}\label{sec:DoubleFields}

So far we have introduced the basic field-theoretical notions of
supergravity,
and explained the importance of T-duality in string theory. It is now time
to
start exploring how the supergravity degrees of freedom can be rearranged in a
T-duality
invariant
formulation of DFT \cite{Hull:2009mi}. For this to occur, we must put everything
in T-duality
representations, i.e., in objects that have well-defined transformation
properties under T-duality.

Let us begin with the fields. As mentioned, we consider on the one hand a
metric
$g_{ij}$ and a two-form $b_{ij}$ which
can combine into a
symmetric generalized metric ${\cal H}_{MN}$
given by
\be
 \label{gennmet}
  {\cal H}_{MN} \  \ = \
  \begin{pmatrix}    g^{ij} & -g^{ik}b_{kj}\\[0.5ex]
  b_{ik}g^{kj} & g_{ij}-b_{ik}g^{kl}b_{lj}\end{pmatrix}\, .
 \ee

Notice that this metric has the same form as the one defined in
(\ref{gennmetTduality})
 but here the fields are non-constant.
This is an $O(D,D)$ element, and its inverse is obtained by raising the
indices
with the $O(D,D)$ metric $\eta^{MP}$ introduced in (\ref{eta})
\be
{\cal H} \in O(D,D)\ , \ \ \ \ \ {\cal H}^{MN} = \eta^{MP} {\cal H}_{PQ}
\eta^{QP} \ , \ \ \ \ \ {\cal H}_{MP} {\cal H}^{PN} = \delta^N_M\, .
\label{constraintsGenMet}
\ee
Actually, all the indices in DFT are raised and lowered with the
$O(D,D)$
invariant metric (\ref{eta}).
On the other hand the dilaton $\phi$ is combined with the determinant of
the
metric $g$ in an $O(D,D)$ scalar $d$
 \be
 e^{-2 d} = \sqrt{g} e^{-2 \phi}\, .
 \ee

 Before showing how these objects transform under local and global
symmetries,
let us mention where these generalized fields are defined. Since
everything must be organized in T-duality representations, the
coordinates
cannot be an exception.
Paradoxically, we only have $D$ of them: $x^i$, while
the lowest dimensional
representation of $O(D,D)$
is the fundamental,
which has
dimension $2D$. We therefore face the question of what should we combine
the
supergravity coordinates with, in order to complete the fundamental
representation. It turns out that there are no such objects in
supergravity, so
we must {\it introduce} new
coordinates $\tilde x_i$. We can now define a generalized notion of
coordinates
\be
X^M = (\tilde x_i, x^i)\, ,
\ee
and demand that the generalized fields depend on this double set of coordinates
\be
{\cal H}_{MN} (X)\ , \ \ \ \ \ d(X)\, .
\ee
From the point of view of compactifications on tori, these coordinates
correspond to the Fourier duals to the generalized momenta
${\cal P}^M$ (\ref{genmomenta}). However, here we will consider more generally a
background independent formulation \cite{Backgroundindependent} in which the
generalized metric \cite{Generalizedmetric} can be defined on more general
backgrounds.

It is important to recall that here the coordinates can either parameterize
compact or non-compact directions indistinctively. Even if non-compact, one can
still formulate a full $O(D,D)$ covariant theory.  In this case, the duals to
the non-compact directions are just ineffective, and one can simply assume that
nothing depends on them. This will become clear later, when we consider DFT in
the context of four-dimensional effective theories. For the moment, this
distinction is irrelevant.

Being in the fundamental representation, the coordinates rotate under
$O(D,D)$
as follows
\be
X^M \to h^M{}_N \ X^N \ , \ \ \ \ \ \ h \in O(D,D)\, ,
\ee
so they mix under these global transformations. Given that $x^i$ and
$\tilde x_i$ are related by T-duality, the later are usually referred to
as {\it
dual coordinates}. Under $O(D,D)$ transformations, the fields change as
follows
\be
{\cal H}_{MN} (X) \to h_M{}^P \ h_N{}^Q \ {\cal H}_{PQ}(h\ X) \ , \ \ \ \
\ d(X)
\to d(h\ X) \, .\label{OnnRotations}
\ee
In the particular case in which $h$ corresponds to  T-dualities
(\ref{Tdualities}) in isometric directions (i.e. in directions in which
the
fields have no coordinate dependence), these transformations reproduce the
Buscher rules (\ref{Buscher}) and (\ref{gendil}) for $g_{ij}$, $b_{ij}$ and
$\phi$. It can be
shown that the different components of (\ref{OnnRotations}) are
equivalent to (\ref{Buscher}), which were derived assuming
T-duality is
performed along an isometry. More generally, the  transformation rules
(\ref{OnnRotations}) admit the possibility of performing  T-duality in
non-isometric directions \cite{doublegeom}, the reason being that DFT is defined
on a double
space, so contrary to what happens in supergravity, if a T-duality hits a
non-isometric direction the result is simply that the resulting
configuration
will depend on the T-dual coordinate.

The reader might be quite confused at this point,   wondering
what
these dual coordinates correspond to in the supergravity picture. Well,
they
simply have no meaning from a supergravity point of view. Then, there must be
some
mechanism to constrain the coordinate dependence, and moreover since we
want a
T-duality invariant formulation, such constraint must be duality
invariant. The
constraint in question goes under many names in the  literature, the most
common
ones being {\it strong constraint} or {\it section condition}. This
restriction
consists of a differential equation
\be
\eta^{MN} \partial_M \partial_N (\dots) = 0
\ee
where $\eta_{MN}$ is the $O(D,D)$ invariant
metric
introduced in (\ref{eta}).
For later convenience we recast it as
\be
Y^M{}_P{}^N{}_Q \ \partial_M \partial_N (\dots)= 0\, ,\label{strongconstraint}
\ee
where we have introduced the tensor
\be
Y^M{}_P{}^N{}_Q = \eta^{MN} \eta_{PQ} \label{Ytensor}
\ee
following the notation in \cite{Uduality2}, which is very useful to explore generalizations of DFT to more general U-duality groups, as we will see in Section \ref{sec:EFT}. The dots in (\ref{strongconstraint}) represent
any
field or gauge parameter, and also products of them. Notice that since the
tensor $Y$ is an $O(D,D)$ invariant, so is the constraint. This means
that if a
given configuration solves the strong constraint, any T-duality
transformation
of it will also do. When written in components, the constraint takes the
form
\be
\tilde \partial^i \partial_i (\dots) = 0\, ,
\ee
 so a possible solution is $\tilde \partial^i (\dots) = 0$, or any
$O(D,D)$
rotation of this. Actually, it can be proven that this is the
only
solution. Therefore, even if formally in this formulation the fields
depend on
the double set of coordinates, when the strong constraint is imposed the
only
possible configurations allowed by it depend on a $D$-dimensional section
of the
space. When this section corresponds to the  $x^i$ coordinates of
supergravity
(i.e., when all fields and gauge parameters are annihilated by $\tilde
\partial^i$), we will say that the strong constraint is solved in the {\it
supergravity frame}.

When DFT is evaluated on tori, a weaker version of the strong constraint can be
related to the LMC (\ref{LMC}). In this case, the generalized fields must be
expanded in the modes of the double torus $\exp(i X^M {\cal P}_M)$, such that when the
derivatives hit the mode expansion, the LMC contraction ${\cal P}^M {\cal P}_M$ makes its
appearance. Here we will pursue background independence, and moreover we will
later deal with twisted double-tori only up to the zero mode, so the level
matching condition should not be identified with the strong constraint (or any
weaker version) in this review. We will be more specific on this point in Section {\ref{sec:Dimensional reductions}}.

 Throughout this review we will not necessarily impose the strong
constraint, and in many
occasions we will explicitly write the terms that would vanish when it is
imposed.
The reader can choose whether she wants to impose it or not. Only when we
intend
to compare with supergravity in $D$ dimensions we will explicitly impose the
strong
constraint and
choose the supergravity frame (in these cases we will mention this
explicitly).
The relevance of dealing with  configurations that violate the strong
constraint will become apparent
when we get to the point of analyzing dimensional reductions of DFT, and
the
risks of going beyond supergravity will be properly explained and
emphasized. Let us emphasize that DFT is a restricted theory though, so one cannot just relax it and consider generic configurations: the consistency constraints of the theory are imposed by demanding closure of the gauge transformations, as we will discuss later. These closure constraints are solved in particular by the solutions to the strong constraint, but other solutions exist, and then it is convenient to stay as general as possible.

\subsection{Generalized Lie derivative}

We have seen that the $D$-dimensional  metric and two-form field transform
under
diffeomorphisms (\ref{diffeossugra}) and that the two-form also enjoys a
gauge
symmetry (\ref{twoformgaugesugra}). These fields have been unified into a
single
object called generalized metric (\ref{gennmet}), and then one wonders
whether
there are generalized diffeomorphisms  unifying the usual diffeomorphisms (\ref{diffeossugra})
 and gauge
transformations (\ref{twoformgaugesugra}). Since the former are parameterized by a $D$-dimensional
vector,
and the latter by a $D$-dimensional one-form, one can think of considering
a
generalized gauge parameter
\be
\xi^M = (\tilde \lambda_i, \lambda^i)\, .
\ee
Then, the generalized diffeomorphisms and gauge transformations of the two-form can be unified as
\bea
{\cal L}_\xi e^{-2d} &=& \partial_M \left(\xi^M e^{-2d}\right)\, ,
\label{density}\\
{\cal L}_\xi {\cal H}_{MN} &=& L_\xi {\cal H}_{MN} + Y^R{}_M{}^P{}_Q\
\partial^Q
\xi_P\ {\cal H}_{RN} + Y^R{}_N{}^P{}_Q\ \partial^Q \xi_P\ {\cal H}_{MR}\, ,
\label{genLie}
\eea
where $L_\xi$ is the Lie derivative (\ref{Lie}) in $2D$-dimensions, and $Y$,  already
defined in (\ref{Ytensor}), measures the departure from conventional Riemannian
geometry. We see here that $e^{-2d}$ transforms as a density, and as such
it
will correspond to the integration measure when we deal with the action.
When the generalized metric is parameterized
as in (\ref{gennmet}) in terms of $g_{ij}$ and $b_{ij}$, and the strong
constraint is imposed in the supergravity frame (i.e., when $\tilde
\partial^i =
0$), the different components of (\ref{genLie}) yield
\bea
{\cal L}_\xi g_{ij} &=& L_\lambda g_{ij}\, , \\
{\cal L}_\xi b_{ij} &=& L_\lambda b_{ij} + 2\ \partial_{[i} \tilde
\lambda_{j]}\, ,
\eea
and then the local transformations of supergravity
(\ref{diffeossugra})-(\ref{twoformgaugesugra}) are recovered. The
generalization
of the usual Lie derivative with the addition of the term with $Y$ is not
only
essential in order to recover the standard transformations of the bosonic NS-NS sector of supergravity, but
also
to preserve the $O(D,D)$ metric
 \be
 {\cal L}_\xi\ \eta_{MN} = 0\, .
 \ee

To end this discussion, we present
the general form of the generalized Lie derivative
with
respect to a vector $\xi$ acting on a tensorial density $V^M$ with weight
$\omega(V)$, which is given by the following gauge transformation
\be
{\cal L}_\xi V^{M} = \xi^P \partial_P V^{M} +
(\partial^M \xi_P - \partial_P \xi^M) V^P + \omega(V) \partial_P \xi^P \
V^M \, .\label{gendiffs}
\ee
This expression is trivially extended to other tensors with different
index structure. In particular, when this is applied to $e^{-2d}$ with
$\omega(e^{-2d}) = 1$ and ${\cal H}_{MN}$ with $\omega({\cal H}) = 0$, the
transformations (\ref{density}) and (\ref{genLie})  are respectively
recovered. As we will discuss in the next section, closure of these
generalized diffeomorphisms
  imposes  differential constraints on the theory.

Let us finally highlight that the action of these generalized diffeomorphisms has been defined when transforming tensorial quantities. Notice however that, for example, the derivative of a vector $\partial_M V^N$ is non-tensorial. It is then instructive to denote its transformation as
\be
\delta_\xi (\partial_M V^N) = \partial_M (\delta_\xi V^N) = \partial_M ({\cal L}_\xi V^N)
\ee
where we have used that the transformation of a vector is dictated by the generalized Lie derivative (\ref{gendiffs}).
One can however extend the definition of the generalized diffeomorphisms ${\cal L}_\xi$, to act on non-tensorial
quantities as if they were actually tensorial. Since $\delta_\xi$ represents the actual transformation,
one can define the failure of any object to transform covariantly as
\be
\Delta_\xi \equiv \delta_\xi - {\cal L}_\xi \, ,\label{failure}
\ee
such that when acting on  tensors, say $V^M$, one finds
\be
\Delta_\xi V^M = \delta_\xi V^M - {\cal L}_\xi V^M = 0\, ,
\ee
or equivalently, for any non-tensorial quantity, say $W^M$, we have
\be
\delta_\xi W^M = {\cal L}_\xi W^M + \Delta_\xi W^M\, .
\ee
This notation is very useful for the analysis of the consistency constraints of the theory, to which we now move.
\subsection{Consistency constraints} \label{sec:consistency}

Given the structure of generalized diffeomorphisms (\ref{genLie}), one must
check
that
they actually define a closed group \cite{Hull:2009zb}. This requires, in
particular, that
two successive gauge transformations parameterized by $\xi_1$ and $\xi_2$,
acting
on a given field $\xi_3$, must reproduce a new gauge transformation
parameterized by some given $\xi_{12} (\xi_1,  \xi_2)$ acting on the same
vector
\be
\Delta_{123}{}^M = - \Delta_{\xi_1} \left({\cal L}_{\xi_2} \xi_3^M\right) =
\left([{\cal L}_{\xi_1}, \ {\cal L }_{\xi_2}] - {\cal
L}_{\xi_{12}}\right) \xi_3^M = 0 \, ,\label{closure}
\ee
where we have defined $\Delta_\xi$ as in (\ref{failure}). In other words, the
generalized Lie derivative must send tensors into tensors. The resulting
parameter is given by
\be
\xi_{12} = {\cal L}_{\xi_1} \xi_2\, ,
\ee
provided the following constraint holds
\be
\Delta_{123}{}^M = Y^P{}_R{}^Q{}_S \ \left(2 \partial_P \xi^R_{[1}\
\partial_Q
\xi^M_{2]}\ \xi_3^S - \partial_P\xi^R_1\ \xi_2^S\ \partial_Q
\xi_3^M\right) = 0\, .
\label{consistency}
\ee
This was written here for vectors with vanishing weight, for simplicity.
The parameter $\xi_{12}$ goes under the name of D-bracket, and its
antisymmetric
part is named C-bracket
\be
\xi_{[12]}^M = [[\xi_1,\ \xi_2]]^M = \frac 1 2 ({\cal L}_{\xi_1} \xi_2^M -
{\cal
L}_{\xi_2} \xi_1^M) = [\xi_1,\ \xi_2]^M + Y^M{}_N{}^P{}_Q\ \xi_{[1}^Q
\partial_P
\xi_{2]}^N\, .
\ee
It corresponds to an extension of the Lie bracket (\ref{Lie}), since it contains a
correction proportional to the invariant $Y$, which in turn corrects the
Lie
derivative. Respectively, the D and C-bracket reduce to the Dorfman and
Courant
brackets \cite{Courant} when the strong constraint is imposed in the supergravity frame.
Under the constraint (\ref{consistency}), the following relation holds
\be
[{\cal L}_{\xi_1}, \ {\cal L }_{\xi_2}] = {\cal L}_{[[\xi_1,\ \xi_2]]}\, .
\label{ConsistencyAntisymm}
\ee
Notice also that symmetrizing (\ref{closure}), we find the so-called
Leibniz
rule (which arises here as a constraint)
\be
{\cal L}_{((\xi_1,\ \xi_2))} = 0\  , \ \ \ \ \ ((\xi_1,\ \xi_2)) =
\xi_{(12)}\, .
\label{Leibniz}
\ee
The  D-bracket, which satisfies the Jacobi identity, then contains a
symmetric
piece that must generate trivial gauge transformations. This fact is important
because on the other hand, the C-bracket, which is antisymmetric, has a
non-vanishing Jacobiator
\be
J(\xi_1, \xi_2,\xi_3) = [[ [[\xi_1,\ \xi_2]],\ \xi_3]] + {\rm cyclic}\, .
\ee
However, using (\ref{ConsistencyAntisymm}) and (\ref{Leibniz}), one can
rapidly
show that \cite{Uduality2}
\be
J(\xi_1, \xi_2,\xi_3) = \frac 1 3 (( [[\xi_1,\ \xi_2]],\ \xi_3)) + {\rm
cyclic}\, ,
\ee
and then the Jacobiator  generates trivial gauge transformations by virtue
of (\ref{Leibniz}).

The condition (\ref{closure})  poses severe consistency constraints on the
generalized
diffeomorphisms (i.e. their possible generalized gauge parameters).
 Therefore, DFT is a {\it constrained} or {\it restricted}
theory.
The generalized gauge parameters cannot be generic, but must be
constrained to
solve (\ref{closure}). Supergravity is safe from this problem, because
the
usual $D$-dimensional diffeomorphisms and two-form gauge transformations do form
a
group. It is then to be expected that under the imposition of the strong
constraint, (\ref{closure}) is automatically satisfied. This is trivial
from
(\ref{consistency}) because all of its terms are of the form
(\ref{strongconstraint}), but more generally these equations leave room
for
strong constraint-violating configurations
\cite{Aldazabal:2011nj,Geissbuhler:2011mx,GDFT}, as we will see later.

\subsection{The action}

The NS-NS sector of DFT has an action from which one can derive equations of
motion. Before
showing
its explicit form, let us introduce some objects that will be useful
later. The
generalized metric can be decomposed as
\be
{\cal H}_{MN} = E^{\bar A}{}_M\ S_{\bar A \bar B}\ E^{\bar B}{}_N\, ,
\ee
with an $O(D,D)$ generalized frame
\be
\eta_{MN} = E^{\bar A}{}_M \ \eta_{\bar A \bar B} \ E^{\bar B}{}_N\, ,
\ee
where $\eta_{\bar A \bar B}$ raises and lowers flat indices and takes the
same form as $\eta_{MN}$ (\ref{eta}). The generalized frame
 $E^{\bar A}{}_M$ transforms as follows under generalized diffeomorphisms
 \be {\cal L}_\xi E^{\bar A}{}_M = \xi^P \partial_P E^{\bar A}{}_M + (\partial_M \xi^P - \partial^P \xi_M)\ E^{\bar A}{}_P\ ,
 \ee
and can be parameterized in terms
of
the vielbein of the $D$-dimensional metric $g_{ij} = e^{\bar a}{}_i
s_{\bar a
\bar b} e^{\bar b}{}_j$, where $s_{\bar a \bar b} = {\rm diag}(-+\dots+)$
is the
$D$-dimensional Minkowski metric, as
\be
E^{\bar A}{}_M = \left(\begin{matrix}e_{\bar a}{}^i &  e_{\bar a}{}^j
b_{ji} \\
0 & e^{\bar a}{}_i\end{matrix}\right)\ , \ \ \ \ \ S_{\bar A \bar B} =
\left(\begin{matrix}s^{\bar a \bar b} & 0 \\ 0 & s_{\bar a \bar
b}\end{matrix}\right) \, .\label{genFrame}
\ee
Since the Minkowski metric is invariant under Lorentz
transformations $O(1,D-1)$, the metric $S_{\bar A \bar B}$ is invariant under
double
Lorentz transformations
\be H = O(1,D-1)\times O(1,D-1) \label{Htransf}\ee
 which correspond to
the
maximal (pseudo-)compact subgroup of $G = O(D,D)$. Therefore, the
generalized metric  is invariant under local double Lorentz
transformations, and
thus it parameterizes the coset $G/H$. The dimension of the coset is
$D^2$,
and this allows to accommodate a symmetric $D$-dimensional metric $g_{ij}$
 and an antisymmetric $D$-dimensional two-form $b_{ij}$, as we have seen. Technically, the triangular parametrization of the generalized frame would break down under a T-duality, and then one has to restore the triangular gauge through an H-transformation.
From the generalized frame $E_{\bar
A}{}^M$ and dilaton $ d$ one can build the {\it generalized fluxes}
\begin{eqnarray}
{\cal F}_{\bar A\bar B\bar C} &=& E_{\bar C M } {\cal L}_{E_{\bar A}}
E_{\bar B}{}^M = 3 \Omega_{[\bar A\bar B\bar
C]}\, ,\label{DynFabc}\\
{\cal F}_{\bar A} &=& - e^{2d}{\cal L}_{E_{\bar A}} e^{-2d} = \Omega^{\bar
B}{}_{\bar B\bar A} + 2 E_{\bar A}{}^M
\partial_M d\, ,\label{DynFa}
\end{eqnarray}
out of the following object
\be
\Omega_{\bar A\bar B\bar C} = E_{\bar A}{}^M \partial_M E_{\bar B}{}^N
E_{\bar C
N} = - \Omega_{\bar A \bar C \bar B}\, ,\label{Weizenbock}
\ee
that will be referred to as the generalized {\it Weitzenb\"ock
connection}.

Since all these objects are written in planar indices, they are manifestly
$O(D,D)$ invariant, so any combination of them will also be. The generalized
fluxes (\ref{DynFabc})-(\ref{DynFa}) depend on the fields and are therefore
dynamical. Later, when we analyze compactifications of the theory, they will play
an important role, for they will be related to the covariant quantities in the
effective action, and will moreover reduce to the usual constant fluxes, or
gaugings in the lower dimensional theory, hence the name {\it generalized fluxes}.

The generalized frame and dilaton enter in the action of DFT only through
the
dynamical fluxes (\ref{DynFabc}) and (\ref{DynFa}).
Indeed, up to total
derivatives, the action takes the form
\bea\label{ActionDFT}
  S \ = \ \int dX e^{-2d}\ {\cal R}\, ,
 \eea
 with
\bea\label{R}
	{\cal R} &=& {\cal F}_{\bar A\bar B\bar C}\ {\cal F}_{\bar D\bar
E\bar
F}\
	\left[\frac{1}{4} S^{\bar A\bar D} \eta^{\bar B\bar E} \eta^{\bar
C\bar
F}-\frac{1}{12} S^{\bar A\bar D} S^{\bar B\bar E} S^{\bar C\bar F} - \frac
1 6
\eta^{\bar A\bar D}\eta^{\bar B\bar E}\eta^{\bar C\bar F}\right] \nn\\
&& + \ {\cal F}_{\bar A } {\cal F}_{\bar B}\  \left[\vphantom{\frac 1
2}\eta^{\bar A \bar B} - S^{\bar A \bar B}\right]
\, .
\eea
In this formulation, it takes the same form as the scalar potential of
half-maximal
supergravity in four dimensions. We will be more specific about this later, but
for the readers who are familiar with gauged supergravities, notice that
identifying here the dynamical fluxes with gaugings and the $S_{\bar A \bar B}$
matrix with the moduli scalar matrix, this action resembles the form of the
scalar potential of \cite{Schon}. This frame formulation was introduced in
\cite{Siegel:1993th}, later related to
other formulations in \cite{framelikegeom}, and also discussed in
\cite{Exploring}.

Written in this form, the $O(D,D)$
invariance
is manifest. However, some local symmetries are hidden and the invariance
of the
action must be explicitly verified. Under generalized diffeomorphisms, the
dynamical
fluxes transform as
\bea
\delta_\xi {\cal F}_{\bar A\bar B\bar C} &=& \xi^{\bar D} \partial_{\bar
D}
{\cal F}_{\bar A\bar B\bar C} + \Delta_{\xi \bar A \bar B \bar C}\, ,\nn\\
\delta_\xi {\cal F}_{\bar A} &=& \xi^{\bar D} \partial_{\bar D} {\cal
F}_{\bar
A} + \Delta_{\xi \bar A} \, ,
\label{gaugetransfsfluxes}
\eea
where
\bea
\Delta_{\xi \bar A \bar B \bar C} &=& 4{\cal Z}_{\bar A\bar B\bar C\bar
D}\xi^{\bar
D}  +  3\partial_{\bar D} \xi_{[\bar A} \Omega^{\bar D}{}_{\bar B\bar
C]}\, ,\nn\\
\Delta_{\xi \bar A} &=&{\cal Z}_{\bar A\bar B}\xi^{\bar B} + {\cal
F}^{\bar B}
\partial_{\bar B} \xi_{\bar A} - \partial^{\bar B} \partial_{\bar B}
\xi_{\bar
A} + \Omega^{\bar C}{}_{\bar A\bar B} \partial_{\bar C} \xi^{\bar
B}\, ,\label{DeltasGaugeInv}
\eea
and we have defined
\bea
    {\cal Z}_{\bar A\bar B\bar C\bar D} &=& \partial_{[\bar A}{\cal
F}_{\bar
B\bar C\bar D]}-\frac{3}{4}{\cal F}_{[\bar A\bar B}{}^{\bar E} {\cal
F}_{\bar
C\bar D]\bar E}\ =\ -\frac{3}{4}\Omega_{\bar E[\bar A\bar B}\Omega^{\bar
E}{}_{\bar C\bar D]}\, ,\label{Zetas}\\
    {\cal Z}_{\bar A \bar B} &=& \partial^{\bar C} {\cal F}_{\bar C\bar
A\bar B}
+2\partial_{[\bar A}{\cal F}_{\bar B]}-{\cal F}^{\bar C} {\cal F}_{\bar
C\bar
A\bar B}\ =\ \left(\partial^M\partial_M E_{[\bar A}{}^N \right)E_{\bar
B]N}
              - 2\Omega^{\bar C}{}_{\bar A\bar B} \partial_{\bar C} d\, .\nn
\eea
The vanishing of (\ref{DeltasGaugeInv})  follows from the closure
conditions (\ref{closure}), precisely because the dynamical fluxes are
defined through generalized diffeomorphisms (\ref{DynFabc}), (\ref{DynFa})
\bea
\Delta_{\xi} {\cal F}_{\bar A \bar B\bar C} &=& \Delta_{\xi  \bar A \bar B
\bar C} = E_{\bar C M } \Delta_\xi ({\cal L}_{E_{\bar A}} E_{\bar B}{}^M)
= 0\, ,\nn \\
\Delta_{\xi} {\cal F}_{\bar A} &=& \Delta_{\xi \bar A} = - e^{2d}
\Delta_\xi ({\cal L}_{E_{\bar A}} e^{-2d})  = 0\, .\label{SclarFluxes}
\eea
Therefore, the dynamical fluxes in flat indices  transform as scalars
under generalized diffeomorphisms.

Let us now argue that due to the closure constraints (\ref{SclarFluxes}),
the action
of DFT is invariant under generalized diffeomorphisms. In fact, since
$e^{-2d}$
transforms as a density (recall (\ref{density}))
 \be
 \delta_\xi e^{-2d} = \partial_P (\xi^P e^{-2d})\, ,\label{e2ddensity}
 \ee
 for the action to be invariant under generalized diffeomorphisms, $\cal R$ must
transform as a scalar. Using the gauge transformation rules for the
generalized
fluxes (\ref{gaugetransfsfluxes}) together with (\ref{SclarFluxes}) one
arrives at the following result
\be
\delta_\xi {\cal R} = {\cal L}_\xi {\cal R} = \xi^P\partial_P {\cal R}\, .
\label{Rtransform}
\ee
 Combining (\ref{e2ddensity}) with
(\ref{Rtransform}), it can be checked that the Lagrangian density $e^{-2d}\
{\cal
R}$ transforms as a total derivative, and then the action
(\ref{ActionDFT}) is
invariant.

We have seen that in addition to generalized diffeomorphisms, the theory must be
invariant under local double Lorentz transformations (\ref{Htransf})
parameterized by an infinitesimal $\Lambda_{\bar A}{}^{\bar B}$. This
parameter
must be antisymmetric $\Lambda_{\bar A\bar B} = - \Lambda_{\bar B\bar A}$
to
guarantee the invariance of  $\eta_{\bar A \bar B}$ and it must also
satisfy $S_{\bar
A}{}^{\bar C}\Lambda_{\bar C \bar B} = \Lambda_{\bar A\bar C} S^{\bar
C}{}_{\bar
B}$ to guarantee the invariance of $S_{\bar A \bar B}$. The frame
transforms as
\be
\delta_\Lambda E_{\bar A}{}^M = \Lambda_{\bar A}{}^{\bar B} E_{\bar B}{}^M\, , \label{BeinTransfH}
\ee
and this guarantees that the generalized metric is invariant. The
invariance of
the action is, however, less clear, and a short computation shows that
\be
\delta_\Lambda S= \int dX \ e^{-2d}\ {\cal Z}_{\bar A\bar C}\
\Lambda_{\bar
B}{}^{\bar C}\ (\eta^{\bar A \bar B} - S^{\bar A\bar B})\, ,
\ee
with ${\cal Z}_{\bar A \bar B}$ defined in (\ref{Zetas}). Then, the
invariance
of the action (\ref{ActionDFT}) under double Lorentz transformations (\ref{BeinTransfH}) is
also guaranteed from closure, since
\be
{\cal Z}_{\bar A\bar B} =  \Delta_{E_{\bar A}} {\cal F}_{\bar B} = 0\, .
\ee
As happens with all the constraints in DFT, which follow from (\ref{closure}),
they are solved by the strong
constraint but admit more general solutions (this can be seen  especially in  
(\ref{consistency}) where cancelations could occur without demanding each
contribution to vanish independently).

This flux formulation of DFT is a small extension of the
generalized metric formulation introduced in \cite{Generalizedmetric}. It
incorporates terms that would
vanish
under the imposition of the strong constraint in a covariant way.
After some algebra, it can be shown that the action (\ref{ActionDFT}) can
be
recast in the form
 \be\label{ActionDFTgenmet}
 \begin{split}
  S = \int dX e^{-2d} \bigg(  & 4\,{\cal H}^{MN}\partial_{M}\partial_{N}d
  -\partial_{M}\partial_{N}{\cal H}^{MN}  -4\,{\cal
H}^{MN}\partial_{M}d\,\partial_{N}d
   + 4 \partial_M {\cal H}^{MN}  \,\partial_Nd\\[1.0ex]
    ~&\!\!\!\!\!\!\!\!\!\!\!+\frac{1}{8}\,{\cal H}^{MN}\partial_{M}{\cal
H}^{KL}\,
  \partial_{N}{\cal H}_{KL}-\frac{1}{2}{\cal H}^{MN}\partial_{M}{\cal
H}^{KL}\,
  \partial_{K}{\cal H}_{NL} + \Delta_{({\rm SC}) } {\cal R}\bigg)\, ,
 \end{split}
 \ee
up to total derivatives. Here, we have separated all terms in
(\ref{ActionDFT})
that vanish under the imposition of the strong constraint
$\Delta_{(SC)}{\cal R}$, to facilitate the
comparison with the generalized metric formulation \cite{Generalizedmetric}.

To conclude this section, we recall that in order to recover the
supergravity
action (\ref{sugraaction}), the strong constraint must be imposed in the
supergravity frame. Then, when $\tilde \partial^i = 0$ is imposed on
(\ref{ActionDFTgenmet}), and the generalized metric is parameterized in
terms of
the $D$-dimensional metric and two-form as in (\ref{gennmet}), the DFT action
(\ref{ActionDFTgenmet})
reproduces (\ref{sugraaction}) exactly.

\subsection{Equations of motion}
The equations of motion in DFT were extensively discussed in \cite{EOMsKwak} for
different formulations
of the theory.
 For the flux formulation we have just presented, the variation of the action
with respect to $E_{\bar A}{}^M$ and to $d$
takes the form
\bea
    \delta_E S &=& \int d X\ e^{-2d}\  {\cal G}^{\bar A\bar B} \delta
E_{\bar
A\bar B}\, ,\\
    \delta_d S &=& \int d X\ e^{-2d}\  {\cal G} \delta d\, ,
\eea
where
\be
    \delta E_{\bar A\bar B} = \delta E_{\bar A}{}^M E_{\bar BM}= - \delta
E_{\bar B\bar A}\, ,
\ee
to incorporate the fact that the generalized bein preserves the $O(D,D)$
metric
(\ref{eta}).
It can easily be checked that the variations of the generalized fluxes are
given by
\bea
    \delta_E {\cal F}_{\bar A\bar B\bar C}     &=& 3\left(\partial_{[\bar
A}
\delta E_{\bar B\bar C]}+ \delta E_{[\bar A}{}^{\bar D} {\cal F}_{\bar
B\bar
C]\bar D}\right) \, ,\\
    \delta_E {\cal F}_{\bar A}       &=& \partial^{\bar B} \delta E_{\bar
B\bar
A}+ \delta E_{\bar A}{}^{\bar B}{\cal F}_{\bar B} \, ,\\
    \delta_d {\cal F}_{\bar A}         &=& 2\partial_{\bar A}\,\delta d\, .
\eea
We then obtain
\bea
    {\cal G}^{[\bar A\bar B]} 	&=& 2 (S^{\bar D [\bar A} - \eta^{\bar
D[\bar A}) \partial^{\bar B]} {\cal F}_{\bar D} + ({\cal F}_{\bar D} -
\partial_{\bar D}) \check{{\cal F}}^{\bar D[\bar A \bar B]} + \check{{\cal
F}}^{\bar C \bar D [\bar A} {\cal F}_{\bar C \bar D}{}^{\bar B]}    \, ,\\
    {\cal G}  &=& -2{\cal R}\, ,
\eea
where
\be
\check{\cal F}^{\bar A \bar B \bar C} =  \frac 3 2 {\cal F}_{\bar
D}{}^{\bar B \bar C} S^{\bar A \bar D} - \frac 1 2 {\cal F}_{\bar D \bar E
\bar F} S^{\bar A \bar D}S^{\bar B \bar E}S^{\bar C \bar F}- {\cal
F}^{\bar A \bar B \bar C}\, .
\ee
 The equations of motion are then
\be
    {\cal G}^{[\bar A \bar B]} = 0\ , \ \ \ \ \
    {\cal G} =0 \, .\label{EOMSbein}
\ee
Upon decomposing these equations in components, and standing in the
supergravity
frame of the strong constraint, one recovers the equations of motion of
supergravity (\ref{eom})-(\ref{eomd}), provided the generalized frame is parameterized
as in
(\ref{genFrame}).

For completeness let us also mention that had we varied the action in the
generalized metric formulation (\ref{ActionDFTgenmet}) with respect to the
generalized metric (and setting to zero the strong-constraint-like terms), we
would have found \cite{Generalizedmetric}, \cite{EOMsKwak}
\be
\delta_{\cal H} S = \int dX e^{-2d} \delta{\cal H}^{MN} {\cal K}_{MN}\, ,
\ee
with
\bea
{\cal K}_{MN} &=& \frac 18 \partial_M {\cal H}^{KL} \partial_N {\cal H}_{KL} -
\frac 1 4 (\partial_L - 2 (\partial_L d)) ({\cal H}^{LK} \partial_K {\cal
H}_{MN}) + 2 \partial_M \partial_N d \\
&& - \frac 1 2 \partial_{(M|} {\cal H}^{KL} \partial_L {\cal H}_{|N)K}  + \frac
1 2 (\partial_L - 2 (\partial_L d)) ({\cal H}^{KL} \partial_{(M} {\cal H}_{N)K}
+ {\cal H}^K{}_{(M|}\partial_K {\cal H}^L{}_{|N)})\, .\nn
\eea
Notice however, that the variations $\delta{\cal H}_{MN}$ are not generic, but
must be subjected to constraints inherited from (\ref{constraintsGenMet}). This
implies that only some projections of ${\cal K}_{MN}$ give the equations of
motion, through a generalized Ricci flatness equation:
\be
\widehat {\cal R}_{MN} = \hat{P}_{(M}{}^P \check{P}_{N)}{}^Q {\cal K}_{PQ} =
0\, , \label{GenMetEOMS}
\ee
where we introduced some projectors that will be useful in the following section
\be
\hat{P}_{MN} = \frac 1 2 (\eta_{MN} - {\cal H}_{MN}) \, , \ \ \ \ \
\check{P}_{MN} = \frac 1 2 (\eta_{MN} + {\cal H}_{MN})\, .\label{PbarP0}
\ee
Finally, imposing the strong constraint to (\ref{EOMSbein}) they can be taken to
the form (\ref{GenMetEOMS}).

These equations of motion will be revisited in the next section from a
geometrical point of view.

\newpage
\section{Double Geometry}
\label{sec: Double Geometry}
We have explored the basics of the bosonic NS-NS sector of DFT, starting from its degrees of freedom,
the
double space on which it is defined, its consistency constraints, the
action and
equations of motion, etc. In particular, the action was tendentiously
written in terms of a generalized Ricci scalar and the equations of motion
were
cast in a generalized Ricci flatness form. But, is there some underlying
geometry? Can DFT be formulated in a more fundamental (generalized)
geometrical way? It turns out that there is such a formulation, but it
differs
from the Riemannian geometry out of which General Relativity is constructed.
We
find it instructive to begin this section with a basic review of the notions of
Riemannian
geometry that will then be generalized for DFT.

\subsection{Riemannian geometry basics} \label{sec:Riemann}

Even though General Relativity follows from an action of the form
\be
S = \int dx\ \sqrt{g}\ R \, ,\label{ActionGR}
\ee
where $R$ is the Ricci scalar and $g$ the determinant of the metric, we
know
that there exists an underlying geometry out of which this theory can be
obtained.
The starting point can be taken to be the Lie derivative (\ref{Lie})
\be
L_\xi V^i = \xi^k \partial_k V^i - \partial_k \xi^i V^k \, .\label{LieRiemann}
\ee
 The derivative
of a
vector is non-tensorial under the diffeomorphisms (\ref{LieRiemann}), so one
starts by introducing a
covariant
derivative
\be
\nabla_i V^j = \partial_i V^j + \Gamma_{ik}{}^j V^k\, ,
\ee
defined in terms of a Christoffel connection $\Gamma$, whose purpose is to
compensate the failure of the derivative to transform as a tensor.
Therefore,
the failure to transform as a tensor under diffeomorphisms parameterized by
$\xi$,
denoted by $\Delta_\xi$, is given by
\be
\Delta_\xi \Gamma_{ij}{}^k =   \partial_i \partial_j \xi^k\, .
\ee
The torsion can be defined through
\be
T_{ij}{}^{k} \xi^i V^j = (L^{\nabla}_{\xi} - L_{\xi})V^k =  2
\Gamma_{[ij]}{}^{k} \xi^i V^j \, .\label{torsionRiemann}
\ee
The superscript $\nabla$ is just notation to indicate that, in the Lie
derivative,
the partial derivatives should be replaced by covariant derivatives. A
condition
to be satisfied in Riemannian geometry is covariant constancy of the
metric
$g_{ij}$. It receives the name of metric compatibility
\be
\nabla_i g_{j k}  =\partial_i g_{jk} - \Gamma_{i j}{}^l g_{l k} -
\Gamma_{i
k}{}^{l} g_{j l} = 0\, .
\ee
This fixes the symmetric part of the connection
\be
\Gamma_{(ij)}{}^k = \frac 1 2 g^{k l} (\partial_i g_{j l} + \partial_j
g_{i l} -
\partial_l g_{ij}) - g_{m (i} T_{j)l}{}^m g^{l k}\, .
\label{metriccompRiemann}
\ee
When the connection is torsionless
\be
T_{ij}{}^k = 2\Gamma_{[ij]}{}^k = 0\, ,
\ee
 it is named Levi-Civita. Notice that the Levi-Civita connection is
symmetric
and completely fixed by metric compatibility (\ref{metriccompRiemann}) in
terms
of the degrees of freedom of General Relativity, namely the metric $g_{ij}$
\be
\Gamma_{ij}{}^k = \frac 1 2 g^{k l} (\partial_i g_{j l} + \partial_j g_{i
l} -
\partial_l g_{ij})\, .\label{levicivitaRiemann}
\ee
Let us note that the Levi-Civita connection satisfies the partial
integration
rule in the presence of the measure $\sqrt{g}$
\be
\int dx\ \sqrt{g} \ U \nabla_i V^i = - \int dx \ \sqrt{g}\ V^i \nabla_i U\, ,
\ee
given that its trace satisfies
\be
\Gamma_{ki}{}^k = \frac 1 {\sqrt {g}}\partial_i \sqrt{g}\, .
\ee

~

In a vielbein formulation, one also introduces a spin connection $W_{i \bar
a}{}^{\bar b}$ so that
\be
\nabla_i e_{\bar a}{}^j  = \partial_i e_{\bar a}{}^j + \Gamma_{i k}{}^j
e_{\bar
a}{}^k - W_{i \bar a}{}^{\bar b}
e_{\bar b}{}^j\, ,
\ee
and compatibility with the vielbein: $\nabla_i e_{\bar a}{}^j  = 0$, relates
the
Christoffel connection with the Weitzenb\"ock connection
\be
\Omega_{\bar a \bar b}{}^{\bar c} = e_{\bar a}{}^{i}\ \partial_i e_{\bar
b}{}^j\
e^{\bar c}{}_j \, ,\label{WeizenbockRiemann}
\ee
through
\be
W_{i \bar a}{}^{\bar b} = \Omega_{\bar c \bar a}{}^{\bar b} e^{\bar c}{}_i
+
\Gamma_{i j}{}^k e_{\bar a}{}^j e^{\bar b}{}_k\, .
\ee
For future reference, we also introduce the notion of dynamical Scherk-Schwarz flux,
defined by the Lie derivative as
\be
e^{\bar c}{}_i  L_{e_{\bar a}} e_{\bar b}{}^i = f_{\bar a \bar b}{}^{\bar
c} = 2
\Omega_{[\bar a \bar b]}{}^{\bar c} \, .\label{spinconfluxes}
\ee
Notice the analogy with the generalized fluxes (\ref{DynFabc}) defined in terms of the generalized Lie derivative
(\ref{gendiffs}). Then, the projection of the torsionless spin connection to the space of
fluxes
(i.e., its antisymmetrization in the first two indices) is proportional to
the
fluxes, given that the projection of the Levi-Civita connection to this
space
vanishes. In fact, it can be shown that in general
\be
e_{\bar a}{}^i W_{i \bar b}{}^{\bar c} = \frac 1 2 \left(f_{\bar a \bar
b}{}^{\bar c} + s_{\bar a \bar d} s^{\bar c \bar e} f_{\bar e \bar
b}{}^{\bar
d}+ s_{\bar b \bar d} s^{\bar c \bar e} f_{\bar e \bar a}{}^{\bar
d}\right)\, .
\ee
Then, the spin connection is fully expressible in terms of dynamical Scherk-Schwarz fluxes.

~

Having introduced the connections and their properties, we now turn to
curvatures. The commutator of two covariant derivatives reads
\be
[\nabla_i,\ \nabla_j] V^k = R_{ijl}{}^k\ V^l - T_{ij}{}^l\ \nabla_l V^k\, ,
\ee
with
\be
R_{ijl}{}^k = \partial_i \Gamma_{jl}{}^k - \partial_j \Gamma_{il}{}^k +
\Gamma_{im}{}^k \Gamma_{jl}{}^m - \Gamma_{jm}{}^k
\Gamma_{il}{}^m\label{Riemann}
\ee
the Riemann tensor, which is covariant under Lie derivatives. It takes the
same
form when it is written in terms of the spin connection
\be
R_{ij\bar a}{}^{\bar b} = R_{ijk}{}^l e_{\bar a}{}^k e^{\bar b}{}_l =
\partial_i
W_{j\bar a}{}^{\bar b} - \partial_j W_{i\bar a}{}^{\bar b} + W_{i\bar
c}{}^{\bar
b} W_{j\bar a}{}^{\bar c} - W_{j \bar c}{}^{\bar b} W_{i\bar a}{}^{\bar c}\, ,
\ee
and it has the following properties in the absence of torsion
\be
R_{ijlk} = R_{ijl}{}^m g_{mk} = R_{([ij][lk])} \ , \ \ \ \ R_{[ijl]}{}^k =
0\, ,
\label{propertiesRiemann}
\ee
the latter known as Bianchi Identity (BI). The Riemann tensor is a
very
powerful object in the sense that it dictates how tensors are
parallel-transported, and
for this reason it is also known as the curvature tensor. Tracing the
Riemann
tensor, one obtains the (symmetric) Ricci tensor
\be
R_{ij} = R_{ikj}{}^k = R_{ji}\, ,
\ee
and tracing further leads to the Ricci scalar
\be
R = g^{ij} R_{ij}\, .
\ee
The later defines the object out of which the action of General Relativity
(\ref{ActionGR})
is
built, while the vanishing of the former gives the
equations of
motion
\be
R_{ij}  = 0\, .
\ee
This equation is known as {\it Ricci flatness}, and the solutions to these
equations are said to be {\it Ricci flat}. Note that the Riemann and Ricci
tensors and the Ricci scalar are completely defined for a torsionless and
metric
compatible connection in terms of the metric.

Before turning to the generalizations of these objects needed for DFT,
let us mention that combining the above results, the action of
General Relativity can be written purely in terms of dynamical Scherk-Schwarz fluxes as
\be
S = \frac 1 4 \int dx\ \sqrt{g} \ f_{\bar a \bar b}{}^{\bar c} f_{\bar d
\bar
e}{}^{\bar f} \ \left[4 \delta_{\bar c}^{\bar a}\delta_{\bar f}^{\bar d}
s^{\bar
b \bar e} - 2 \delta_{\bar f}^{\bar b}\delta_{\bar c}^{\bar e} s^{\bar a
\bar d}
- s^{\bar a\bar d} s^{\bar b \bar e} s_{\bar c \bar f}\right] \,
.\label{ActionFluxesRiemann}
\ee
This is also analog to the situation in DFT (\ref{R}).

\subsection{Generalized connections and torsion}

Some of the ingredients discussed in the last subsection, already found their
generalized analogs in
 previous sections. For example, the Lie derivative (\ref{LieRiemann}) has
already been extended to
its generalized version in double geometry in
(\ref{gendiffs}). The  Weitzenb\"ock connection (\ref{WeizenbockRiemann}) has
also been generalized in
(\ref{Weizenbock}), and out of it, so have the fluxes (\ref{DynFabc}) been
extended to (\ref{spinconfluxes}).
Moreover, the actions (\ref{ActionFluxesRiemann}) and (\ref{R}) were both shown
to be expressible in terms of fluxes.
 So, how far can we go? The aim of this section is to continue with the
comparison, in order to find
similarities and differences between the usual Riemannian geometry and double
geometry. This is
 mostly based on
\cite{Siegel:1993th,framelikegeom,GeometryPark,TypeIIWaldram,GeometryZwiebach}.

Having defined a generalized Lie derivative, it is natural to seek  a
covariant derivative. We consider one of the form
\be
\nabla_M V_{\bar A}{}^N  = \partial_M V_{\bar A}{}^N + \Gamma_{MP}{}^N
V_{\bar
A}{}^P - W_{M\bar A}{}^{\bar B} \ V_{\bar B}{}^N\, ,
\ee
with trivial extension to tensors with more indices. Here we have
introduced a
Christoffel connection $\Gamma$ and a spin connection $W$ whose
transformation
properties must compensate the failure of the partial derivative of a
tensor to
transform covariantly both under generalized diffeomorphisms and double Lorentz
transformations.

We can now demand some properties on the connections, as we did in the
Riemannian
geometry construction. Let us analyze the implications of the following
conditions:
\begin{itemize}
\item Compatibility with the generalized frame
\be
\nabla_M E_{\bar A}{}^N = 0 \, . \label{compgenframe}
\ee
As in conventional Riemannian geometry, this simply relates the Christoffel
connection
with the spin connection through
\be
W_{M \bar A}{}^{ \bar B} = E^{\bar C}{}_M\ \Omega_{\bar C\bar A}{}^{\bar
B} +
\Gamma_{MN}{}^P \ E_{\bar A}{}^N \ E^{\bar B}{}_P\, ,
\label{RelatedConnections}
\ee
where we have written the Weitzenb\"ock connection defined in
(\ref{Weizenbock}), which
is
totally determined by the generalized frame. Then, this condition simply
says
that if some components of the spin (Christoffel) connection were determined,
the
corresponding components of the Christoffel (spin) connection would also be.

\item Compatibility with the $O(D,D)$ invariant metric
\be
\nabla_M \eta^{PQ} = 2 \Gamma_M{}^{(PQ)} = 0 \, .\label{etacompatibility}
\ee
This simply states that the Christoffel connection must be antisymmetric
in its
two last indices
\be
\Gamma_{MNP} = - \Gamma_{MPN}
\ee
Notice that, since we have seen in (\ref{Weizenbock}) that the Weitzenb\"ock
connection satisfies this property as
well, due to (\ref{RelatedConnections}) so does the spin
connection
\be
W_{M\bar A \bar B} = - W_{M \bar B \bar A}\, .
\ee
\item Compatibility with the generalized metric
\be
\nabla_M {\cal H}^{PQ} = \partial_M {\cal H}^{PQ} + 2 \Gamma_{MR}{}^{(P}
{\cal
H}^{Q)R}  = 0\, .
\ee
Its planar variant $\nabla_M S^{\bar A \bar B} = 0$ is then automatically
guaranteed if compatibility with the generalized frame
(\ref{compgenframe}) is imposed.

The implications of the combined $O(D,D)$ and generalized metric
compatibilities is better understood through the introduction of the
following two
projectors (\ref{PbarP0})
\be
\hat{P}_{MN} = \frac 1 2 (\eta_{MN} - {\cal H}_{MN}) \, , \ \ \ \ \
\check{P}_{MN} = \frac 1 2 (\eta_{MN} + {\cal H}_{MN})\, ,\label{PbarP}
\ee
which satisfy the properties
\be
\hat{P}_M{}^Q \hat{P}_Q{}^N = \hat{P}_M{}^N \ , \ \ \ \ \ \check{P}_M{}^Q
\check{P}_Q{}^N = \check{P}_M{}^N \ , \ \ \ \ \ \hat{P}_M{}{}^N +
\check{P}_M{}^N = \delta_M{}^N\, .
\ee
Compatibility with both metrics then equals compatibility with these
projectors
\be
\nabla_M \hat{P}_N{}^Q = 0 \ , \ \ \ \ \nabla_M  \check{P}_N{}^Q = 0\, ,
\ee
which in turn implies
\be
\check{P}_N{}^R\ \hat{P}_S{}^Q\ \Gamma_{MR}{}^{S} = \hat{P}_{R}{}^Q \partial_M
\check{P}_N{}^R\, .
\ee
Then, compatibility with the generalized metric and $O(D,D)$ metric
combined
imply that only these projections of the connection are
determined.

\item Partial integration in the presence of the generalized density $e^{-2d}$ (\ref{e2ddensity})
\be
\int e^{-2d} U \nabla_M V^M = -  \int e^{-2d} V^M \nabla_M U \ \ \ \
\Rightarrow
\ \ \ \ \Gamma_{PM}{}^P = -2 \partial_M d\, .
\ee
Notice that if the generalized frame were compatible, this would imply in
turn
that
\be
E_{\bar C}{}^N \ W_{N\bar A}{}^{\bar C} = - {\cal F}_{\bar A}\, .
\label{traceconnection}
\ee

This requirement can also be considered as compatibility with the measure
$e^{-2d}$, provided a trace part in the covariant derivative is added when
acting on tensorial densities.

\item Vanishing torsion. The Riemannian definition of torsion
(\ref{torsionRiemann}) is tensorial with respect to the Lie derivative,
but not
under generalized diffeomorphisms. In order to define a covariant notion
of
torsion, one can mimic its definition in terms of the Lie derivative, and
replace it with the covariant derivative \cite{TypeIIWaldram}
\be
({\cal L}^{\nabla}_\xi - {\cal L}_\xi)V^M = {\cal T}_{PQ}{}^M \xi^P V^Q \
, \ \
\ \ \ {\cal T}_{PQ}{}^M = 2 \Gamma_{[PQ]}{}^M + Y^M{}_Q{}^R{}_S
\Gamma_{RP}{}^S\, .\label{GenTorsion}
\ee
This defines a covariant {\it generalized torsion}, which corrects the
usual
Riemannian definition through the invariant $Y$ defined in (\ref{Ytensor}),
which
in turn corrects the Lie derivative. Vanishing generalized torsion has the
following consequence on the Christoffel connection
\be
2\Gamma_{[PQ]}{}^M + \Gamma^M{}_{PQ} = 0\, .
\ee
If this is additionally supplemented with the $O(D,D)$ metric
compatibility
(\ref{etacompatibility}), one gets that the totally antisymmetric part of
the
Christoffel connection vanishes
\be
\Gamma_{[MNP]} = 0\ \ \ \ \ \ \Leftrightarrow \ \ \ \ \ \ 3 W_{[\bar A
\bar B
\bar C]} =  {\cal F}_{\bar A \bar B \bar C}\, ,\label{torsionless}
\ee
where the implication assumes generalized frame compatibility.
\item Connections determined in terms of physical degrees of freedom. Typically, under the imposition of the above constraints on the connections, only some of their components get determined in terms of the physical fields. In \cite{GeometryPark}, the connections were further demanded to live in the kernel of some projectors, allowing for a full determination of the connection. The prize to pay is that under these projections the derivative is ``semi-covariant'', i.e. only some projections of it behave covariantly under transformations.
 \end{itemize}
As we reviewed in the previous section, compatibility with the $O(D,D)$
metric
is absent in Riemannian geometry. There,  metric compatibility and
vanishing
torsion determine the connection completely, and moreover guarantee
partial
integration in the presence of the measure $\sqrt{g}$. Here, the measure
contains a dilaton dependent part, and then one has to demand in addition,
compatibility with the generalized dilaton. An agreement between
Riemannian
geometry and double geometry is that   vanishing (generalized) torsion
implies
that the projection of the spin connection to the space of fluxes is
proportional to the fluxes (\ref{spinconfluxes}) and (\ref{torsionless}).

Despite the many coincidences between Riemannian and double geometry,
there is a
striking difference. While in the former demanding  metric compatibility
and
vanishing  torsion determines the connection completely, in double
geometry
these requirements turn out to leave undetermined components of the
connection.
Only some projections of the connections are determined, such as the trace
(\ref{traceconnection}) and its full antisymmetrization
(\ref{torsionless}),
among others.

To highlight the differences and similarities between Riemannian and double
geometry,
we list in Table (\ref{RiemannVsDouble}) some of the quantities appearing in both
frameworks.

\begin{table}[h!]
\begin{center}
\scalebox{1}[1]{
\begin{tabular}{| c | c | c | }
\hline
 & {\rm Riemannian geometry} & {\rm Double geometry}  \\[1mm]
\hline \hline
Frame compatibility & $W = \Omega + \Gamma$ & $W= \Omega + \Gamma$
\\[1mm]\hline
$O(D,D)$ compatibility  & $-------$ & $\Gamma_{MNP} = - \Gamma_{MPN}$
\\[1mm]
  &  & $W_{M\bar A \bar B} = - W_{M \bar B \bar A}$  \\[1mm]\hline
Metric compatibility & $\partial_i g_{jk} = 2 \Gamma_{i (j}{}^l g_{k)l} $
&
$\partial_M {\cal H}_{PQ} = 2 \Gamma_{M (P}{}^N {\cal H}_{Q)N}$
\\[1mm]\hline
Vanishing torsion & $\Gamma_{[ij]}{}^{k} = 0$ & $\Gamma_{[MNP]} = 0$
\\[1mm]
  & $W_{[\bar a \bar b]}{}^{\bar c} = 2 f_{\bar a \bar b}{}^{\bar c}$ &
$W_{[\bar A \bar B \bar C]} = 3 {\cal F}_{\bar A \bar B \bar C}$
\\[1mm]\hline
Measure compatibility & $\Gamma_{ki}{}^k = \frac 1 {\sqrt g}\partial_i \sqrt{g}$
&
$\Gamma_{PM}{}^P = e^{2d} \partial_M e^{-2d}$ \\[1mm]
 & $W_{\bar b \bar a }{}^{\bar b} = f_{\bar b \bar a}{}^{\bar b}$ &
$W_{\bar B
\bar A}{}^{\bar B} = -{\cal F}_{\bar A}$ \\[1mm]\hline
Determined part & Totally fixed & Only some \\[1mm]
 & $\Gamma_{ij}{}^k = \frac 1 2 g^{k l} (\partial_i g_{j l} + \partial_j
g_{i l}
- \partial_l g_{ij})$ & projections \\[1mm]\hline
Covariance failure & $\Delta_\xi \Gamma_{ij}{}^k = \partial_i\partial_j
\xi^k$ &
$ \Delta_\xi \Gamma_{MNP} = 2\partial_{M}\partial_{[N} \xi_{P]} $ \\[1mm]
& & $\ \ \ \ \ +\  \Omega_{R NP}\Omega^R{}_{MS }\xi^S$ \\[1mm]
\hline
\end{tabular}
}
\end{center}
{\it \caption{A list of conditions is given for objects in
Riemannian
and double geometry, with their corresponding implications on the
connections.
Every line assumes that the previous ones hold. } \label{RiemannVsDouble}}
\end{table}
\subsection{Generalized curvature}
In this section we will assume that the generalized Christoffel and spin
connections satisfy all the conditions listed in Table \ref{RiemannVsDouble}. We
would
now like to seek  a generalized curvature. The first natural guess
would be
to consider the conventional definition of Riemann tensor (\ref{Riemann}) and
extend
it straightforwardly to the double space, namely
\be
R_{MNP}{}^Q = 2 \partial_{[M} \Gamma_{N]P}{}^Q +2 \Gamma_{[M|R}{}^Q
\Gamma_{|N]P}{}^R\, .
\ee
However, this does not work because this expression is non-covariant under
generalized
diffeomorphisms
\be
\Delta_\xi R_{MNP}{}^Q = 2 \Delta_\xi \Gamma_{[MN]}{}^R \Gamma_{RP}{}^Q +
{\rm
strong\ constraint}\, .\label{FailureGenRiemann}
\ee
In Riemannian geometry, this would be proportional to the failure of the
torsion
to be covariant, which is zero. Here however $\Gamma_{[MN]}{}^P$ is not
the
torsion, because as we have seen, it is non-covariant. This in turn
translates
into the non-covariance of the Riemann tensor. As explained above, one has to
resort
to a generalized version of torsion (\ref{GenTorsion})
 \be
 {\cal T}_{PQ}{}^M = 2 \Gamma_{[PQ]}{}^M + Y^M{}_Q{}^R{}_S \Gamma_{RP}{}^S
= 0\, .
\label{GenTorsion2}
 \ee
In addition, even if the first term in (\ref{FailureGenRiemann}) were
zero, we
would have to deal with the other terms taking the form of the strong
constraint if we were not imposing it from the beginning. For the moment, let us
ignore them, and we will come back
to them
later.
Notice that vanishing torsion (\ref{GenTorsion2}) implies
\be
\Delta_\xi R_{MNPQ} = - \Delta_\xi \Gamma_{RMN } \Gamma^R{}_{PQ} + {\rm
strong\
constraint}\, ,
\ee
and then it is trivial to check that the following combination
\be
{\cal R}_{MNPQ} = R_{MNPQ} + R_{PQMN} + \Gamma_{RMN}
\Gamma^{R}{}_{PQ} + {\rm strong\ constraint}
\ee
is tensorial up to terms taking the form of the strong constraint:
\be
\Delta_\xi {\cal R}_{MNPQ} = {\rm strong\ constraint}\, .
\ee

Taking the strong constraint-like terms into account, the full generalized
Riemann tensor is given by
\be
{\cal R}_{MNPQ} =  R_{MNPQ} + R_{PQMN} + \frac 1 {2D} Y^{R}{}_L{}^{SL}
\left(\Gamma_{RMN} \Gamma_{SPQ} - \Omega_{RMN} \Omega_{SPQ}\right)\, ,
\ee
and is now covariant up to the consistency constraints of the theory discussed in Section \ref{sec:consistency}.

Since the connection has undetermined components, so does this generalized
Riemann tensor. This combination of connections and derivatives does not
project
the connections to their determined part, so we are left with an
undetermined
Riemann tensor. The projections of the Riemann tensor with the projectors
(\ref{PbarP}) turn out to be either vanishing or unprojected as well. This
situation marks a striking difference with Riemannian geometry.

We can now wonder whether some traces (and further projections) of this
generalized Riemann tensor lead to sensible quantities, such as some
generalized
Ricci tensor related to the equations of motion of DFT (\ref{EOMSbein}),
or some
generalized Ricci scalar related to (\ref{R}). For this to occur, the traces
must
necessarily project the connections in the Riemann tensor in such a way
that
only their determined part survives.

Tracing the generalized Riemann tensor with the projector $\hat{P}$
(\ref{PbarP}), one can define a generalized notion of Ricci tensor
\be
 {\cal R}_{MN} = \hat{P}^{PQ} {\cal R}_{MPNQ}\, ,
\ee
from which the action of DFT and its equations of motion can be obtained from
traces and projections. Taking another trace one recovers the (already defined)
generalized Ricci
scalar
\be
{\cal R} = \frac 1 4 \hat{P}^{MN} {\cal R}_{MN}
\ee
that defines the action of DFT (it actually gives this tensor up to terms
that
constitute total derivatives when introduced in the action
(\ref{ActionDFT})).
On the other hand, the following projections of this new generalized Ricci
tensor contain the information on the equations of motion (\ref{EOMSbein})
\be
{\cal G}_{[MN]} =   \hat{P}_{[M}{}^{P} \check{P}_{N]}{}^Q \  {\cal
R}_{PQ} = 0\, .
\label{RicciEOMS}
\ee
It might be quite confusing that the projections of the generalized Ricci
tensor
yielding the equations of motion correspond to the vanishing of an
antisymmetric
tensor. However, there is a remarkable property of  matrices of the
form
(\ref{RicciEOMS})
\be
\begin{matrix}
\hat{P}_{M}{}^R  \check{P}_{N}{}^S {\cal R}_{RS}= 0&\Rightarrow &
\hat{P}_{[M}{}^R  \check{P}_{N]}{}^S {\cal R}_{RS} = 0& \Rightarrow & \hat
P_{Q}{}^M  \hat{P}_{[M}{}^R  \check{P}_{N]}{}^S {\cal R}_{RS} = 0\\
\Uparrow & &\Updownarrow & & \Downarrow\\
\hat P_Q{}^M\hat{P}_{(M}{}^R  \check{P}_{N)}{}^S {\cal R}_{RS}= 0&
\Leftarrow&
\hat{P}_{(M}{}^R  \check{P}_{N)}{}^S {\cal R}_{RS} = 0& \Leftarrow &
\hat{P}_{M}{}^R  \check{P}_{N}{}^S {\cal R}_{RS} = 0
\end{matrix}\, .
\ee
Therefore, the vanishing of the antisymmetric part of $\hat{P}_{M}{}^R
\check{P}_{N}{}^S {\cal R}_{RS}$ contains the same information as the
vanishing of the symmetric part. Then, one can alternatively define a
symmetric
generalized Ricci tensor whose vanishing yields the equations of motion as
well
\be
\widehat {\cal R}_{MN} =   \hat{P}_{(M}{}^R  \check{P}_{N)}{}^S\ {\cal
R}_{RS}
= 0\, .
\ee

We summarize some differences between the geometric quantities in
Riemannian and
double geometry in Table (\ref{RiemannVsDouble2}).
\begin{table}[h!]
\begin{center}
\scalebox{1}[1]{
\begin{tabular}{| c | c | c | }
\hline
 & {\rm Riemannian geometry} & {\rm Double geometry}  \\[1mm]
\hline \hline
Torsion & $T_{ij}{}^{k} =  2 \Gamma_{[ij]}{}^{k}$ & ${\cal T}_{MN}{}^{P} =
 2
\Gamma_{[MN]}{}^P + \Gamma^P{}_{MN}$  \\[1mm]\hline
Riemann tensor  & Determined & Undetermined  \\[1mm]
  & $R_{ijl}{}^k = 2\partial_{[i} \Gamma_{j]l}{}^k$ & ${\cal R}_{MNPQ}=
R_{MNPQ} + R_{PQMN} $ \\[1mm]
& $\ \ \ \ \  + 2 \Gamma_{[i|m}{}^k \Gamma_{|j]l}{}^m$& $+ \Gamma_{RMN}
\Gamma^R{}_{PQ} - \Omega_{RMN} \Omega^R{}_{PQ}$  \\[1mm]\hline
Ricci tensor  & Determined & Undetermined  \\[1mm]
  & $R_{ij} = R_{ikj}{}^k$ & ${\cal R}_{MN} = \hat{P}_P{}^Q {\cal R}_{MQN}{}^P$
\\[1mm]\hline
EOM & $R_{ij} = 0$ & $  \hat{P}_{(M}{}^R  \check{P}_{N)}{}^S\ {\cal
R}_{RS} =
0$  \\[1mm]\hline
Ricci Scalar & $R = g^{ij} R_{ij}$ & ${\cal R} = \frac 1 4 \hat{P}^{MN} {\cal
R}_{MN}$\\[1mm]
\hline
\end{tabular}
}
\end{center}
{\it \caption{A list of definitions of curvatures is given for Riemannian
and
double geometry. } \label{RiemannVsDouble2}}
\end{table}

An alternative to this approach was considered in \cite{GeometryBerman}, where
 only the Weitzenb\"ock connection is non-vanishing and  the spin connection is
set to zero.
The Weitzenb\"ock connection is torsionful, and the torsion coincides with the
generalized fluxes (\ref{DynFabc}). This connection is flat, and then the
Riemann tensor vanishes, but the dynamics is encoded in the torsion and one can
still build the DFT action and equations of motion from it, by demanding
$H$-invariance (\ref{Htransf}). Since the connection and torsion are fully determined, this
approach has the advantage of the absence of unphysical degrees of freedom.
This also has a general relativity analog with its corresponding similarities
and differences.

\subsection{Generalized Bianchi identities}

The generalized Riemann tensor satisfies the same symmetry properties as
in
Riemannian geometry (\ref{propertiesRiemann})
\be
R_{MNPQ} = R_{([MN][PQ])}\, ,
\ee
plus a set of Generalized BI
\be
{\cal R}_{[\bar A \bar B \bar C \bar D]} = {\cal Z}_{\bar A\bar
B\bar
C\bar D} = \partial_{[\bar A}{\cal F}_{\bar B\bar C\bar
D]}-\frac{3}{4}{\cal
F}_{[\bar A\bar B}{}^{\bar E} {\cal F}_{\bar C\bar D]\bar E}\, ,
\ee
which under the strong constraint in the supergravity frame simply become
the
BI of supergravity (\ref{threeform}) and
(\ref{propertiesRiemann}), as we will see later. Notice that due to the
consistency constraints (\ref{closure}) this vanishes as a in the usual
Riemannian case
\be
{\cal Z}_{\bar A \bar B \bar C \bar D} = \Delta_{E_{\bar A}} {\cal F}_{\bar B
\bar C \bar D} = E_{\bar D M} \Delta_{\bar A} {\cal L}_{E_{\bar B}} E_{\bar
C}{}^M = 0\, .
\ee
BI in DFT were extensively discussed in \cite{GeometryZwiebach}.
\newpage

\section{Dimensional reductions}
\label{sec:Dimensional reductions}
In order to make contact with four-dimensional physics, we have to address the
dimensional
reduction of DFT. Strictly speaking, we were already assuming that some directions were compact, but now make the distinction between compact and non-compact directions precise, and evaluate the dynamics around particular backgrounds. We
begin this section with a brief review of Scherk-Schwarz (SS) compactifications
of
supergravity \cite{Scherk:1979zr}, and then extend these ideas to
dimensionally reduce DFT to
four dimensions, following \cite{Aldazabal:2011nj}, \cite{Geissbuhler:2011mx}.
We show that the resulting effective action corresponds to the electric bosonic sector of
half-maximal gauged supergravity \cite{Schon} containing all duality orbits of electric fluxes,
including the non-geometric ones \cite{Dibitetto:2012rk}.

\subsection{Scherk-Schwarz compactifications} \label{secSS}

Let us briefly recall how    geometric fluxes emerge in
 SS compactifications of supergravity,
along the lines of  \cite{Kaloper:1999yr}.

Consider the NS-NS sector of
supergravity containing a $D$-dimensional metric $g_{ij} =
e_i{}^{\bar a} s_{\bar a \bar b} e_j{}^{\bar b}$, a two-form field $b_{ij}$ and
a dilaton $\phi$, all depending on $D$ space-time coordinates $x^i$ (we are
thinking of $D = 10$). We will refer to the $D-$dimensional theory  as the {\it
parent
theory}.
When dimensionally reduced to  $d =D-n$ dimensions,
the resulting lower
dimensional theory will be referred to as the {\it effective theory}.

 SS reductions
can be introduced as the following set of steps to be
performed in order to obtain the effective theory:
\begin{itemize}
\item Split coordinates
\be
x^i = (x^\mu, y^m)\, .
\ee
The coordinates $y^m$, $m = 1,\dots,n$ correspond to the  compact space
directions, while $x^\mu$, $\mu = 1,\dots,d$ are the space-time directions
of the effective theory. The former (latter) are called {\it internal}
({\it external}).

\item Split indices in fields and parameters. The original $D$-dimensional
theory enjoys a set of
symmetries and the fields belong to representations of these symmetries.
Upon compactification,  the parent symmetry groups will be broken to those
of the effective theory. The fields must then be decomposed into the
representations of the symmetry group in the lower dimensional theory
\be
g_{ij} = \left(\begin{matrix}g_{\mu \nu} + g_{pq} A^p{}_\mu A^q{}_\nu &
A^p{}_\mu g_{p n} \\ g_{m p} A^p{}_\nu & g_{mn}\end{matrix}\right)\, ,
\ee
\be
b_{ij} = \left(\begin{matrix}b_{\mu\nu} - \frac{1}{2} (A^p{}_\mu V_{p \nu}
- A^{p}{}_\nu V_{p \mu})  + A^{p}{}_\mu A^{q}{}_\nu b_{pq}& V_{n \mu} -
b_{n p } A^{p}{}_\mu\\ - V_{m \nu}+ b_{mp} A^p{}_\nu & b_{mn}
\end{matrix}\right)\, ,
\ee
i.e. into internal, external and mixed components.   Notice that here there is
an abuse of notation in that $g_{\mu\nu}$ are not the $\mu\nu$ components
of $g_{ij}$.

Also the parameters of gauge transformations must be split
\be
\lambda^i = (\epsilon^\mu,\ \Lambda^m) \ , \ \ \ \ \tilde \lambda_i =
(\epsilon_\mu,\ \Lambda_m)\, .
\ee

\item Provide a reduction ansatz. The
particular
dependence of the fields on the external and internal coordinates is of the form
\bea
g_{\mu\nu}  &=& \widehat g_{\mu\nu} (x)\ , \ \ \ \ \ \ \ \ \  \ \ \ \ \ \
\ \ \ \ \ \ b_{\mu\nu}  \ =\  \widehat b_{\mu\nu} (x) \, ,\label{ansatzSS}\\
A^m{}_\mu &=& u_a{}^m (y) \widehat A^a{}_\mu (x)\ , \ \ \ \ \ \ \ \ \ \ \
V_{m\mu} \ =\  u^a{}_m (y) \widehat V_{a\mu} (x) \, ,\nn\\
g_{mn} &=& u^a{}_m (y) u^b{}_n(y) \widehat g_{ab}(x)\ , \ \ \ \ \ b_{mn} \
=\ u^a{}_m (y) u^b{}_n(y) \widehat b_{ab}(x) + v_{mn}(y)\, ,\nn
\eea

and similarly for the dilaton $\phi = \widehat \phi(x)$. The procedure
even tells you what form this ansatz should have. If there is a global
symmetry in the theory, such as a shift in the two-form $b \to b + v$,
then one simply ``gauges'' the global symmetry by making it depend on the
internal coordinates $v \to v(y)$.
The $y$-dependent elements $u(y)$ and $v(y)$ are called twists. Once the
procedure is over, the dependence on internal coordinates will disappear,
but the information on the twists will remain in the form of
structure-like constants that will parameterize the possible deformations
of the effective action. For this reason, the twist matrices are taken to
be constant in the external directions,
because otherwise  Lorentz invariance would be explicitly broken by
these constants in the effective action. The hatted fields on the other
hand depend only on the external coordinates, and will therefore correspond to
the dynamical degrees of freedom in the effective action.
These are a $d$-dimensional metric $\widehat g_{\mu \nu}$ and a two-form
$\widehat b_{\mu \nu}$, plus $2n$ vectors $(\widehat A^a{}_{\mu},\
\widehat V_{a \mu})$, plus $n^2 + 1$ scalars $(\widehat g_{ab},\ \widehat
b_{ab},\ \widehat \phi)$.

The gauge parameters must be twisted as well
\be
\lambda^i = \left(\widehat \epsilon^\mu(x),\ u_a{}^m(y)\widehat
\Lambda^a(x)\right) \ , \ \ \ \ \tilde \lambda_i = \left(\widehat
\epsilon_\mu(x),\ u_m{}^a(y) \widehat\Lambda_a(x)\right)\, .
\ee

\item Identify residual gauge transformations.
The gauge transformations of the
parent supergravity theory are given by Lie derivatives (\ref{diffeossugra})
\be
L_\lambda V^i = \lambda^j \partial_j V^i - V^j \partial_j \lambda^i  \, ,
\ee
   plus gauge transformations of the two-form (\ref{twoformgaugesugra}).
Plugging the fields and gauge parameters with the SS form into these, one
obtains the resulting gauge transformations of the effective theory. For
example, taking $V^i = (\widehat v^\mu(x),\ u_a{}^m(y) \widehat v^a(x))$, one
gets
\be
L_\lambda V^\mu = \widehat \epsilon^\nu \partial_\nu \widehat v^\mu -
\widehat v^\nu \partial_\nu \widehat \epsilon^\mu \equiv \widehat
L_{\widehat \epsilon} \widehat v^\mu\, ,
\ee
and then the $d$-dimensional Lie derivative of the effective action is obtained.
Similarly,
\be
L_\lambda V^m = u_a{}^m \ \widehat L_{\widehat \lambda} \widehat V^a\, ,
\ee
where the resulting transformation is {\it gauged}
\be
\widehat L_{\widehat \lambda} \widehat V^a  =  L_{\widehat \lambda}
\widehat V^a + f_{bc}{}^a \widehat \Lambda^b \widehat v^c\, ,
\ee
 since it receives the contribution from the following combination of
twist matrices
 \begin{eqnarray}
   f_{ab}{}^c & =& u_a{}^m\ \partial_m u_b{}^n\ u^c{}_n \ - \  u_b{}^m\
\partial_m u_a{}^n\ u^c{}_n\, ,
 \end{eqnarray}
which takes the same form as the SS flux (\ref{spinconfluxes}).
 Even if these objects are defined in terms of the twist $u^a{}_m$,  which
is $y$-dependent, given that they appear in the residual transformations
and we look for a $y$-independent theory, one must impose the constraint
that they are constant. In the literature, these constants are
known as {\it metric fluxes}, since they correspond to the background
fluxes of the metric (notice that the twist $u^a{}_m(y)$ corresponds to
the internal coordinate dependence of the metric (\ref{ansatzSS})).

Pursuing this procedure with all the components of all the gauge
transformations, we find the gauge transformations for all the fields in
the effective action. To render the result readable, let us rearrange things in
a compact language. The gauge
parameters are taken to be of the form
 \be
 \widehat \xi = (\widehat \epsilon_\mu, \widehat \epsilon^\mu, \widehat
\Lambda^A) \ , \ \ \ \ \widehat \Lambda^A = (\widehat  \lambda_a,\
\widehat \lambda^a)\, ,
 \ee
 and similarly the vector fields
 \be
 \widehat A^A{}_\mu = (\widehat V_{a\mu}, \ \widehat
A^a{}_\mu)\label{Aeffective}
 \ee
and the scalars
\be
\widehat {\cal M}_{AB} \  \ = \
  \begin{pmatrix}    \widehat g^{ab} & -\widehat g^{ac}\widehat
b_{cb}\\[0.5ex]
 \widehat  b_{ac}\widehat g^{cb} & \widehat g_{ab}-\widehat b_{ac}\widehat
g^{cd}\widehat b_{db}\end{pmatrix}\, .\label{Meffective}
\ee
Then, the different gauge transformations, parameterized by the different
components of $\widehat \xi$ are inherited from the parent gauge
transformations, and take the form
\bea \delta_{\widehat \xi}\  \widehat g_{\mu\nu}&=&L_{\widehat \epsilon}\
\widehat g_{\mu\nu}\, ,\\
\delta_{\widehat \xi}\ \widehat b_{\mu\nu} &=&L_{\widehat \epsilon}\
\widehat b_{\mu\nu} + \left(\partial_\mu {\widehat \epsilon}_\nu -
\partial_\nu {\widehat \epsilon}_\mu \right)\, ,\\
\delta_{{\widehat \xi}}\ \widehat{A}^A{}_{\mu} &=&L_{\widehat \epsilon}\
\widehat{A}^A{}_{\mu} -\partial_\mu{\widehat \Lambda}^A  +
f_{BC}{}^{A}~{\widehat \Lambda}^B \widehat {A}^C{}_{\mu}\, ,\\
\delta_{{\widehat \xi}}\ \widehat {\cal M}_{AB} &=& L_{\widehat \epsilon} \
\widehat {\cal M}_{AB} +
f_{AC}{}^D~\widehat {\Lambda}^C \widehat {\cal M}_{DB} +
f_{BC}{}^D ~\widehat {\Lambda}^C \widehat {\cal M}_{AD}\, .\eea
Hence, we can readily identify the role of the different components of $\widehat
\xi$: $\widehat \epsilon^\mu$ are the diffeomorphism parameters, $\widehat
\epsilon_\mu$ generate gauge transformations of the two-form, and $\widehat
\Lambda^A$ the parameters of the gauge transformations for vectors. While here
we have made a great effort to unify all these transformations, in DFT this
unification is there from the beginning, as we will see later.

Here, we have introduced the ``gaugings'' or ``fluxes'' $f_{AB}{}^C$,
which  have the following non-vanishing components
\bea
f_{abc} &=& 3(\partial_{[a} v_{bc]} + f_{[ab}{}^d v_{c]d})\, ,\nn \\
f_{ab}{}^c &=& u_a{}^m\ \partial_m u_b{}^n\ u^c{}_n \ - \  u_b{}^m\
\partial_m u_a{}^n\ u^c{}_n\, ,\label{geometric fluxes}
\eea
while the rest of them vanish
\be
f_a{}^{bc} = 0 \ , \ \ \ \ f^{abc} = 0 \, .\label{NongeometricSS}
\ee
This compact way of writing the results assumes that indices are raised
and lowered with an $O(n,n)$ metric
\be
\eta_{AB} = \left(\begin{matrix} 0 & \delta^a{}_b\\ \delta_a{}^b & 0
\end{matrix}\right)\, .\label{etaOdd}
\ee
When written in the form $f_{ABC} = f_{AB}{}^D \eta_{DC}$ they are totally
antisymmetric $f_{ABC} = f_{[ABC]}$.

\item Obtain the $d$-dimensional effective action. When the SS ansatz is plugged
in the supergravity
action, the result is
\bea
S&=&\int dx \sqrt{\widehat g}e^{-2\widehat \phi} \left( {\bf R} ~+ ~
4 ~\partial ^\mu \widehat \phi\partial_\mu \widehat \phi -~
\frac14  \widehat {\cal M}_{AB}{\cal F}^{A\mu\nu}
{\cal F}^{B}{}_{\mu\nu} \right. \label{ActionSS}\\
&&~~~~~~~~~~~~~~~~~~~~~~~
 \left.-~\frac1{12}{\cal G}_{\mu\nu\rho}
{\cal G}^{\mu\nu\rho}
+~ \frac 18 D_{\mu} \widehat {\cal M}_{AB}D^{\mu}\widehat {\cal M}^{AB} +
V \right)\, .
\nn\eea
Here ${\bf R}$ is the
$d$-dimensional Ricci scalar, and we have defined the field
strengths as \bea {\cal F}^{A}{}_{\mu\nu}&=&\partial_\mu
\widehat A^{A}{}_{\nu}-\partial_\nu \widehat A^{A}{}_{\mu} -f_{BC}{}^A
\widehat A^B{}_\mu~\widehat A^C{}_\nu \, ,\cr {\cal
G}_{\mu\rho\lambda}&=&3
\partial_{[\mu} \widehat b_{\rho\lambda]}  -f_{ABC} \widehat A^A{}_\mu
\widehat A^B{}_{\rho}
\widehat A^C{}_{\lambda}
 + 3 \partial_{[\mu} \widehat A^A{}_{\rho}\widehat A_{\lambda]A} ,
\label{CurvaturesSS}
\eea
and a covariant derivative for scalars as
\bea D_{\mu} \widehat {\cal
M}_{AB}=\partial_\mu \widehat {\cal M}_{AB}- f_{AD}{}^C \widehat A^D{}_\mu
\widehat {\cal
M}_{CB} - f_{BD}{}^C \widehat A^D{}_\mu \widehat{\cal M}_{AC}\, .
\label{CovariantDerivativeSS}\eea
Also, due to the gaugings, a scalar potential arose
\be
V =
-~~ \frac14 f_{DA}{}^C~f_{CB}{}^{D}
\widehat {\cal M}^{AB}
 - \frac1{12}f_{AC}{}^E~f_{BD}{}^F \widehat{\cal M}^{AB} ~\widehat{\cal
M}^{CD} ~
\widehat{\cal M}_{EF} \ - \ \frac 1 6 f_{ABC} f^{ABC}
\label{ScalarPotSS}\, ,\ee
\end{itemize}
which strongly resembles the form of the DFT action (\ref{R}). Let us mention that we have actually considered a simplified ansatz. Lorentz invariance is also preserved if a warp factor is included in the reductions ansatz, which would turn on additional flux backgrounds of the form $f_A$, in which case the effective action would exactly coincide with the DFT action.

This concludes the introduction to the basic notions of SS
compactifications of supergravity. We should say that there exist different related Scherk-Schwarz compactifications, and
their distinction goes beyond the scope of this review. Also, the consistency of these reductions is subtle and by
no means automatic, and we refer to the literature for detailed discussions on these points (see for example
\cite{Samtleben:2008pe,Grana:2013ila}).

\subsection{Geometric fluxes} \label{secGeometricSS}

In the SS reduction  defined in (\ref{ansatzSS}),
we restricted to
the zero modes and truncated all the states of
the infinite tower of  Kaluza-Klein (KK) modes.
Had
we conserved them, the effective action would have been  more involved
and would have had towers of KK degrees of freedom. Typically, these modes
are neglected because their masses scale proportionally to the order of
each mode. If they were kept, other stringy states with comparable masses
should be kept as well, and the effective theory would have to be completed
with the corresponding contributions.

This can be more clearly seen in a toroidal compactification.
Indeed, notice that a
compactification on a torus with vanishing background of the two-form
corresponds to
taking $u^a{}_m = \delta^a_m$ and $v_{ab}
= 0$ in the SS procedure. In this case, (\ref{geometric fluxes})
 would give $f_{ABC} = 0$, i.e. we get an ungauged theory.
Recalling the mass spectrum of closed strings on tori (\ref{massformulaT})
and the fact that the winding modes decouple in the field theory limit,
we see that the zero mode of such a compactification is massless for
the fields
considered in supergravity  (with $N=\tilde N=1$). Had we kept states with $p\ne
0$,
to be consistent
we should have also taken into account other string excitations with comparable
masses.
Since all the fluxes vanish in this case, no masses can
be generated in the effective theory.

These compactifications on tori with vanishing form fluxes (i.e., configurations
with $f_{ABC} = 0$) present many phenomenological problems:

\begin{itemize}
\item The scalar potential vanishes, so {\it any} configuration of scalars
corresponds to a possible minimum of the theory. The moduli space is then
fully degenerate, and all scalars are massless. This poses a problem
because, on the one hand, there are no massless scalars in nature, and on
the other hand, the theory loses all predictability since one has the freedom to
choose any
vacuum of the effective theory.

\item Since the scalar potential vanishes, there is no way to generate a
cosmological constant in the lower-dimensional theory. This is contrary to
experimental evidence, which indicates that our universe has a tiny
positive cosmological constant, i.e. it is a de Sitter (dS) universe.

\item The gravitinos of the supersymmetric completion of the theory are
massless as well.  If we start with an ${\cal N } = 1$ theory in $D = 10$,
we would end with an ${\cal N} = 4$ theory in $d = 4$. This is too much
supersymmetry
and we have no possibility to break it.

\item Since the fluxes play the role of structure constants, their
vanishing implies that the gauge symmetries are abelian. Then, Standard-Model
like interactions
are not possible.
\end{itemize}

It is then clear that a torus compactification is not interesting from a
phenomenological point of view. The situation changes when the twists
$u^a{}_m (y)$ and $v_{ab}(y)$ are such that $f_{ABC}\neq 0$. We have seen
that they allow to turn on metric fluxes $f_{ab}{}^c$ (through $u^a{}_m$) and
two-form fluxes $f_{abc}$ (through $v_{ab}$) in (\ref{geometric fluxes}). The
appearance of these fluxes now generates a scalar potential (\ref{ScalarPotSS})
that
classically lifts the moduli space. This in turn generates masses for
scalars and gravitinos, renders the gauge symmetries non-abelian and
allows for the possibility of a cosmological constant. However, although
the phenomenological perspectives improved, it turns out that geometric
fluxes seem not to be enough for moduli stabilization and dS vacua, and then one
has to go beyond them. There
are a number of no-go theorems and evidence  \cite{dS} pointing in this
direction.

In the literature, the two-form and metric fluxes both go under the name
of  {\it geometric fluxes}, and are denoted
\bea
H_{abc} &\equiv& f_{abc}\  =\ 3(\partial_{[a} v_{bc]} +  f_{[ab}{}^d
v_{c]d}) \, ,\label{geometric fluxes2} \\
\omega_{ab}{}^c &\equiv& f_{ab}{}^c\ =\ u_a{}^m\ \partial_m u_b{}^n\
u^c{}_n \ - \  u_b{}^m\ \partial_m u_a{}^n\ u^c{}_n\, \nn
\eea
respectively.
Since T-dualities exchange metric and two-form components
(\ref{Buscher}),
they exchange these fluxes as well
\begin{equation}
H_{abc}\ \ {\stackrel{h^{(c)}} {\longleftrightarrow}}\ \ \omega_{ab}{}^c\, .
\label{TchainGeom}
\end{equation}

Let us now devote a few lines to give an interpretation of the SS procedure
in terms of a compactification. The SS ansatz (\ref{ansatzSS}) can be
interpreted as follows. The twists $u^a{}_m (y)$ correspond to the metric
background in the compact space, and $\widehat g_{ab}$ amount to
perturbations. The full internal metric reads
\be
g_{mn} = u^a{}_m (y) \widehat g_{ab}(x) u^b{}_n(y)\, .
\ee
When plugging this in the supergravity action, one obtains an effective
theory for the perturbations $\widehat g_{mn}$, that is deformed by
parameters that only depend on the background. Then, freezing the
perturbations as
\be
\widehat g_{ab} (x) = \delta_{ab} \   \ \ \ \ \Rightarrow \ \ \ \ \
g_{mn} = u^a{}_m (y) \delta_{ab} u^b{}_n(y)\, ,
\ee
gives the background on which one is compactifying, and the effective action
dictates the dynamics of the perturbations around the background. Similarly, the
perturbations of the two-form are given by $\widehat b_{ab}$, and freezing
them gives the corresponding two-form background
\be
\widehat b_{ab} (x) = 0 \   \ \ \ \ \Rightarrow \ \ \ \ \  b_{mn}
=v_{mn}(y)\, .
\ee
The twist matrices $u^a{}_m$ and $v_{mn}$ can then be interpreted as the
backgrounds associated to the vielbein and the two-form of the compact
space. From now on, when referring to backgrounds we shall assume that the
perturbations are frozen.

Let us now explore a very simple setting that gives rise to a flux for the
two-form,  $H_{abc}$
(later, we will consider all its T-duals).
This is the canonical example in the literature on (non-)geometric fluxes, and
it is nicely discussed in \cite{Wecht:2007wu}. Most of the terminology related
to (non-)geometric fluxes is taken from this example, so we find it instructive
to revisit it here. For simplicity, we consider a three-dimensional
 internal space, which
can be embedded in the full internal six-dimensional space. Consider a
compactification on a three-torus with a non-trivial two-form
\be
g_{mn} = \delta_{mn} \ , \ \ \ b_{23} = N y^1  \ \ \ \ \Leftrightarrow \ \
\ \ u_m{}^a = \delta^a_m \ , \ \ \ v_{23} = N y^1 \, .\label{backgroundT3H}
\ee
Plugging this in (\ref{geometric fluxes2}), we obtain
\be
H_{123} = N \ , \ \ \ \ \omega_{12}{}^3 = \omega_{23}{}^1 =
\omega_{31}{}^2 = 0\, ,
\ee
so a compactification on a torus with a non-trivial two-form field turns on a
$H$-flux in the effective action.

Since this background has  isometries  in the directions $y^2, y^3$, we can
perform a T-duality in one of these directions, lets say $h^{(3)}$, through the
Buscher rules
(\ref{Buscher}). Then, we obtain the background
\be
ds^2 = g_{mn} dy^m dy^n = (dy^1)^2 +(dy^2)^2 + (dy^3 + N y^1 dy^2)^2 \ , \
\ \ \ \ b_{mn} = 0 \, .\label{twisted torus}
\ee
This corresponds to
\be
u^a{}_m = \left(\begin{matrix} 1& 0 & 0 \\ 0& 1&0 \\ 0&  N y^1 &
1\end{matrix}\right) \ , \ \ \ \ v_{mn} = 0\, .
\ee
Plugging this in (\ref{geometric fluxes2}), we find that the fluxes turned
on in the effective action are now
\be
H_{123} = \omega_{23}{}^1 = \omega_{31}{}^2 = 0 \ , \ \ \ \
\omega_{12}{}^3 = N\, ,
\ee
in agreement with the T-duality chain (\ref{TchainGeom}). The background
(\ref{twisted torus}) is called {\it twisted torus}, and it generates metric
fluxes $\omega_{ab}{}^c$ upon compactifications.  In more general backgrounds,
SS
compactifications allow to turn on form and metric fluxes simultaneously,
provided the compactification is done on a twisted torus with a non-trivial
two-form background. Examples of different Scherk-Schwarz compactifications in
different scenarios, and their relation to gauged supergravity can be found in
\cite{FreedWitten}.

\subsection{Gauged supergravities and duality orbits}

The effective action (\ref{ActionSS}), obtained by means of a SS
compactification, is a
particular gauged supergravity. For a review of gauged supergravity see
\cite{Samtleben:2008pe}. These kind of dimensional reductions
preserves all the supersymmetry of the parent theory, and are therefore
highly constrained. When the starting point is $D = 10$
supergravity with ${\cal N} = 1$ supersymmetries ($16$ supercharges), the
$d = 4$ effective theory preserves all the supercharges and has
therefore ${\cal N} = 4$ supersymmetries. This corresponds to half the
maximal allowed supersymmetries, and so they are called {\it half-maximal gauged
supergravities}. These theories have been widely studied irrespectively of
their stringy higher dimensional origin, and the full set of possible
deformations have been classified in \cite{Schon}. Let us here review the basics
of the bosonic sector of $d =
4$ half-maximal gauged supergravity, so that we can then identify
particular gaugings as specific reductions in different backgrounds.

The bosonic field content of half-maximal gauged supergravity in four
dimensions consists of a metric $\widehat g_{\mu \nu}$, 12 vector fields
$\widehat A^A{}_\mu$ and 38 scalars, arranged in two objects: a complex
parameter $\tau
= e^{-2\widehat \phi} + i \widehat B_0$ and a scalar matrix $\widehat {\cal
M}_{AB}$ with 36 independent components parameterizing the coset
$O(6,6)/O(6)\times O(6)$.

There is an additional freedom to couple an arbitrary number $N$ of vector
multiplets but, for simplicity,  we will not consider this possibility (otherwise the global symmetry group would have to be extended to $O(D,D+N)$ \cite{Aldazabal:2011nj}).
Also, the global symmetry group contains an $SL(2)$ factor as well, related to S-duality, which
mixes electric and magnetic sectors. This is not captured by DFT (unless the global symmetry group is further extended to include S-duality) and then one can only obtain the electric sector.

The ungauged theory is invariant under an $O(6,6)$ global symmetry group,
and by ``ungauged'' we mean that the gauge group is the Abelian $U(1)^{12}$.
This group can however be rendered non-Abelian by gauging a subgroup of
$O(6,6)$. Given the $O(6,6)$ generators $(t_\alpha)_A{}^B$, with $\alpha =
1,\dots,66$; $A = 1,\dots,12$,  there is a powerful object named embedding
tensor
$\Theta_A{}^\alpha$ that dictates the possible gaugings of the theory. The
gauge group generators are given by $\Theta_A{}^\alpha (t_\alpha)_B{}^C$,
so $\Theta_A{}^\alpha$ establishes how the gauge group is embedded in the
global symmetry group. The ${\bf 12} \otimes {\bf 66}$ components of
$\Theta_A{}^\alpha$ are restricted by a linear constraint that leaves only
${\bf 12} + {\bf 220}$ components, parameterized by
\be
\xi_A \ , \ \ \ \ \ f_{ABC} = f_{[ABC]}\, ,
\ee
and these are further restricted by quadratic constraints
\bea
\xi_A \xi^A&=&0\, ,\\
\xi^C  f_{ABC}&=&0 \\
f_{E[AB} f^E{}_{CD]} &=& \frac 1 3 f_{[ABC} \xi_{D]}\, ,
\eea
necessary for gauge invariance of the embedding tensor (and closure of the algebra). The $O(6,6)$ indices are raised and
lowered with the invariant metric (\ref{etaOdd}).

In four dimensions, two-forms are dual to scalars. Dualizing the scalar
$\widehat B_0 \to \widehat b_{\mu \nu}$, the action of the electric
bosonic sector of half-maximal gauged supergravity then takes the form
(\ref{ActionSS}) when $\xi_A = 0$. We then see that the SS
compactification of $D = 10$  supergravity on a twisted torus
with two-form flux leads to a particular half-maximal gauged supergravity
in $d = 4$. The only possible deformations in that theory are given by
(\ref{geometric fluxes2}). From now on we will restrict to the gaugings
$f_{ABC}$ and set the rest of them to zero, i.e. $\xi_A = 0$, for simplicity.

The global symmetries of the ungauged theory amount to  $O(6,6)$
transformations
\be
\widehat A^A{}_\mu \to h_B{}^A \ \widehat A^B{}_\mu \ , \ \ \ \ \ \widehat
{\cal M}_{AB} \to h_A{}^C \ \widehat {\cal M}_{CD} \ h_B{}^D\, ,
\label{FieldRedefinition}
\ee
where the elements $h \in O(6,6)$ were introduced in Section
\ref{secTduality}. When the gaugings are turned on, the global symmetry
group is broken by them. However, $O(6,6)$ transformations do not change
the physics. In fact, given a configuration of gaugings $f_{ABC}$ with
their corresponding action (\ref{ActionSS}), any $O(6,6)$ rotation of them,
\be
f_{ABC} \to h_A{}^D h_B{}^E h_C{}^F f_{DEF}\, ,
\ee
would yield a different configuration with a corresponding different
action. However, through a field redefinition of the form
(\ref{FieldRedefinition}), this action can be taken to the original form.
In other words, we have the relation
\be
S\left[h_A{}^D h_B{}^E h_C{}^F f_{DEF}, \ \widehat A^A{}_\mu,\ \widehat
{\cal M}_{AB}\right] = S\left[f_{ABC},\ h_B{}^A   \widehat A^B{}_\mu,\
h_A{}^C   \widehat {\cal M}_{CD}  h_B{}^D\right]\, ,
\ee
and so an $O(6,6)$ transformation of the gaugings just amounts to a field
redefinition. Then, it corresponds to the same theory. For this reason, it is
not convenient to talk about configurations of gaugings, but rather of {\it
orbits} of gaugings. An orbit is a set of configurations  related by
duality transformations, so that different theories correspond to different
duality orbits of gaugings.

An intriguing feature of gauged supergravities is that they admit more
deformations than those that can be reached by means of geometric
compactifications on twisted tori with two-form flux. In fact, for generic
configurations, the embedding tensor has components
\be
Q_a{}^{bc} = f_a{}^{bc} \ , \ \ \ \ R^{abc} = f^{abc}\, ,
\ee
that cannot be turned on through the canonical SS compactification
(\ref{NongeometricSS}). Since the other set of gaugings $f_{abc}, f^a{}_{bc}$
were identified with the geometrical fluxes,
these are said to be {\it non-geometric} gaugings. Here we have named
them $Q$ and $R$ to match the standard parlance in the literature of flux
compactifications. One then wonders to what kind of backgrounds or
compactifications these gaugings would correspond to. As we will see, T-duality
has a very
concrete answer to this question.

Before moving to a discussion on non-geometric fluxes, let us briefly review the
arguments of \cite{stw,acfi} to invoke non-geometric fluxes from a string theory
perspective. In \cite{stw,acfi}, all supergravities in $D=10$ and
$11$ dimensions are compactified in a geometric sense to four dimensions. These
higher dimensional supergravities are the low energy limit of duality related
string theories, like for instance Type IIA and Type IIB
strings.
Each compactification gives rise to a fluxed effective action containing only
geometric fluxes.

When duality transformations are applied at the level of the four
dimensional effective action, one finds that,  although the parent theories are
connected by dualities, the effective theories are  not \cite{stw,acfi}. Thus, new
non-geometric fluxes have to be invoked so that the theories  match.
In this process, gaugings (or fluxes) that look geometric in one picture (duality frame), are non-geometric
in others, and all of them should be included in string compactifications in order
to preserve all the {\it stringy} information at the level of the effective
action. Moreover, when all the gaugings are considered together in the effective action, the
resulting  (super-)potential  includes {\it all} the possible deformations
(gaugings) of gauged supergravity.
All the T-dual deformations are captured by
generalized geometric  compactifications of DFT, as we will see.

We have seen in Section \ref{secGeometricSS} that starting with a toroidal
background with a two-form flux $H_{123}$ (\ref{backgroundT3H}), a $h^{(3)}$
T-duality can be performed in the direction $y^3$
leading to a twisted
torus with metric flux $\omega_{12}{}^3$ (\ref{twisted torus}). The latter
still has an isometry in the direction $y^2$,
so nothing prevents us from
doing a new T-duality, namely $h^{(2)}$. At the level of fluxes, the chain
would go as
\begin{equation}
H_{abc}\ \ {\stackrel{h^{(c)}} {\longleftrightarrow}}\ \ \omega_{ab}{}^c\
\ {\stackrel{h^{(b)}} {\longleftrightarrow}}\ \ Q_a{}^{bc}\, ,
\end{equation}
and so a compactification on the resulting background would turn on a
$Q_1{}^{23}$ flux in the effective action. Instead of using the Buscher
rules, we find it more instructive to T-dualize via the construction of a
generalized metric. For the twisted torus (\ref{twisted torus}) it takes
the form
\be
{\cal H}_{MN} = \left(\begin{matrix}  g^{mn} & - g^{mp} b_{pn}\\[0.5ex]
b_{mp} g^{pn} &  g_{mn}- b_{mp} g^{pq} b_{qn}\end{matrix}\right) =
\left(\begin{smallmatrix}  1 & 0 & 0 & 0 & 0 & 0 \\ 0 & 1 & -Ny^1 & 0 & 0 & 0
\\ 0 & -Ny^1 & 1 + (N y^1)^2 & 0 & 0 & 0 \\ 0 & 0 & 0 & 1 & 0 & 0 \\ 0 & 0
& 0 & 0 & 1 + (Ny^1)^2& N y^1 \\ 0 & 0 & 0 & 0 & N y^1 & 1
\end{smallmatrix}\right)\, .\nn
\ee
Now acting on this twisted torus background with a T-duality in the
direction $y^{2}$
\be
{\cal H}_{MN} \to h^{(2)}{}_M{}^P h^{(2)}{}_N{}^Q {\cal H}_{PQ} \ , \ \ \
\ \ h^{(2)} = \left(\begin{smallmatrix}  1 & 0 & 0 & 0 & 0 & 0 \\ 0 & 0 & 0 & 0 & 1
& 0 \\ 0 & 0 & 1 & 0 & 0 & 0 \\ 0 & 0 & 0 & 1 & 0 & 0 \\ 0 & 1 & 0 & 0 & 0
& 0 \\ 0 & 0 & 0 & 0 &0 & 1 \end{smallmatrix}\right)\, ,
\ee
we get
\be
h^{(2)}{}_M{}^P h^{(2)}{}_N{}^Q {\cal H}_{PQ} = \left(\begin{smallmatrix}  1 & 0 & 0
& 0 & 0 & 0 \\ 0 & 1 + (N y^1)^2 & 0 & 0 & 0 &  Ny^1 \\ 0 & 0 & 1 + (N
y^1)^2 & 0 & -Ny^1 & 0 \\ 0 & 0 & 0 & 1 & 0 & 0 \\ 0 & 0 & -Ny^1 & 0 & 1 &
0 \\ 0 & Ny^1 & 0 & 0 &0 & 1 \end{smallmatrix}\right) \, ,\label{Qbackground}
\ee
and from here we can obtain the  background metric
\be
ds^2 = g_{mn} dy^m dy^n = (dy^1)^2 + \frac 1 {1 + (Ny^1)^2} [(dy^2)^2 +
(dy^3)^2]
 \ee
 and the two-form
 \be
 b_{23} = - \frac{Ny^1}{1 + (N y^1)^2}
 \ee
associated to the $Q_{1}{}^{23}$ flux. This background only depends on $y^1$ in
the directions orthogonal to $y^1$, so this corresponds to a base coordinate. When
undergoing a monodromy $y^1 \to y^1 + 1$, the solution does not come back to
itself, but rather to an $O(2,2) \in O(3,3)$ rotation of it. Since this duality
element mixes the metric and the two-form in a non-trivial way, it is called a
T-fold \cite{doublegeom}. From a supergravity point of view, these backgrounds
are globally ill-defined because the T-duality element needed to ``glue'' the
two different coordinate patches is not an element of the geometric (i.e. diffeos + shifts (\ref{Tdualities})) subgroup of
$O(3,3)$. This background is then said to correspond to a {\it globally
non-geometric flux} $Q_a{}^{bc}$. Notice however that from the double-space
point of view there is no such global issue provided one allows
transitions with the full $O(3,3)$ symmetry group (including T-dualities (\ref{Tdualities})). In this case, the
identifications between the coordinates under monodromies involve the dual ones,
and the generalized bein is globally well defined on the double space.

If we intended to do a further T-duality \cite{stw}, say in the direction $y^{1}$,
 \begin{equation}
H_{abc}\ \ {\stackrel{h^{(c)}} {\longleftrightarrow}}\ \ \omega_{ab}{}^c\
\ {\stackrel{h^{(b)}} {\longleftrightarrow}}\ \ Q_a{}^{bc}\ \
{\stackrel{h^{(a)}} {\longleftrightarrow}}\ \ R^{abc}\, ,
\label{TdualityChain}
\end{equation}
 we would face the problem that we ran out of isometries. Therefore the
resulting background would have to depend on a ``dual'' coordinate and we
would lose any notion of locality in terms of the usual coordinates on
which supergravity is defined. For this reason, the fluxes $R^{abc}$ are usually
named {\it locally  non-geometric}. Clearly, again, this form of non-geometry is
not a problem in the double space either.

Notice that the chain (\ref{TdualityChain}) connects different
configurations of gaugings via T-duality. By definition, they all
correspond to the same orbit, so the four dimensional theory really
does not distinguish between compactifications on tori with two-form flux,
twisted tori or T-folds, that are connected by T-dualities. In this sense,
the orbit itself is basically geometric: if we were given an action with a
single flux, either $H$, $\omega$, $Q$ or $R$, we would always find a geometric
uplift and face its corresponding phenomenological problems. A different
situation would be that of an action containing both geometric {\it and}
non-geometric fluxes {\it simultaneously} turned on. T-duality would
exchange geometric with non-geometric fluxes, and it would never be able to get rid
of the non-geometric ones. This kind of configurations are said to
belong to a duality orbit of non-geometric fluxes \cite{Dibitetto:2012rk}, and they
cannot be reached by
means of a standard SS compactification of supergravity. They are actually
the most interesting
orbits since they circumvent all the no-go theorems preventing moduli
fixing, dS vacua, etc. \cite{dS}.

\begin{figure}[h]
\label{SSdiagram}
\centering
\begin{minipage}[b]{\linewidth}
\centering
\begin{displaymath}
\xymatrix@R=60pt{
\mbox{\framebox{\makebox[1.5\width]{Supergravity $D$-dim}}}
\ar@{-->}[rr]^-{O(D,D)}
\ar@{->}[d]_-{\mbox{Geometric SS}\ \ }
&& \mbox{\framebox{\makebox[6\width]{DFT}}}
\ar@{..>}[d]^-{\ \ \mbox{Generalized SS}} \\
\framebox{\makebox[1.2\width]{$\begin{matrix}\mbox {Gauged supergravity} \\
d\mbox{-dim (geometric fluxes)}\end{matrix}$}} \ar@{~>}[rr]_-{O(n,n)} &&
\framebox{\makebox[1.4\width]{$\begin{matrix}\mbox {Gauged supergravity} \\
d\mbox{-dim (all fluxes)}\end{matrix}$}} }
\end{displaymath}
\caption{We picture the logic of DFT compactifications \cite{Aldazabal:2011nj}.
While standard SS reductions from supergravity in $D = d + n$ dimensions (solid line)
give rise to gauged supergravity involving only geometric fluxes in $d$ dimensions, invoking
duality arguments at the level of the effective action one can conjecture the
need for dual fluxes \cite{stw,acfi} to complete all the deformations of gauged
supergravity (waved line). More fundamentally, DFT is the $O(D,D)$-covariantization of
supergravity (dashed line), and generalized SS compactifications of DFT give
rise to gauged supergravities with all possible deformations (dotted line).}
\end{minipage}
\end{figure}
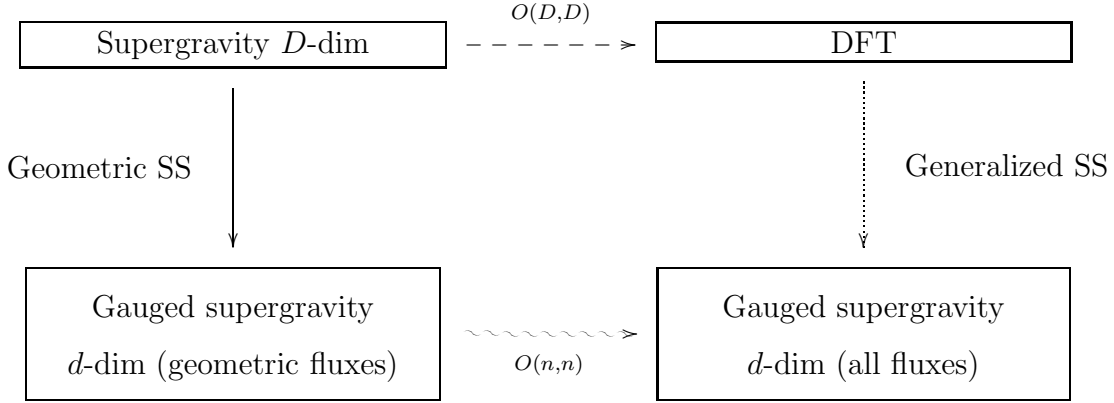

As we will see, being T-duality invariant and defined on a double space,
DFT is free from  global and local issues. Generalized SS compactifications
of DFT will be the topic of the forthcoming subsections. We will see that
DFT provides a beautiful geometric uplift of all duality orbits, including
the non-geometric ones. We anticipate the final picture in Figure
\ref{SSdiagram}.

\subsection{Generalized Scherk-Schwarz compactifications} \label{secGenSS}

Here we generalize the SS procedure in a duality
covariant way by applying it
 to DFT. This and the following sections are mostly based on
\cite{Aldazabal:2011nj,Geissbuhler:2011mx,GDFT}. Let us then follow the steps
introduced in Section
\ref{secSS}, although in a different order for convenience.

\begin{itemize}
\item In the double space we have coordinates $X^M = (\tilde x_i, x^i)$, so we split them as
follows $\tilde x_i = (\tilde x_\mu, \tilde y_m)$ and $x^i = (x^\mu,
y^m)$. As before, $m = 1, \dots, n$ are indices denoting internal directions and $\mu = 1, \dots, d$ are space-time
indices. Then, we have a {\it double} external space
and a {\it double} internal one with coordinates $\mathbb{X} =
(\tilde x_\mu, x^\mu)$ and $
\mathbb{Y}^A = (\tilde y_m, y^m)$, respectively.

\item Next we propose a reduction  ansatz for the fields and gauge
parameters in the theory, inspired in the global symmetries of DFT. For
the generalized bein and dilaton we have
\be
E^{\bar A}{}_M (X) = \widehat E^{\bar A}{}_I (\mathbb{X}) \ U^I{}_M
(\mathbb{Y}) \ , \ \ \ \ \ d(X) = \widehat d(\mathbb{X}) +
\lambda(\mathbb{Y})\, ,\label{ans}
\ee
and for the gauge parameters we have
\be
\xi^M (X) = \widehat \xi^I(\mathbb{X})\ U_I{}^M (\mathbb{Y})\, .
\ee
Here  $M, N=1, \dots , 2D$ are curved indices in the parent theory and $I, J=1, \dots , 2D$
are curved indices in the effective theory.
Again we use the notation that hatted objects are $\mathbb{X}$-dependent,
and all the (double) internal $\mathbb{Y}$-dependence enters through the
twists $U_I{}^M \in O(n,n)$ and $\lambda$.

\item We plug this ansatz in the generalized fluxes, and get
\bea
{\cal F}_{\bar A\bar B\bar C} &=& \widehat F_{\bar A\bar B\bar C} +
f_{IJK}\ \widehat E_{\bar A}{}^I\widehat E_{\bar B}{}^J\widehat E_{\bar
C}{}^K  \, ,\label{F3reduced}\\
{\cal F}_{\bar A} &=& \widehat F_{\bar A} + f_I\ \widehat E_{\bar A}{}^I\, ,
\label{F1reduced}
\eea
where we have split the coordinate dependence in $\mathbb{X}$-dependent
quantities
\bea
\widehat F_{\bar A\bar B\bar C}& =& 3 \widehat \Omega_{[\bar A\bar B\bar
C]}\ , \ \ \ \ \ \ \ \ \ \ \ \ \ \  \ \ \ \ \widehat \Omega_{\bar A\bar
B\bar C} = \widehat E_{\bar A}{}^I \partial_I\widehat E_{\bar
B}{}^J\widehat E_{\bar CJ} \, ,\label{GenFluxesEffective}\\
\widehat F_{\bar A} &=& \widehat \Omega^{\bar B}{}_{\bar B\bar A} + 2
\widehat E_{\bar A}{}^I \partial_I \widehat d\, ,
\eea
and $\mathbb{Y}$-dependent ones
\bea
f_{IJK} &=& 3 \tilde \Omega_{[IJK]} \ , \ \ \ \ \ \ \ \ \ \ \ \ \ \  \ \ \
\  \tilde \Omega_{IJK} = U_I{}^M \partial_M U_J{}^N
U_{KN}\, ,\label{gaugingsDFT}\\
f_{I} &=& \tilde \Omega^J{}_{JI} + 2 U_I{}^M \partial_M \lambda\, .
\eea

This splitting is possible provided one imposes the following constraint
on the duality twist $U_I{}^M$
\be
U_I{}^M \partial_M \widehat g = \partial_I \widehat g \ , \ \ \ \ \ \
\partial^M U_I{}^N\ \partial_M \widehat g = 0 \label{gaugingsLorentz1}
\ee
This restriction on the duality twist, implies that it must be trivial in
the $\mathbb{X}$-directions. There is a very important physical reason for
this constraint to hold. The quantities $f_{IJK}$ and $f_{I}$ are named
{\it gaugings}, and we take them to be constant
\be
f_{IJK} = {\rm constant} \ , \ \ \ \ \ f_{I} = {\rm constant}\, .
\ee
This is due to the fact that they appear in the action through the
generalized fluxes ${\cal F}_{\bar A\bar B\bar C}$ and ${\cal F}_{\bar A}$, and
since we look for
 a $\mathbb{Y}$-independent effective Lagrangian, they  must be
$\mathbb{Y}$-independent because their dependence comes only through the
gaugings, which were requested to be constant. This in turn implies that the internal space is paralellizable, namely,  the twist must be  globally defined. The constraint
(\ref{gaugingsLorentz1}) can be recast as
\be
f_{IJ}{}^K \partial_K \widehat g= 0\ , \ \ \ \ \ f^I \partial_I \widehat
g= 0\, .
\ee
Its effect is to protect Lorentz invariance in the reduced theory. Notice
that $\partial_I \widehat g$ is only non-vanishing in the external
directions, and then if the gaugings had legs in these directions they
would explicitly break Lorentz symmetry. Therefore, $f_{IJK}$ and $f_I$ can
only be non-vanishing along the double internal space. For simplicity here
we will only analyze the case $f_{I} = 0$, since consistency of the theory
would otherwise require a slightly modified reduction ansatz. A discussion on
how to turn the gaugings on in the usual supergravity picture can be found in
\cite{Derendinger:2007xp}, and in a duality covariant way in DFT in
\cite{Geissbuhler:2011mx}.

\item We next plug  (\ref{F3reduced}) and (\ref{F1reduced}) into the
action of DFT (\ref{ActionDFT}) to obtain the effective theory
\bea
S_{GDFT} = v \int d\mathbb{X} e^{-2 \widehat d} \bigg[&&
\!\!\!\!\!\!\!\!\!-\frac 1 4 \bigg(\widehat F_{IK}{}^L + f_{IK}{}^L \bigg)
\bigg(\widehat F_{JL}{}^K + f_{JL}{}^K \bigg) \widehat {\cal H}^{IJ}
\label{ActionGDFT}\\ && \!\!\!\!\!\!\!\!\!-\frac 1 {12} \bigg(\widehat
F_{IJ}{}^K + f_{IJ}{}^K \bigg) \bigg(\widehat F_{LH}{}^G + f_{LH}{}^G
\bigg) \widehat {\cal H}^{IL}\widehat {\cal H}^{JH}\widehat {\cal
H}_{KG}\nn\\
&&\!\!\!\!\!\!\!\!\! - \frac 1 6 \bigg(\widehat F_{IJK} + f_{IJK} \bigg)
\bigg(\widehat F^{IJK} + f^{IJK} \bigg) + \bigg(\widehat {\cal H}^{IJ} -
\eta^{IJ}\bigg) \widehat F_I \widehat F_J\bigg]\, .\nn
\eea
The internal coordinate dependence factorized, and it just amounts to an
overall constant factor
\be
v = \int d\mathbb{Y} e^{-2 \lambda}\, .
\ee
If the gaugings vanished $f_{IJK} = 0$, one recovers the usual DFT
action (\ref{ActionDFT}) in less dimensions. This then corresponds to a
{\it gauged} DFT (GDFT) \cite{heteroticHohm,GDFT} (see \cite{Berman:2013cli} for a geometric and supersymmetric treatment of GDFT), which has been obtained
through a generalized SS
compactification of a higher dimensional {\it parent} DFT.

\item The symmetries of the GDFT are inherited from those of the parent DFT.
For instance, the generalized Lie derivative induces the gauge
transformations in the effective action
    \be  {\cal L}_{\xi} V^M =  U_I{}^M\ \widehat{\cal L}_{\widehat \xi}
\widehat{V}^I  \, ,\label{DefGauge}\ee
namely
\bea \widehat{\cal L}_{\widehat{\xi}} \widehat{V}^I &=& {\cal
L}_{\widehat{\xi}} \widehat{V}^I - f^I{}_{JK}{\widehat{\xi}}^J
\widehat{V}^K  \, .\label{twistedGaugetransf}\eea
The first term is the usual generalized Lie derivative, and the second one
amounts to a deformation due to the gaugings. These
induced transformations now close (in the sense of (\ref{consistency})) when the following quadratic constraints
are imposed on the gaugings
    \be
      f_{H[IJ} f_{KL]}{}^H = 0 \, ,\label{effectiveconstraints1}
    \ee
    and the strong constraint holds in the external space
    \be
     \partial_I \widehat V\  \partial^I \widehat W = 0
\label{effectiveconstraints2}
    \ee
    for any hatted quantity, such as effective fields or gauged
parameters. The action of GDFT (\ref{ActionGDFT})
is invariant under (\ref{twistedGaugetransf}) up to these constraints.
Moreover, it can be checked that compactifying the constraints of the
parent DFT gives the same result that one would obtain by directly
computing the consistency conditions of the effective GDFT. These amount
to even more relaxed versions of (\ref{effectiveconstraints1}) and
(\ref{effectiveconstraints2}).
\end{itemize}

\subsection{From gauged DFT to gauged supergravity}
\label{secDFTtogaugedsugra}

Now that we have built a covariant formulation of the effective theory, we
can choose to solve the {\it effective} strong constraint
(\ref{effectiveconstraints2}) in the usual frame of a gauged supergravity
$\partial_I \widehat V \partial^I \widehat W = 0 \ \rightarrow \ \tilde
\partial^\mu \widehat V = 0$, i.e. we solve the strong constraint in the
effective action by demanding that the effective fields and gauged
parameters only depend on $x^\mu$. Due to the coordinate splitting $X \to
\mathbb{X} , \mathbb{Y}$  a convenient re-parameterization of the
effective generalized metric $\widehat{\cal H}^{IJ}$ is in order. The
$O(D,D)$ group is now broken to $O(d,d) \times O(n,n)$, and then it is
convenient to rotate the group metric to the form
\be
\eta_{IJ} = \left(\begin{matrix} & \delta^\mu{}_\nu &  \\
\delta_{\mu}{}^\nu &  &  \\   &
  & \eta_{AB}\end{matrix}\right) \, ,\label{GroupMetricEffective}
\ee
where $\eta_{AB}$ was defined in (\ref{etaOdd}). This amounts to a
re-parameterization of the generalized bein
\be
\widehat E^{\bar A}{}_I = \left(\begin{matrix}\widehat e_{\bar a}{}^\mu &
-\widehat e_{\bar a}{}^\rho \widehat c_{\rho \mu} & -\widehat e_{\bar
a}{}^\rho \widehat A_{A \rho} \\ 0 & \widehat e^{\bar a}{}_\mu & 0 \\ 0 &
\widehat \Phi^{\bar A}{}_B \widehat A^B{}_{\mu } & \widehat \Phi^{\bar
A}{}_A \end{matrix}\right)\, ,\label{GenBeinEffective}
\ee
which is now associated to the following generalized metric
\be
\widehat {\cal H}_{IJ} = \left(\begin{matrix} \widehat g^{\mu \nu} & -
\widehat g^{\mu \rho}\widehat c_{\rho \nu} & - \widehat g^{\mu \rho}
\widehat A_{A\rho} \\
- \widehat g^{\nu\rho}\widehat c_{\rho \mu} & \widehat g_{\mu \nu} +
\widehat A^C{}_\mu \widehat {\cal M}_{CD} \widehat A^D{}_\nu  + \widehat
c_{\rho\mu} \widehat g^{\rho \sigma} \widehat c_{\sigma \nu} & \widehat
{\cal M}_{AC} \widehat A^C{}_\mu + \widehat A_{A\rho} \widehat
g^{\rho\sigma} \widehat c_{\sigma \mu} \\ - \widehat g^{\nu \rho} \widehat
A_{B \rho} & \widehat {\cal M} _{BC} \widehat A^C{}_{\nu} + \widehat A_{B
\rho}\widehat g^{\rho \sigma} \widehat c_{\sigma \nu}& \widehat {\cal
M}_{AB} + \widehat A_{A \rho} \widehat g^{\rho \sigma}\widehat
A_{B\sigma}\end{matrix}\right) \, .\label{GenMetEffective}
\ee
Here we have introduced the combination $\widehat c_{\mu\nu} = \widehat
b_{\mu \nu} + \frac{1}{2} \widehat A^B{}_\mu \widehat A_{B\nu}$. Also,
$\widehat A^A{}_\mu$ are the vectors (\ref{Aeffective}) and $\widehat
\Phi^{\bar A}{}_A$ is the scalar bein for the scalar metric $\widehat
{\cal M}_{AB}$ defined in (\ref{Meffective}). Notice that we have run out
of indices, so we are denoting with the same letter $\bar A$ the {\it full}
flat index, and the {\it internal} one, the distinction should be clear
from the context. Also due to the splitting, now the gaugings are only
non-vanishing in the
internal components
\be
f_{IJK} = \left\{ \begin{matrix} f_{ABC} \  \ \ \ \ (I,J,K) = (A,B,C) \\
\!\!\!\!\!\!\!\!\!\!\!\!\!\!\!\!\!\!\!\!\!\!0 \ \ \ \ \ \ \ \ \ {\rm
otherwise}\end{matrix}\right. \, .\label{FluxesEffective}
\ee

Then, plugging (\ref{GenBeinEffective}) and (\ref{FluxesEffective}) into
(\ref{GenFluxesEffective}), and taking into account that the indices are
now raised and lowered with (\ref{GroupMetricEffective}), we can readily
identify some of the components of the compactified generalized fluxes
with covariant quantities in the effective action
(\ref{CurvaturesSS})-(\ref{CovariantDerivativeSS}), namely
\bea
{\cal F}_{\bar a \bar b \bar c} &=& \widehat e_{\bar a}{}^\mu \widehat e_{\bar
b}{}^\nu \widehat e_{\bar c}{}^\rho\ {\cal G}_{\mu\nu\rho}\, ,\\
{\cal F}_{\bar a \bar b}{}^{\bar C} &=&\widehat e_{\bar a}{}^\mu \widehat e_{\bar
b}{}^\nu \widehat \Phi^{\bar C}{}_C\ F^C{}_{\mu\nu} \, ,\\
{\cal F}_{\bar a\bar B}{}^{\bar C} &=& \widehat e_{\bar a}{}^\mu \widehat
\Phi^{\bar C}{}_C\ D_\mu \widehat \Phi_{\bar B}{}^C\, ,
\eea
where
\be
D_\mu \widehat \Phi_{\bar B}{}^C = \partial_\mu \widehat \Phi_{\bar B}{}^C
- f_{AB}{}^C \widehat A_\mu{}^A \widehat \Phi_{\bar B}{}^B
\ee
is the covariant derivative of the scalar bein. Finally, plugging
(\ref{GenBeinEffective}) and (\ref{FluxesEffective}) in the action
(\ref{ActionGDFT}) of GDFT, one recovers the effective action of gauged
supergravity (\ref{ActionSS}). Therefore, gauged supergravities are particular
examples of GDFT.
\subsection{Duality orbits of non-geometric fluxes}

Even if it looks like the generalized SS procedure discussed in Sections
\ref{secGenSS} and \ref{secDFTtogaugedsugra} leads to the same action
(\ref{ActionSS}) obtained from the usual geometric SS compactification of
Section \ref{secSS}, this is not correct. The difference
resides in the gaugings. While the geometric SS reduction only allows to
turn on the fluxes (\ref{geometric fluxes}) and the others
(\ref{NongeometricSS})  vanish, the generalized SS reduction of DFT allows, in
principle, to turn on {\it all} the gaugings simultaneously.

We have defined the gaugings or fluxes in (\ref{gaugingsDFT}) in terms of
a duality valued twist matrix $U(\mathbb{Y}) \in O(n,n)$. This generalizes
the usual SS gaugings in two ways:
\begin{itemize}
\item Global extension. The geometric SS gaugings are generated through
$u^a{}_m$ and $v_{mn}$ in (\ref{geometric fluxes}), which respectively
correspond to the metric and two-form background. They can both be
combined into the $O(n,n)$ duality twist matrix in the form
    \be
    U^A{}_M = \left(\begin{matrix} u_a{}^m & u_a{}^n v_{nm} \\ 0 &
u^a{}_m\end{matrix}\right)\, .
    \ee
    A T-duality transformation would break the triangular form of this
matrix, into a new element of $O(n,n)$ containing a South-West component.
Therefore, the only backgrounds that are allowed in the usual SS
compactification of supergravity are those that
come
back to themselves under monodromies, up to $u$ and/or $v$-transformations only, i.e. the
{\it geometric} subgroup of the full $O(n,n)$. In a generalized SS
compactification, we now allow the duality twist to be a generic element
of $O(n,n)$. This includes, in addition to the elements $u^a{}_m$ and $v_{[mn]}$,
 a new component usually dubbed $\beta^{[mn]}$
    \be
    U^A{}_M = \left(\begin{matrix} u_a{}^m & u_a{}^n v_{nm} \\ u^a{}_n
\beta^{nm} & u^a{}_m + u^a{}_n \beta^{np} v_{pm}\end{matrix}\right)\, .
\label{TwistParam}
    \ee
    The effect of this extension is now to allow for backgrounds that
come back to themselves under monodromies, up to a generic $O(n,n)$
transformation. This is the case of the T-folds discussed before. Then,
the generalized SS compactification allows for new backgrounds that are
globally ill defined from the usual (geometric) supergravity point of view.

\item Local extension. The fluxes (\ref{gaugingsDFT}) are now not only
defined in terms of an extended duality twist, but also in terms of a
generalized derivative with respect to all coordinates. This would allow
for more richness in the space of gaugings, if the duality twist violated
the strong constraint. In this case, the dual coordinate dependence would
make no sense from a supergravity point of view, altering the standard notion of
locality.
\end{itemize}

Let us now show that the quadratic constraints
(\ref{effectiveconstraints1}) are weaker than the strong
constraint. For the duality twist, the strong constraint implies
\be
\tilde \Omega_{EAB} \tilde \Omega^E{}_{CD} = 0\, .\label{SCWeitz}
\ee
where $\tilde\Omega_{ABC}$ was defined in (\ref{gaugingsDFT}).
On the other hand, similar to the BI (\ref{Zetas}), one can show that
\be
\partial_{[A} f_{BCD]}-\frac{3}{4}f_{[AB}{}^E f_{CD]E}\ =\
-\frac{3}{4}\tilde \Omega_{E[AB}\tilde \Omega^E{}_{CD]}\, .
\ee
For constant gaugings the first term drops out, and then we see that the
quadratic constraints correspond to a relaxed version of the strong
constraint (\ref{SCWeitz}), because they only require the totally antisymmetric
part of (\ref{SCWeitz}) to vanish.

\begin{figure}[h!]
\begin{center}
\begin{tikzpicture}[scale=0.5,>=latex']
        \path[P_3] \ellip;
        \path[P_4] \ellipp;
        \draw[very thick] (-9,4) -- (-5,-5.45);
        \draw[very thick] (-4,5.65) -- (1,-6);
        \path (-3,2.25) node[left] {\textbf{A}}
            (-1.5,-3.5) node [right] {\textbf{B}};
        \path (-7,-1) node[left] {Orbits (1)}
            (.5,-4.5) node [right] {Orbits (2)};
        \path (16.5,-5) node[left] {\textbf{Flux space}}
            (-.5,.5) node [right] {\textbf{Geometric fluxes}};
\end{tikzpicture}
\caption{We picture the space of gaugings (or fluxes) in half-maximal
supergravities in $d=7,8$ \cite{Dibitetto:2012rk}. A point in this diagram corresponds to a
given configuration. If two points lie in the same diagonal line (orbit) they
are related by a duality transformation. Different theories are classified by
orbits (lines) rather than configurations (points). The configuration space
splits in a subgroup of geometric
(i.e. only involving fluxes
like $H_{abc}$ and $\omega_{ab}{}^{c}$) and non-geometric (involving fluxes
$Q_a{}^{bc}$ and $R^{abc}$) configurations. The space of orbits then splits in two: (1)
non-geometric orbits (truly half-maximal) and (2) geometric orbits (basically
maximal) that intersect the geometric space (between A and B).
}\label{FigureOrbits}
\end{center}
\end{figure}
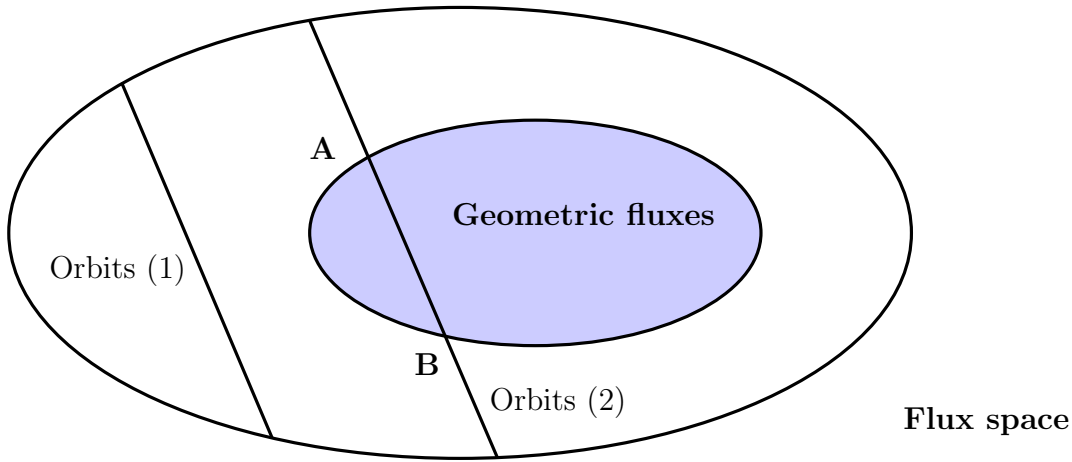

This is however not the end of the story. One has to show that there exist
solutions to the quadratic constraints that violate the strong
constraint.
As we explained, the gauged supergravities we are dealing with are
half-maximal. Half-maximal gauged supergravities split into two different
groups: those that can be obtained by means of a truncation of a maximal
supergravity, and those that cannot (in $d = 4$ see \cite{ExceptionalFluxComp}).
The former inherit the quadratic
constraints of the maximal theory, which in the language of the
electric half-maximal gaugings take the form (for simplicity we take $\lambda =
0$ in (\ref{ans}))
\be
f_{ABC} f^{ABC} = 3 \tilde \Omega_{ABC} \tilde \Omega^{ABC} =
0\, .\label{maximalconstraint}
\ee
Therefore, the genuine half-maximal theories, which violate the above
constraint, must be necessarily generated through a truly doubled duality
twist \cite{Dibitetto:2012rk}. On the other hand, when the strong constraint
holds, the only
reachable theories are those that admit an uplift to a maximal
supergravity. In Figure \ref{FigureOrbits}  we have pictured the kind of orbits
that one finds in half-maximal supergravities (the case $d = 4$ should be
analyzed separately due to the extra $SL(2)$ factor \cite{ExceptionalFluxComp}). Let us stress that the notion of non-geometry discussed in \cite{Dibitetto:2012rk} is local, and then a duality orbit of non-geometric fluxes contains {\it all} fluxes simultaneously turned on $H_{abc}$, $\omega_{ab}{}^c$, $Q_a{}^{bc}$  and $R^{abc}$. However, one could also define a notion of globally non-geometric orbit, which could admit a representative without $R^{abc}$-flux.

The idea of combining geometric with non-geometric fluxes simultaneously is
usually considered with some precaution. It is common to  find objections
against these configurations mostly based on scaling arguments. The
non-geometric fluxes are sometimes associated to windings (since they are mostly
generated through dual coordinate dependence), while the geometric ones are
related to momentum. A quick look at the mass formula (\ref{massformulaT}) shows
that, for a given radius $R$, when momentum (winding) modes are heavy, the
winding (momentum) modes are light. Considering both of them simultaneously then
leads to unavoidable heavy modes in the spectrum. This enters in conflict with
the fact that one is truncating
the heavy mass levels of the
string from the beginning. The conclusion of this argument is that one should then impose the
strong constraint, so as to truncate the heavy part of the spectrum, and this in
turn only permits geometric orbits.

Notice however that these arguments are purely based on the KK-mode expansions
of fields on
tori. The relation between (winding) momentum and (dual) coordinates, is given by the
Fourier transforms of the KK-modes of the torus. Moreover, the mass formula
(\ref{massformulaT}) only holds for tori. In this section, we do not consider
tori, and moreover, we do not consider KK excitations. We only consider the
zero-modes of the fields on twisted-double-tori. The only connection between these two
situations is when the duality twist is taken to be constant, in which case we
would be dealing with the (massless) zero-modes on a torus, as discussed in section 5.2.  When the twist
matrix is non-constant, the effective theory becomes {\it massive}, but these
masses are corrections to the {\it massless} modes on the torus through a twist.
We are then correcting massless modes, through a procedure that truncates all
the problematic (KK) modes. When dealing with moduli fixing, one has to make
sure that for a given non-geometric orbit, the masses of the scalar fields
(which are totally unrelated to (\ref{massformulaT})) in a given vacuum are small
compared to the scales of the modes we are neglecting.

A similar reasoning prevents us from relating the strong constraint (or a weaker
version of it) with the LMC condition (\ref{LMC}), as we explained in Section
\ref{sec:DoubleFields}. If we had considered tori compactifications, and kept the tower of excited states, whenever the derivatives in strong-constraint like terms acted on the mode expansion they would form contractions like ${\cal P}^M {\cal P}_M$ related to the LMC. Here, we are considering the zero-modes, for which ${\cal P}^M {\cal P}_M = 0$ is trivially satisfied  because ${\cal P}^M = 0$. Then, in SS compactifications in which the tower of KK modes is truncated, it might not be correct to identify the LMC with strong-like constraints. Instead, the consistency constraints are given by the quadratic constraints for gaugings (\ref{effectiveconstraints1}).

Combining fluxes and their derivatives, it is possible to construct three
quantities that vanish upon imposition of the strong constraint \cite{Exploring}:
\bea
   \partial_{[A} f_{BCD]}-\frac{3}{4}f_{[AB}{}^E f_{CD]E}  &=&  {\cal Z}_{ABCD}
\, ,\label{FirstBI2}\\
    \partial^E f_{EAB} + 2\partial_{[A}f_{B]}-f^E f_{EAB} &=&  {\cal Z}_{AB} \,
,\label{SecondBI2}\\
    \partial^E f_E-\frac{1}{2} f^E f_E +\frac{1}{12}f_{ABC}f^{ABC} &=& {\cal Z}
\, .\label{ThirdBI2}
\eea
They correspond to duality orbits of {\it generalized BI} for all the dual
fluxes. The first two (\ref{FirstBI2}), (\ref{SecondBI2}) are related to the
constraints of the theory, and are obtained from compactifications of
(\ref{Zetas}). The last one (\ref{ThirdBI2}) was associated to the embedding of
the theory into a maximal theory (\ref{maximalconstraint}).

The fluxes $f_{ABC}$ encode the standard T-dual fluxes, as we reviewed. This can
be seen by splitting the indices as
\be
f_{abc} = H_{abc}\ , \ \ \ f^a{}_{bc} = \omega_{bc}{}^a\ , \ \ \ f^{ab}{}_c =
Q_c{}^{ab}\ , \ \ \ f^{abc} = R^{abc}\, .
\ee
Through T-dualities they are related according to the chain
(\ref{TdualityChain}). Recall that we are only dealing here with the case $f_A =
0$.
 Splitting in components equation (\ref{FirstBI2}), we find
\bea
\partial_{[a}H_{bcd]} -\frac{3}{2}H_{e[ab} \omega_{cd]}{}^e &=&{\cal Z}_{abcd}\
,\nn\\
3\partial_{[a}\omega_{bc]}{}^d - \partial^d H_{abc}
+3\omega_{[ab}{}^e\omega_{c]e}{}^d
-3Q_{[a}{}^{de}H_{bc]e} &=& {\cal Z}_{abc}{}^d\ ,\nn\\
2\partial_{[a}Q_{b]}{}^{cd} +2 \partial^{[c}\omega_{ab}{}^{d]}
- \omega_{ab}{}^e Q_e{}^{cd} - H_{abe}R^{ecd} +
4Q_{[a}{}^{e[c}\omega_{b]e}{}^{d]} &=&{\cal Z}_{ab}{}^{cd}\ ,\label{obi}\\
3\partial^{[a}Q_d{}^{bc]} - \partial_d R^{abc} +3Q_e{}^{[ab}Q_d{}^{c]e}
-3\omega_{de}{}^{[a}R^{bc]e} &=& {\cal Z}^{abc}{}_d\ ,\nn\\
\partial^{[a}R^{bcd]} -\frac{3}{2}R^{e[ab} Q_e{}^{cd]} &=&{\cal Z}^{abcd}\ .\nn
\eea
These reduce to those of \cite{stw} for constant
fluxes under the strong constraint,
and to those of \cite{Blumenhagen:2012pc} for non-constant fluxes. Equation
(\ref{ThirdBI2}), on the other hand, reads in components
\be
\frac{1}{6}H_{abc}R^{abc}
+ \frac{1}{2} \omega_{ab}{}^c Q_c{}^{ab} = {\cal Z}\ .\label{nnnn}
\ee
This corresponds to an orthogonality-like condition between geometric ($H_{abc},
\omega_{ab}{}^c$) and non-geometric ($Q_a{}^{bc}, R^{abc}$) fluxes, as expected.
And this is the reason why the failure of this equation to vanish {\it requires}
non-geometric fluxes. Therefore, (\ref{nnnn}) can  be used to classify duality orbits of
non-geometric fluxes.

Using the extended parameterization of the twist matrix (\ref{TwistParam})
\be
 U^A{}_M = \left(\begin{matrix} u_a{}^m & u_a{}^n v_{nm} \\ u^a{}_n
\beta^{nm} & u^a{}_m + u^a{}_n \beta^{np}
v_{pm}\end{matrix}\right)\, ,\label{TwistParam2}
\ee
and inserting this in the definition of the fluxes (\ref{gaugingsDFT}),
(\ref{FluxesEffective})
\be
f_{ABC} = 3\  U_{[A|}{}^M\ \partial_M U_{|B}{}^N U_{C]N}\, ,
\ee
we find in components
\bea\label{nf}
H_{abc} &=& 3\left[ \nabla_{[a} v_{bc]}  - v_{d[a} \tilde\nabla^d v_{bc]}
\right]\, ,\\
\omega_{ab}{}^{c} &=& 2 \Gamma_{[ab]}{}^c + \tilde \nabla^c v_{ab} + 2
\Gamma^{mc}{}_{[a} v_{b]m} + \beta^{cm}H_{mab}\, ,\nn\\
Q_c{}^{ab} &=& 2 \Gamma^{[ab]}{}_c +\partial_c\beta^{ab}+  v_{cm} \tilde
\partial^m\beta^{ab} + 2 \omega_{mc}{}^{[a}\beta^{b]m}  -H_{mnc}
\beta^{ma}\beta^{nb}\, ,\nn\\
R^{abc} &=& 3\left[\beta^{[\underline am}\nabla_m \beta^{\underline b\underline
c]}
+ \tilde \nabla^{[a} \beta^{bc]} +
 v_{mn} \tilde \nabla^n\beta^{[ab} \beta^{c]m} + \beta^{[\underline
am}\beta^{\underline bn} \tilde\nabla^{\underline
c]}v_{mn}
 \right ] +
\beta^{am}\beta^{ bn}\beta^{cl}H_{mnl}\, ,\nn
\eea
where we have used the following relations and definitions
\be
u_a{}^m u^a{}_n = \delta^m_n \ , \ \ \ \ \ \ u_a{}^m u^b{}_m = \delta_a^b\,
,\quad
v_{ab} = u_a{}^m u_b{}^n v_{mn} \ , \ \ \ \ \ \ \beta^{ab} = u^a{}_m u^b{}_n
\beta^{mn}\, ,\nn
\ee
\be
\partial_a = u_a{}^m \partial_m  \ , \ \ \ \ \ \ \tilde \partial^a = u^a{}_m
\tilde \partial^m\, ,\nn
\ee
\bea
\nabla_a v_{bc}=\partial_a v_{bc}-\Gamma_{ab}{}^d v_{dc}-\Gamma_{ac}{}^d
v_{bd}\, ,\quad
\tilde \nabla^a v_{bc}=\tilde\partial^a v_{bc}+\Gamma^{ad}{}_{b}
v_{dc}+\Gamma^{ad}{}_{c} v_{bd}\, ,\nn\\
\nabla_a
\beta^{bc}=\partial_a\beta^{bc}+\Gamma_{ad}{}^b\beta^{dc}+\Gamma_{ad}{}^c\beta^{
bd}\, ,\quad
\tilde \nabla^a
\beta^{bc}=\tilde\partial^a\beta^{bc}-\Gamma^{ab}{}_{d}\beta^{dc}-\Gamma^{ac}{}_
{d}\beta^{bd}\, ,\nn
\eea
and
\be
\Gamma_{ab}{}^c = u_a{}^m \partial_m u_b{}^n e^c{}_n \ , \ \ \ \ \ \
\Gamma^{ab}{}_c = u^a{}_m \tilde \partial^m u^b{}_n u_c{}^n\, .\label{Gammas}
\ee

The expressions (\ref{nf}) are very useful to explore the uplifting of fluxes to higher dimensional theories. Notice that while $H_{abc}$ and $\omega_{ab}{}^c$ can be generated through geometric twists $u^a{}_m$ and $v_{mn}$, the non-geometric fluxes $Q_a{}^{bc}$ and/or $R^{abc}$ require $\beta$-twists and/or dual coordinate dependence. This also serves to show that the distinction between ``globally'' and ``locally'' non-geometric fluxes is just a terminology inherited from the toy example discussed before, since $Q_a{}^{bc}$ can arise from dual coordinate dependence, and $R^{abc}$ can arise from globally non-geometric  compactifications with non-trivial $\beta$-twist. Setting $\beta^{ij} = 0$ and $\tilde \partial^i = 0$, the expressions (\ref{nf}) reduce to (\ref{geometric fluxes2}).

\newpage

\section{U-duality and extended geometry}
\label{sec:EFT}

The compactification of $D = 11$ supergravity and M-theory on an $n$-dimensional
torus, enjoys a U-duality symmetry $E_{n(n)}$ (see for example
\cite{Uduality,Cremmer:1997ct,Obers:1998fb}). The idea of extending the
space-time and/or the tangent space so as to accommodate such symmetries was introduced in
\cite{HullU}, \cite{PachecoU} and more recently considered in
\cite{BermanPerry,Uduality1,Udualities3}. In this section, we review
some of the approaches to replace the T-duality group by the U-duality group, in order
to incorporate all the extra fields (like R-R in Type II theories or the three-form of
M-theory) in a duality covariant manner, much under the same philosophy as that of DFT.

\subsection{Generalized diffeomorphisms and the section condition}

We have seen that the generalized diffeomorphisms of DFT (\ref{gendiffs})
discussed in the previous sections enjoy the following properties:
\begin{itemize}
\item They preserve the duality group invariant, in that case the
$O(D,D)$ metric $\eta_{MN}$.
\item They are defined in terms of an invariant $Y$-tensor related to the
definition of the strong constraint.
\item They reproduce the gauge transformations of the $D$-dimensional
metric and two-form, upon application of the strong constraint.
\item When ``twisted'', they give rise to fluxes or gaugings in the
representations allowed by supersymmetry.
\item Their closure imposes a set of constraints that, on the one hand are
solved by the strong constraint, and on the other, reproduce the quadratic
constraints of the supergravity gaugings upon ``twisting''.
\end{itemize}
Clearly, the inclusion of other fields, like R-R sector in Type
II string theory or the more general three-form of M-theory, requires an
enlargement and further generalization of the already generalized Lie
derivative, to be compatible now with U-duality. All the transformation
properties of gravitational and tensorial degrees of freedom, which mix
under U-duality, must now be accommodated (and unified) in a new
generalized Lie derivative. We will see that such a generalized
transformation enjoys the U-duality extension of the properties listed
above.

This generalization is a little more involved since the U-duality group
jumps with dimension. For the $n$ internal dimensions of M-theory, it
corresponds to exceptional groups $E_{n(n)}$, and in $n > 8$ one encounters the
complication of infinite-dimensional Kac-Moody type algebras.
Given the disconnected structure of the groups for different dimensions,
it is convenient to work case by case. As in DFT, where the space is
doubled to account for the winding degrees of freedom of the string, here
the space is further enlarged to account for the wrapping states of
M-branes. The internal space is then replaced by an {\it extended}
mega-space with extended dimensions, and here for simplicity we neglect
the external space-time. Relevant representations for the different
U-duality groups are given in Table \ref{RepsU}. The mega-space associated
to each of them is ${\rm dim}(R_1)$-dimensional.

\begin{table}[h!]
\begin{center}
\scalebox{1}[1]{
\begin{tabular}{| c | c | c |}
\hline
 & $R_1$ & $R_2$    \\[1mm]
\hline \hline

$E_{4(4)} = SL(5)$ & $\bf 10$ & $\overline{\bf 5}$
\\[1mm]\hline

$E_{5(5)} = SO(5,5)$ & $\bf 16$ & $\bf 10$   \\[1mm]\hline

$E_{6(6)}$ & $\bf 27$  & $\overline{\bf 27}$  \\[1mm]\hline

$E_{7(7)}$ & $\bf 56$ & $\bf 133$  \\[1mm]\hline
\end{tabular}
}
\end{center}
{\it \caption{Some relevant representations of U-duality groups
\cite{Uduality2}. }
\label{RepsU}}
\end{table}

The generalized diffeomorphisms \cite{localsymm,Uduality1} formally preserve the structure
of those
analyzed before for DFT (we are using the notation of \cite{Uduality2})
\be
{\cal L}_\xi V^M = L_\xi V^M + Y^M{}_N{}^P{}_Q\ \partial_P \xi^Q\ V^N\, ,
\label{genDifU}
\ee
where  $Y$ is a U-duality invariant tensor ``measuring'' the departure
from the usual Lie derivative. It can be generically decomposed as
\be
Y^M{}_N{}^P{}_Q = \delta^M_Q \delta^P_N - \alpha P_{(adj)}{}^M{}_N{}^P{}_Q
+ \beta \delta^M_N \delta^P_Q \, ,\label{YU}
\ee
where $P_{(adj)}$ is a projector to the adjoint representation of the
U-duality group, $\alpha$ is a group-theoretical quantity that depends on
the dimension, and $\beta$ is a weight for tensorial densities that also
depends on the group. Indices $M,N$ are in the $R_1$ reps of Table
\ref{RepsU}, and $P_{(adj)}$ corresponds to the adjoint projection
contained in the tensor product $R_1 \otimes \overline{R}_1$. It can be
checked that these generalized Lie derivatives preserve the invariants of
each group. The appearance of the last $\beta$-term is due to the fact
that in the U-duality case one usually considers $E_{n(n)} \times \mathbb{R}^+$
tensorial densities rather than just tensors (we will be more specific
later).

Before showing the general results, to warm up let us first see how the
DFT $O(n,n)$ case fits in this language. The projector to the adjoint representation  is
given by
\be
O(n,n) \ \ :  \ \ \ \ \ \ \ P_{(adj)}{}^M{}_N{}^P{}_Q = \frac 1 2
(\delta^M_Q \delta^P_N - \eta^{MP}\eta_{NQ})\, ,
\ee
and then, introducing this in (\ref{YU}) and comparing with
(\ref{Ytensor}), we find that the correct value of the proportionality
constants is given by $(\alpha, \beta) = (2,0)$ for un-weighted tensors. More generally, the
expression for $Y$ in the different duality groups is given in Table
\ref{YtensorsU}.
\begin{table}[h!]
\begin{center}
\scalebox{1}[1]{
\begin{tabular}{| c | c | c | c | }
\hline
 & $Y^M{}_Q{}^N{}_P$ & $\alpha$ & $\beta$   \\[1mm]
\hline \hline
$O(n,n)$ & $\eta^{MN}\eta_{PQ}$ & $2$  & $0$ \\[1mm]\hline

$E_{4(4)} = SL(5)$ & $\epsilon^{iMN} \epsilon_{iPQ}$ & $3$  & $\frac 1 5$
\\[1mm]\hline

$E_{5(5)} = SO(5,5)$ & $\frac 1 2 (\gamma^i)^{MN}(\gamma_i)_{PQ}$ & $4$  &
$\frac 1 4$ \\[1mm]\hline

$E_{6(6)}$ & $10 d^{MNR} \bar d_{PQR}$  & $6$  & $\frac 1 3$ \\[1mm]\hline

$E_{7(7)}$ & $12 K^{MN}{}_{PQ} + \delta^{(M}_P \delta^{N)}_Q + \frac 1 2
\epsilon^{MN} \epsilon_{PQ}$ & $12$  & $\frac 1 2$ \\[1mm]\hline
\end{tabular}
}
\end{center}
{\it \caption{Invariant $Y$-tensor and proportionality constants for
different dimensions. Here $\eta_{MN}$ is the $O(n,n)$ invariant metric,
$\epsilon_{iMN}$ is the $SL(5)$ alternating tensor, $(\gamma^i)^{MN}$ are
$16 \times 16$ MW representation of the $SO(5,5)$ Clifford algebra,
$d^{MNR}$ and $K^{MNPQ}$ are the symmetric invariant tensors of $E_{6(6)}$ and
$E_{7(7)}$ respectively, and $\epsilon^{MN}$ is the symplectic invariant in
$E_{7(7)}$. These results were taken from \cite{Uduality2}, we refer to that paper for more details.} \label{YtensorsU}}
\end{table}

In the U-duality case, there is also an analogue of the strong constraint,
also known as section condition \cite{localsymm,Uduality1}
\be
P_{({\bf R}_2)  MN}{}^{PQ} \partial_P \partial_Q (\dots)= 0\, , \label{sectioncondition}
\ee
which again acts on any product of fields and gauge parameters.
Generically, any solution to this condition picks out an $n$-dimensional
subspace of the mega-space, which can be associated to the physical space
in M-theory compactifications. Again, when analyzing closure of these
generalized diffeomorphisms, one finds that closure is achieved
automatically when restricted to configurations satisfying the section
condition.

Analogously to the DFT case, when a duality twist reduction of these
generalized Lie derivatives is performed, they induce an effective gauge
transformation giving rise to the embedding tensor components of the
different maximal gauged supergravities for different dimensions. The analogies
don't stop here, since it is also possible to construct an extended
geometrical formalism, introducing generalized connections, torsion and
covariant Ricci-like tensors for these generalized transformations. Lets
now review how this works,  specializing to the $E_{7(7)}$ case for a detailed
exposition.

The cases $n\leq 7$ were studied in \cite{Uduality1} and \cite{Uduality2}. The
cases $n\leq 2$ just reduce to ordinary Riemannian geometry. The case $n = 4$
was studied in \cite{SSMtheory} in the context of gauged supergravities, and a
geometry for it was considered in \cite{SL5Uduality}. The cases $n = 5$ and 6 were
related to SS compactifications in \cite{Musaev}, and  $n = 7$ in
\cite{Extended geometry}, where an extended geometry was also formulated. A
unified geometric description for $n \leq 7$ was considered in \cite{Cederwall}.
The case $n = 8$ was explored in \cite{E8}, and for $n > 8$ the groups are much
more involved. An ambitious programme intended to encompass all formulations
under $E_{11}$ can be found in \cite{E11programme}.

\subsection{The $E_{7(7)}$ case and maximal gauged supergravity}

$E_{7(7)}$ is the U-duality group of gauged maximal supergravity in four
dimensions \cite{de Wit:2007mt}, the ungauged theory being obtained through
compactifications
of M-theory on a seven-torus \cite{de Wit:1982ig}. The idea here is to replace
the internal
seven-space by a $56$-dimensional mega-space, and accommodate the internal
degrees of freedom in a generalized metric ${\cal H}_{MN}$ defined on the
mega-space. This idea was first considered in \cite{doubletorus}, and here we
will present the results of \cite{Extended geometry}. Since we are ignoring the
four-dimensional space time, the
generalized metric should be identified with the scalar degrees of freedom
of the gauged supergravity. The generalized metric transforms covariantly
under $G = E_{7(7)} \times \mathbb{R}^+$, and is invariant under the maximal
compact subgroup $H = SU(8)$. It can be written in terms of a generalized
bein $E_{\bar A}{}^M$ taking values in the quotient $G/H$
\be
E_{\bar A}{}^M = e^{-\Delta} \tilde E_{\bar A}{}^M\, ,
\ee
where we have separated a conformal factor $\Delta$ corresponding to the
$\mathbb{R}^+$ components. Here the flat indices $\bar A, \bar B,\dots$
take values in $H$ and the curved ones $M,N,\dots$ in $G$. Then, $\tilde
E_{\bar A}{}^M$ lives in the quotient $E_{7(7)} / SU(8)$, and the tilde refers
to the $E_{7(7)}$ part of $G$ only. $G$ has a (weighted) symplectic invariant
$\omega_{MN}$ that raises and lowers indices, and a quartic invariant
$K_{MNPQ}$ which is totally symmetric. The fundamental representation of
$E_{7(7)}$ is $\bf 56$ and the adjoint is $\bf 133$. Given a tensorial density
$V^M$, the generalized Lie derivative (or equivalently the exceptional
Dorfman bracket) reads
\be
{\cal L}_\xi V^M = \xi^P \partial_P V^M - 12 P_{(adj)}{}^M{}_N{}^P{}_Q\
\partial_P \xi^Q\ V^N - \frac 1 2 \partial_P \xi^P\ V^M\, .\label{GenLieE7}
\ee
Here $P_{(adj) ((MN)(PQ))}$ is the projector to the adjoint
representation, defined in terms of the $E_{7(7)}$ invariants
\be
P_{(adj) MNPQ} = (t_\alpha)_{NM} (t^\alpha)_{PQ} = \frac 1 {12}
\omega_{M(P}\omega_{Q)N} + K_{MNPQ}\, .
\ee

As we did before in the DFT case, here we can compute the closure of these
transformations
\be
\Delta_{123}{}^M = - \Delta_{\xi_1} {\cal L}_{\xi_2} \xi_3^M = ([{\cal
L}_{\xi_1} , \ {\cal L}_{\xi_2}] - {\cal
L}_{{\cal L}_{\xi_1}\xi_2}) \xi_3^M = 0\, ,
\ee
and get
\bea \Delta_{[12]3}{}^M & = &   Y^Q{}_L{}^O{}_I\ \partial_O \xi_{[2}^I\
\xi_{1]}^L\ \partial_Q \xi_3^M  +\  A^M{}_N{}^J{}_L
Y^Q{}_J{}^O{}_I\ \partial_Q \xi_{[2}^I\ \partial_O \xi^L_{1]}\
\xi_3^N\nn\\ && +\
Q^M{}_N{}^{QO}{}_{LI}\ \partial_Q \partial_O\xi_{[2}^I\ \xi_{1]}^L \
\xi_3^N=0 \, ,\nn\\
\Delta_{(12)3}{}^M &=& -Y^Q{}_L{}^O{}_I \
\partial_Q \xi_{(1}^I\  \xi_{2)}^L \ \partial_O \xi_3^M  +\
Q^M{}_N{}^{QO}{}_{LI} \partial_Q (\xi_{(1}^L \ \partial_O \xi_{2)}^I)\
\xi_3^N \nn\\ && +\
\frac 1 4 \omega_{LI}\omega^{QO} \ \partial_Q\xi_1^L \ \partial_O\xi_2^I \
\xi_3^M =0  \, ,\label{closureE7}\eea
where we have defined
\be Q^M{}_N{}^{QO}{}_{LI} =
Y^Q{}_J{}^O{}_{(L} A{}^J{}_{I)}{}^M{}_N + \frac 1 2 \omega_{IL}
Y^{QMO}{}_N - \frac 1 2 Y^Q{}_L{}^O{}_I \delta^M_N\, . \ee

Notice that all derivatives are contracted as
\be
Y^{M}{}_P{}^N{}_Q \partial_M \partial_N = \bigg(\frac 1 2 \omega^{MN}
\omega_{PQ} - 12 P_{(adj)PQ}{}^{MN}\bigg) \partial_M \partial_N\, .
\ee
As we mentioned in the previous section, when the so-called section
condition (\ref{sectioncondition})
\be
P_{(adj)PQ}{}^{MN} \partial_M \partial_N (\dots) = 0\label{sectionE7}
\ee
is imposed, the closure condition $\Delta_{123}{}^M = 0$ is automatically
satisfied. In fact, it can be seen that any solution to this condition
selects a seven-dimensional subspace of the full $56$-dimensional
mega-space, permitting to make contact with the physical internal compact
directions. When (\ref{sectionE7}) holds, it can also be proven that
$\omega^{MN}\partial_M\partial_N (\dots) = 0$, and therefore also
$Y^{M}{}_P{}^N{}_Q \partial_M \partial_N (\dots) = 0$, in analogy with DFT (\ref{strongconstraint}).

Following the DFT logic (\ref{DynFabc}), we can now define a dynamical flux
\be
{\cal L}_{E_{\bar A}} E_{\bar B} = F_{\bar A \bar B}{}^{\bar C} E_{\bar C}\, ,
\ee
with
\be F_{ \bar A \bar B}{}^{ \bar C} = \Omega_{ \bar A  \bar B}{}^{ \bar
C} - 12 P_{(adj)}{}^{\bar C}{}_{\bar B}{}^{\bar D}{}_{\bar E} \Omega_{\bar
D \bar A}{}^{\bar E}+ \frac 1 2 \Omega_{ \bar D  \bar A}{}^{\bar D}
\delta_{\bar B}^{\bar C}\
,\label{fluxesplanar} \ee where \be \Omega_{\bar A \bar B}{}^{\bar C} =
E_{\bar A}{}^M \partial_M E_{\bar B}{}^N (E^{-1})_N{}^{\bar C} \
,\label{Weizenbockplanar} \ \ee is the $G$-generalized
Weitzenb\"ock connection. Rotating these expressions with the bein we can
define the fluxes with curved indices \be F_{M  N}{}^{ P} = \Omega_{ M
N}{}^{ P} - 12 P_{(adj)}{}^{P}{}_{
N}{}^{ R}{}_{ S} \Omega_{ R  M}{}^{ S}  + \frac 1 2 \Omega_{RM}{}^R
\delta_{ N}^{P}\ ,
 \label{gaugings}
\ee and the corresponding Weitzenb\"ock connection in curved indices takes
values in the algebra of $G$
\be \Omega_{MN}{}^P = - \partial_M\Delta\
\delta_N^P + \tilde \Omega_{MN}{}^P = \Omega_M{}^0 (t_0)_N{}^P + \tilde
\Omega_{M}{}^{\alpha} (t_\alpha)_N{}^P\, . \label{WeizAlgebra} \ee  Here,
$(t_0)_N{}^P = -\delta_N^P$ is the generator of $\mathbb{R}^+$. The $\bf
56 \times 133$ part  \be
\tilde \Omega_{MN}{}^P  = (\tilde E^{-1})_N{}^{\bar B} \partial_M \tilde
E_{\bar B}{}^P \ , \ee contains the irreducible representations $\bf 56
\times 133 = {\bf 56} + {\bf
912} + {\bf 6480}$. The projectors onto the first two representations in
this product are given by \cite{deWit:2002vt} \bea
P_{({\bf 56})M}{}^\alpha,^N{}_\beta &=& \frac {56}{133} (t^\alpha t_\beta)_M{}^N\, ,
\nn\\ P_{({\bf 912})M}{}^\alpha,^N{}_\beta &=& \frac 1 7 \delta^\alpha_\beta
\delta_M^N - \frac{12}{7} (t_\beta t^\alpha)_M{}^N+ \frac {4}{7} (t^\alpha
t_\beta)_M{}^N\label{projectorsE7} \ . \eea

Equations (\ref{gaugings}) to (\ref{projectorsE7}) imply that the fluxes
are in the ${\bf 912}$ and ${\bf 56}$ representations only. More precisely
\be
F_{MN}{}^P = X_{MN}{}^P + D_{MN}{}^P\ ,\label{embedding tensor}
 \ee
with \be X_{MN}{}^P  = \Theta_M{}^\alpha (t_\alpha)_N{}^P \ \ {\rm with}\
\ \ \ \Theta_M{}^\alpha = 7 P_{({\bf 912})M}{}^\alpha,^N{}_\beta\ \tilde
\Omega_N{}^\beta\ , \label{gaugings912}
 \ee
 and
 \be
 D_{MN}{}^P = - \vartheta_M \delta_N^P + 8 P_{(adj)}{}^P{}_N{}^Q{}_M
\vartheta_Q  \ , \ \ \ \ \vartheta_M = - \frac 12 (\tilde\Omega_{PM}{}^P -
3 \partial_M \Delta)\ . \label{gaugings 56} \ee The fluxes $F$ involve
therefore
a projection onto the ${\bf 912}$ given by the gaugings $X_{MN}{}^P$ plus
contributions from the gaugings $\vartheta_M$.  As in the DFT case, in the
language of gauged supergravity they
correspond to the gauge group generators, i.e. they are  contractions of the
embedding tensor with the generators of the global symmetry group. For
this reason, we will sometimes call them ``gaugings''. The $X_{MN}{}^P$
piece in (\ref{embedding tensor}) corresponds to the $\bf 912$ component
of the fluxes, satisfying the properties
\be
X_{M[NP]}  = X_{MP}{}^{P} = X_{(MNP)} = X_{PM}{}^P  =0\ , \ee which are
the well
known conditions satisfied by gaugings in  maximal supergravity. The $D_{MN}{}^P$
piece (\ref{gaugings 56}), on the other hand, contains two terms: one
belonging to the $\bf 56$ associated to $\mathbb{R}^+$, and another one
belonging to the $\bf 56$ in $\bf
56\times 133$. Notice, however, that both terms contain the same degrees of
freedom in terms of $\vartheta_M$ and are therefore not independent.
With these results, we are able to express the gauge group generators
$(F_M)_N{}^P$   as in \cite{LeDiffon:2008sh} \be F_M =  \vartheta_M { t}_0 +
(\Theta_M{}^\alpha + 8 \vartheta_P (t^\alpha)_M{}^P) t_\alpha \
.\label{gaugegenerators} \ee

In  terms of $F_{\bar A\bar B}{}^{\bar C}$, the closure conditions (\ref{closureE7})
evaluated on frames read \bea
\label{Delta4} \Delta_{\bar  A\bar B\bar C}{}^{\bar D} &=&  -\left([F_{\bar
A},F_{\bar B}] + F_{\bar A\bar B}{}^{\bar E}F_{\bar E}\right){}_{\bar
C}{}^{\bar
D}\\ && - 2 {\partial}_{[\bar A} F_{\bar B]}{}_{\bar C}{}^{\bar D} - 12
P_{(adj)}^{\bar D}{}_{\bar C}{}^{\bar E}{}_{\bar F} {\partial}_{\bar E}
F_{\bar A\bar B}{}^{\bar F}
+ \frac 12 {\partial}_{\bar E} F_{\bar A\bar B}{}^{\bar E} \delta_{\bar
C}^{\bar D}  = 0 \, .\nn \eea
When the fluxes are constant, we recover the quadratic constraints of
maximal gauged supergravity. Notice that, as it happens in DFT, these
constraints can be satisfied through configurations that violate the
section condition. This implies necessarily going beyond supergravity, and then
gives rise to a novel description of non-geometry in maximal supergravity. This
might be useful, for instance, to find an extended geometrical uplift of the new
$SO(8)$ gaugings \cite{Dall'Agata:2012bb}, which seem to find obstructions when
it comes to uplifts to $D = 11$ supergravity \cite{deWit:2013ija}. The
conditions (\ref{Delta4}), in turn, imply that the dynamical
fluxes in flat indices behave as scalars
under the following generalized diffeomorphisms with respect to
frame-vectors
\be
\delta_\xi F_{\bar A \bar B}{}^{\bar C} = \xi^{\bar D} \partial_{\bar D}
F_{\bar A \bar B}{}^{\bar C} + \xi^{\bar D} \Delta_{\bar D \bar A \bar
B}{}^{\bar C}\, .\ee

We can now proceed as in the DFT case, and look for a geometric
construction that gives the {\it action} from traces of some generalized
Ricci tensor. Of course, since we only deal with scalars here, the action
will be the scalar potential of the maximal theory. Having defined the
generalized notion of Lie derivative in (\ref{GenLieE7}), it is
natural to look for derivatives that behave covariantly under
such  transformations. We begin by defining the covariant derivative of a
bein $E_{\bar A}{}^M$ as \be \nabla_M E_{\bar A}{}^N = \omega_{M\bar A}{}^{\bar
B}
E_{\bar B}{}^N=
\partial_M E_{\bar A}{}^N  + \Gamma_{MP}{}^N E_{\bar A}{}^P\, ,
\label{RelatedConnectionsE7} \ee in terms of a Christoffel connection
$\Gamma$, or alternatively a spin connection $\omega$.
They are related to the Weitzenb\"ock connection defined in
(\ref{Weizenbockplanar}), which takes values in the algebra of $G$. In addition, one can relate the gaugings to the Weitzenb\"ock
connection through projections, as in equation (\ref{fluxesplanar}). These
connections must also transform properly so as to compensate the failure
of the derivative to transform as a tensor. Given that the covariant
derivative is requested to transform covariantly, so must the spin
connection.

We can define the generalized torsion through \cite{Uduality1} \be {\cal
T}_{\bar A \bar
B}{}^{\bar C} \equiv (E^{-1})_M{}^{\bar C} ({\cal
L}^{\nabla}_{E_{\bar A}}-{\cal L}_{E_{\bar A}})  E_{\bar B}{}^M\ ,
\label{torsion} \ee
 where ${\cal L}^{\nabla}$ is defined as in (\ref{GenLieE7}), but
with a partial replaced by a covariant derivative. Using
(\ref{RelatedConnectionsE7}) we arrive at \be \label{torsionGamma} {\cal
T}_{  \bar A   \bar B}{}^{ \bar C} =
\omega_{\bar A\bar B}{}^{\bar C} - 12 P_{(adj)}{}^{\bar C}{}_{\bar
B}{}^{\bar P}{}_{\bar Q} \omega_{\bar P\bar A}{}^{\bar Q}+\frac 1 2
\omega_{\bar D \bar A}{}^{\bar D} \delta^{\bar C}_{\bar B} - F_{\bar A
\bar B}{}^{\bar C}\ . \ee

Since $\sqrt{\cal  H }$ does not transform as a density under the
generalized diffeomorphisms (\ref{genDifU}),  the proper measure is
given by $(\sqrt{{\cal H}})^{-1/28} = e^{-2\Delta}$ since \be \delta_\xi
e^{-2\Delta} = \partial_P (e^{-2\Delta} \xi^P)\ .\label{measure} \ee
This can be used to impose compatibility with the determinant of the
generalized metric, and together with vanishing torsion they determine the
spin connection (which lives in $\bf 56 \times 133$) up to a
 $\bf 6480$ piece. This piece remains undetermined under these conditions,
but a part of it (corresponding to the $\bf 63$ in $\bf 133 = 63 + 70$)
can be fixed through metric compatibility.

It can then be shown that a torsionless and metric compatible spin
connection has in particular the following determined components
 \bea W_{PM}{}^P &=& - 2 \vartheta_M \nn\\ P_{({\bf 912})QR}{}^S,^{MN}{}_P\
W_{MN}{}^P &=& \frac 1 7 X_{QR}{}^S\nn\\
P_{({\bf 56})QR}{}^S,^{MN}{}_P\ W_{MN}{}^P &=& - \frac {16}{19}
P_{(adj)}{}^S{}_R{}^T{}_Q\ \vartheta_T \ , \label{projsSpin} \eea
where the projectors here are those of (\ref{projectorsE7}) contracted
with the $E_{7(7)}$ generators. This is analog to (\ref{traceconnection}) and (\ref{torsionless}) in DFT, where the projections there simply amounted to tracing and antisymmetrizing the spin connection.

Finally, following the DFT geometrical construction, a generalized Ricci
tensor can be constructed \cite{Uduality1} (unlike the DFT case, the definition
of a
Riemann tensor is less clear)
\be
R_{MN} = \frac 1 2 \left(R_{MN} + R_{NM} + \Gamma_{RM}{}^P Y^R{}_P{}^S{}_Q
\Gamma_{SN}{}^Q - \Omega_{RM}{}^P Y^R{}_P{}^S{}_Q \Omega_{SN}{}^Q \right)
\ee
which is covariant for solutions to the closure constraints. When tracing
it with the generalized metric we can then define a generalized Ricci
scalar
\be
{\cal R} = \frac 1 4 {\cal H}^{MN} {\cal R}_{MN}
\ee
which, for any torsionless and metric compatible connection, can be cast in
the form
\be
 {\cal R} =  \frac 1 {672}  \left(X_{MN}{}^P X_{QR}{}^S {\cal
H}^{MQ} {\cal H}^{NR} {\cal H}_{PS} + 7 {\cal H}^{MN} X_{MP}{}^Q
X_{NQ}{}^P\right) \label{RicciScalarpot}
\ee
provided the gaugings $\vartheta_M = 0$. Remarkably, this is exactly the
scalar potential of maximal supergravity with the very exact overall
factor. The relative factor $7$ is related to that in (\ref{projsSpin})
and can be traced back to the fact that the generalized Lie derivatives
are consistent with supersymmetry, in that they generate fluxes in
accordance to the linear constraints of the maximal theory.

Finally, notice that the Ricci scalar (\ref{RicciScalarpot}) can be
combined with the measure (\ref{measure}) to render a gauge invariant
action
\be
S =  \int dX e^{-2 \Delta}{\cal R}\, .
\ee

We have left some important points uncovered here. One of them is the coupling
between this ``scalar'' internal sector with the rest of the theory, i.e. the
external space-time, its metric and vectors (plus the $p$-form hierarchy of
maximal supergravities \cite{deWit:2008ta}). Steps in this direction were recently performed in \cite{Hohm:2013jma}. The other one is  the relation between this setup with
string or M-theory degrees of freedom. To establish the correspondence, one then
has to provide a proper parameterization of the generalized bein or generalized
metric (see for instance \cite{BermanPerry,Uduality1,Udualities3}).

\newpage

\section{Worldsheet motivations and approaches to DFT}
\label{sec: WSDFT}

As discussed in the previous sections,
DFT was formulated with the purpose of
incorporating T-duality, an essentially  {\it stringy} effect,
 into a {\it particle} field
theory. Clearly, it would be interesting  to deduce DFT from a
world-sheet action, much in the same way as supergravity is
obtained as the low energy effective field theory from the
two-dimensional
 description of the string dynamics.
In this section, we briefly review some of the attempts
that have been followed to construct an $O(D,D)$ covariant two-dimensional
world-sheet theory,
from which DFT might be explicitly derived.

Before proceeding to DFT,
we  briefly recall the process leading from
the world-sheet theory to supergravity.
We refer the reader to the string theory books \cite{books}
and references therein for a more detailed and complete discussion
of this issue. We then discuss
 how the procedure has been implemented
for DFT.

\subsection{The string spacetime action}

Strings propagating in backgrounds  of massless
closed string states
are described by an interacting two dimensional field theory,
obtained
by exponentiating the vertex operators creating those states. The action is
 given by
\bea
S=\frac 1{4\pi\alpha '}\int d^2\sigma \sqrt h
\left [\left (h^{ab}g_{ij}(x)+i\epsilon^{ab}b_{ij}(x)\right )
\partial_ax^i\partial_b x^j+\alpha ' {\bf R}\phi(x)\right ]\, .\label{sm}
\eea
This is a non-linear sigma model
where $\alpha '$ is the square of the string length scale,
$\sigma^a, a=0, 1$ refer to the worldsheet coordinates $\tau, \sigma$,
respectively,
$g_{ij}$ is the spacetime metric, $b_{ij}$ is
the antisymmetric
tensor, the dilaton involves $\phi$ and the trace of $g_{ij}$, and ${\bf R}$
is the curvature scalar of the worldsheet.

A consistent string theory
requires the two-dimensional quantum field theory to have
 local Weyl and Lorentz
invariance. This implies that the trace and $\epsilon^{ab}$
contraction of  the energy-momentum
tensor, respectively, should vanish on-shell, which imposes rather
non-trivial conditions
on the admissible background fields.
Actually, to regulate divergences in a quantum theory, one has  to
introduce a UV cut-off,
and therefore,
physical quantities typically depend
on the scale of a given process after renormalization. Conformal invariance is
achieved
 if the coupling constants do not depend on
the scale of the theory. In this case, the couplings are
$g, b$ and $\phi$ and the scale dependence is
described by
the $\beta-$functions of the renormalization group.

The $\beta-$functions are computed perturbatively. One first expands the
 fields $x^i(\tau, \sigma)$ around a classical solution
$x^i=x_{cl}^i+\pi^i$, where $\pi^i$ is the quantum fluctuation.
The expansion of the Lagrangian then gets quadratic kinetic terms
plus interactions of  the fluctuations. The theory
has an infinite
number of coupling constants: all order derivatives of the
background fields evaluated at  $x_{cl}^i$.
When all the couplings are small, the theory  is then weakly coupled. Assuming
the target space has a
characteristic length scale $R_c$, the effective dimensionless couplings
are of the order
$\alpha'^{1/2} R_c^{-1}$,
and then perturbation theory makes sense if $R_c$ is
much greater than the string scale.
Up to terms involving two spacetime derivatives, the $\beta-$functions are given
by
\bea
\beta_{ij}^g &=& \alpha ' {R}_{ij}+2\alpha '\nabla_i\nabla_j\phi -
\frac{\alpha '}4 H_{ikl}H_j^{kl} + O(\alpha ')\, ,\nn\\
\beta_{ij}^b&=&-\frac {\alpha '}2\nabla^k H_{kij}+\alpha '
H_{kij}\nabla^k \phi
+O(\alpha '^2)\, ,\nn\\
\beta^\phi&=&\frac{D-D_{crit}}4-\frac{\alpha '}2\nabla^2\phi +\alpha
'\nabla_k\phi
\nabla^k\phi-\frac{\alpha '}{24}H_{ijk}H^{ijk}+O(\alpha '^2)\, ,
\label{EomsSugraBeta}
\eea
where ${R}_{ij}$  is the spacetime Ricci tensor, to be distinguished from the
worldsheet
Ricci tensor  ${\bf R}_{ab}$. Terms with more derivatives are higher order in
$\alpha '^{1/2}R_c^{-1}$.  Combining (\ref{EomsSugraBeta}) one then recovers
(\ref{eom}). The term $D-D_{crit}$ in $\beta^\phi$ is the classical Weyl anomaly,
which vanishes in the critical dimension
$D_{crit}=26$ ($D_{crit}=10$) in
(super)string theory in flat
spacetime, ensuring that the negative norm states decouple.

The vanishing  $\beta-$functions equations, determining Weyl invariance and
UV finiteness of the theory,
can  be
interpreted as the equations of motion derived from the following spacetime
action
\bea
S=\int d^Dx\sqrt ge^{-2\phi}\left [-\frac{2(D-D_{crit})}{3\alpha '}
+{R}-\frac 1{12}H_{ijk}H^{ijk}+4\partial_i\phi\partial^i\phi
+O(\alpha ')\right ]\, .
\eea
We recognize here the action for the bosonic  universal
 gravity sector  introduced in (\ref{sugraaction}).\footnote{The equation
of motion  (\ref{eomd}) combined with the trace of
(\ref{eomg}) reproduce $\beta^\phi$ here when $D=D_{crit}$.}

This action can
alternatively be obtained from
 the low energy limit of scattering amplitudes of massless
string modes. Low energies  here refer to energies much smaller than the string
scale,
 i.e. $E<<(\alpha ')^{-\frac 12}$, which is equivalent to fixing $E$ and
taking the limit $\alpha '\rightarrow 0$. Recalling the mass spectrum of closed
strings
\be
M^2 = \frac 2{\alpha '}(N+\tilde N-2)\, ,
\ee
where $N, \tilde N$ are the number operators for the left and right moving
string modes,
we see that it is in this regime that massive modes decouple and
backgrounds of
 massive string states
can be consistently neglected.

The T-duality symmetry of  string scattering amplitudes  suggests
that a T-duality covariant formulation of the string worldsheet action should
exist, from which one could derive a T-duality covariant effective action,
following a procedure analogous to the one we have just described for
conventional string theory.
In the
rest of this section, we review various proposals that have been worked out in
the literature in
order to obtain such
formulation and we then discuss their connection with DFT.

\subsection{Double string sigma model}

Originally, T-duality on the worldsheet was
 implemented in two-dimensional nonlinear sigma models
in backgrounds with $n$ compact dimensions in which the
 metric and two-form
 fields
 have an isometry along the compact directions
\cite{buscher1,buscher2,siegel1}.
By gauging the isometry through a gauge connection
and adding to the action a Lagrange multiplier constraining it
to be pure gauge, so that
the number of
worldsheet degrees of freedom remains the same, one obtains the dual theory.

Specifically,
suppose the non-linear sigma model (\ref{sm})
describing the string dynamics in a metric and two-form
 background
\be
S=\frac 1{2\pi} \int d^2z ~(g_{ij}
 +b_{ij})\partial x^i\bar\partial x^j \, ,\label{sa}
\ee
is invariant under an isometry acting by translation of $x^\kappa$
and the fields $g_{ij}$ and $
b_{ij}$ are independent of $x^\kappa$ (here $x^\kappa$ refers to one or more
of the spacetime coordinates). Here, we disregard the dilaton and
use  complex coordinates
$z=\sigma +i\tau, \bar z=\sigma -i\tau$ on a flat worldsheet in units in
which $\alpha ' =1/2$.
The dual theory can be found from the extended action
\bea
S ' &=&\frac
1
{2\pi} \int d^2z \left [
(g_{ij}+b_{ij})Dx^i\bar Dx^j
+\tilde x_\kappa (\partial\bar A^\kappa
-\bar\partial A^\kappa)\right ]\, , \label{sma}
\eea
where $Dx^\kappa=\partial x^\kappa+A^\kappa$, and the Lagrange multiplier
$\tilde x_\kappa$
enforces the pure gauge condition $\partial \bar A^\kappa -\bar\partial A^\kappa
=0$.
Gauge fixing $x^\kappa =0$, one obtains the
dual model
\be
\tilde S =\frac 1{2\pi} \int d^2z
(\tilde g_{ij} +\tilde b_{ij})\partial\tilde x^i
\bar\partial\tilde x^j
\ee
 by integrating out the gauge fields. In this new theory,
$\tilde g_{ij} $ and $\tilde b_{ij}$
are given by Buscher's rules (\ref{Buscher})
\bea
&& \tilde g_{\kappa\kappa}= \frac 1{g_{\kappa\kappa}} \ , \ \ \ \ \tilde
g_{\kappa i} =  \frac {b_{\kappa i }}{g_{\kappa\kappa}}
\ , \
\ \ \ \tilde g_{ij} = g_{ij} - \frac{g_{\kappa i} g_{\kappa j} -
b_{\kappa i}b_{\kappa j}}{g_{\kappa \kappa}}\ , \nn\\\nn\\
&& \tilde b_{\kappa i} =  \frac{g_{\kappa i }}{g_{\kappa\kappa}}\ , \ \ \ \
\tilde b_{ij} = b_{ij} +
\frac{g_{\kappa i} b_{\kappa j} - b_{\kappa i}g_{\kappa j}}{g_{\kappa \kappa}}
\, .
\eea
Clearly,
the background fields are
in general completely changed by the duality transformation.

In \cite{rv}, the (abelian) T-duality transformations were reformulated
in terms of chiral Noether currents associated with the isometries, and
it was  shown that any dual pair of sigma models can be
obtained by gauging different combinations of chiral currents. The
equivalence of dual  sigma models
at the quantum level was
analyzed in \cite{frj}, where it was shown that, while one Lagrangian
representation is
I.R. free, the dual one is asymptotically free.

This initial approach to deal with T-duality in the worldsheet theory
allows to map a sigma model action to its
T-dual one, but neither of them is manifestly $O(D,D)$ covariant.
However, as mentioned above, the T-duality symmetry of string theory suggests
that an $O(D,D)$ covariant  worldsheet action should exist.
A natural guess for such  formulation
would be a sigma model where the target space coordinates are doubled.
Actually,
a democratic treatment of momentum and winding modes
 leads to
consider independently
the ordinary target space coordinates
 $x^i=x^i_+(\sigma + \tau)+x^i_-(\sigma - \tau)$ associated to momentum
and the dual ones
 $\tilde x^i= x^i_+(\sigma + \tau)
-x^i_-(\sigma - \tau)$ associated to winding, or equivalently,
 the left- and right-moving closed string fields.
 Moreover, since T-duality mixes the metric and two-form fields,
it is reasonable to expect that
these
fields combine to form
the generalized metric ${\cal H}_{MN}$ in (\ref{gennmet}).
These heuristic arguments lead to a world-sheet action of the form
\be
S = \int
dX^M \wedge \star dX^N {\cal H}_{MN} \, ,\qquad
X^M =\left(\begin{matrix}
\tilde x_i\\
x^i
\end{matrix}\right )\, ,
\ee
where $\star$ is the Hodge dual operation on the worldsheet, which is manifestly
duality covariant and two dimensional Lorentz invariant. However, this action
describes twice as many degrees of freedom as the  action (\ref{sa}), and then,
it has
to be supplemented with additional constraints in order to eliminate
the extra coordinates and be able to reproduce the same physics.

A different approach was followed by A. Tseytlin,
who
was able to construct a
manifestly $O(D,D)$ covariant worldsheet action
for chiral bosons  \cite{tseytlin}. Evidencing  that T-duality is
a canonical transformation of the phase space of string theory \cite{sfl},
$O(D,D)$ covariance is achieved through a first order action in time
derivatives:
\bea
S
&=&\frac 1{2}\int d^2\sigma \left ({\cal H}_{MN} \partial_1 X^M\partial_1 X^N-
\eta_{MN}\partial_0 X^M \partial_1 X^N
\right )\, ,\label{ta}
\eea
which  can be naturally interpreted as being
  expressed in terms of phase space variables, with the dual
fields playing the role of the integrated canonical momenta.
This action is invariant under  the following sigma model type symmetry
\be
 X\rightarrow \eta X\, ,\quad {\cal H}\rightarrow \eta^T {\cal H}\eta\, ,
\ee
transforming both the fields
and the couplings.

The price for having duality as a symmetry of the action is the
lack of  two-dimensional Lorentz invariance.
Indeed, introducing the left and right moving parts of the
string coordinates as independent off-shell fields, one has to face
the issue of having to deal with a non-Lorentz invariant action.
This is actually the
case in any Lagrangian description of chiral scalars, as originally discussed in
\cite{fj}, and in general, Lorentz invariance is recovered on-shell \cite{duff, tseytlin, schs}.
Local Lorentz invariance
is achieved here if  ${\cal H}_{MN}$ is either
constant or
depends only on half of the coordinates (in the language of DFT, Lorentz invariance requires the strong constraint \cite{copland}).
The equivalence of the equations of motion following from (\ref{ta}) with those of the ordinary sigma-model (\ref{sa})
was shown in \cite{copland}.

One way to obtain the action (\ref{ta})
starting from  (\ref{sm}) (setting $\phi=0$), is to
 write the Hamiltonian density  in a manifestly
$O(D,D)$ invariant form, in terms of the canonical momenta $p_i$ conjugate to
$x^i$:
$p_i=-g_{ij}\partial_0x^j+b_{ij}\partial_1x^j$, namely
\bea
{H}=\frac 12 \Psi^M{\cal H}_{MN}\Psi^N\, ,\qquad {\rm with}\qquad\Psi^M=\left
(\begin{array}c
p_i\\
\partial_1x^i\end{array}
\right )\quad
\, .
\eea
Identifying the momenta $p_i$
with the dual coordinates $\tilde x_i$ as $p_i=\partial_1\tilde x_i$,
and rewriting the Lagrangian as $L=p_i\partial_0x^i-H$,
the  action (\ref{sm}) can be recast in terms of the double coordinates $X^M$ as
\be
S=\frac 12\int d^2\sigma\left ({\cal H}_{MN}\partial_1X^M\partial_1X^N-
\eta_{MN}\partial_0X^M\partial_1X^N-\Omega_{MN}\partial_0X^M\partial_1X^N\right
)\label{re}
\ee
where
 \bea
\Omega_{MN}=\left (\begin{array}{cc}
0&\delta_i{}^j\\
-\delta^i{}_j&0
\end{array}\right )\, .
\eea
The $\Omega$-term does not contribute to the
field equations and does not affect the classical theory, but it is necessary in
the
quantum theory \cite{doublegeom}
and, in particular, to show the equivalence of
the doubled to the conventional partition function \cite{bt}.\footnote{Reference \cite{bt} also shows
the modular invariance of the one loop double string theory.} The correspondence with the standard
formulation of critical string theory
only
appears
after
integrating out one of the dual fields. Then, either
 the original or the dual Lorentz invariant
action is recovered.

A similar procedure was followed by W. Siegel in the so-called
 two-vielbein formalism \cite{siegel1},
where the metric and antisymmetric tensor are combined in two independent
vielbeins.
Alternatively, as demonstrated
in \cite{rt}, the  non-Lorentz invariant doubled action (\ref{ta}) can be
obtained by fixing the axial
 gauge in the
duality and Lorentz invariant extended action (\ref{sma}).
More recently  in \cite{patalong}, a
 manifestly Lorentz invariant action was obtained based on a path integral analysis at the quantum level.

Tseytlin's formulation can be generalized
 to allow for background fields with arbitrary dependence on the double
coordinates,
i.e. other than a generalized metric
${\cal H}_{MN}(X)$ generically depending on the double coordinates,
one can include a symmetric matrix ${\cal G}_{MN}(X)$
generalizing $\eta_{MN}$,
and an antisymmetric tensor ${\cal C}_{MN}(X)$:
\be
S=\frac 12\int d^2\sigma \left [-({\cal C}_{MN}(X)+{\cal G}_{MN}(X))
\partial_0X^M\partial_1X^N+{\cal H}_{MN}(X)\partial_1X^M\partial_1X^N\right ]\,
.\label{dpa}
\ee
Demanding on-shell Lorentz symmetry of this action
gives constraint equations for the background fields.\footnote{The term with coupling ${\cal
C}_{MN}(X)$
is clearly Lorentz invariant.}  Classical  solutions of these equations
were found in \cite{avra}.
More general non-linear sigma models of this form, in which the generalized metric is
replaced by a generic symmetric matrix,
were analyzed in \cite{sfetsos}, and it was shown that the solutions to the Lorentz
invariance constraints give an action with the form of the Poisson-Lie T-duality action
introduced in \cite{klimcik}.

For completeness we list here other approaches that have been followed in the
literature to construct
double sigma models.

\begin{itemize}

\item
In backgrounds with a toroidal fibre, the string dynamics
can be described
by the (partially) doubled formalism
introduced by C. Hull in \cite{doublegeom}.
This formalism describes a worldsheet
embedding into backgrounds that are locally $T^n$ bundles,
 with coordinates
$(Y^i, {\mathbb X}^A)$,
where $Y^i$ are the coordinates of the base and ${\mathbb X}^A$
are the coordinates
of the doubled torus fibre.
The Lagrangian
is a sum of
an ordinary sigma model Lagrangian $L(Y)$  like that in (\ref{sm}) plus
a sigma
model  for the generalized metric ${\cal H}_{AB}$ of the doubled fibres, which
crucially only depends on $Y^i$
 and there is isometry in all
the
fibre directions:
\be
{\cal L}=\frac 14{\cal H}_{AB}(Y)d{\mathbb X}^A\wedge \star d{\mathbb X}^B+\frac
12\Omega_{AB}d{\mathbb X}^A\wedge d{\mathbb X}^B
+L(Y)\, .
\ee

The action must be supplemented with the chirality constraint
\be
d{\mathbb X}^A=\eta^{AB}{\cal H}_{BC}\star d{\mathbb X}^C\, , \label{cc}
\ee
ensuring that the fibre directions can be thought of as chiral bosons on the
worldsheet, so that the doubling does not increase the number of
physical degrees of
freedom.

This doubled formalism has been very useful
 in elucidating the structure of non-geometric
backgrounds, such as T-folds.

\item
In \cite{WorldsheetDFT} the constraint (\ref{cc}) was incorporated into the
action,
which then reads:
\be
S=\frac 12\int d^2\sigma\left [-{\cal G}_{\alpha\beta}\partial_1X^\alpha
\partial_1 X^\alpha+{\cal L}_{\alpha\beta}\partial_1X^\alpha\partial_0X^\beta
+{\cal K}_{\alpha\beta}\partial_0X^\alpha\partial_0X^\beta\right ]\, ,
\ee
where $X^\alpha = (Y^i, {\mathbb X}^A)$ and
\bea
{\cal G}_{\alpha\beta}=\left (\begin{array}{cc}
g_{ij}
&0\\ 0&{\cal H}_{AB}
\end{array}\right )\, ,\quad
{\cal L}_{\alpha\beta}=\left (\begin{array}{cc}
0
&0\\ 0&{\eta}_{AB}\end{array}\right )\, ,\quad
{\cal K}_{\alpha\beta}=\left (\begin{array}{cc}
g_{ij}&0\\ 0&0
\end{array}\right )\, ,\quad\eea
$g_{ij}$ being the standard sigma model metric for the base.

\item
For
doubled backgrounds  which are locally  a  group manifold, the non-linear
Poisson Lie sigma
model proposed in \cite
{klimcik}
was rewritten in
\cite{re} as
\be
S=\frac 12\oint d^2\sigma \left ({\cal
H}_{AB}\partial_1X^A\partial_1X^B-\eta_{AB}\partial_1X^A\partial_0X^B
\right )+\frac 1{12}\int_V
{t}_{IJK}{dX}^I\wedge{dX}^J\wedge{dX}^K\, ,\ee
where $V$ is the volume of the membrane whose boundary is the string worldsheet
and $t_{IJK}$ are the structure constants of the gauge algebra.

\end{itemize}

\subsection{DFT from the double sigma model}

We have seen that
the vanishing $\beta-$functions equations can be interpreted as
equations of motion derived from the
string low energy  effective field theory. It is then through the background
field equations of the double sigma model that one expects to make the
connection between
string theory and  DFT. But this raises some conceptual questions. For instance,
perturbation theory
in the non-linear sigma model (\ref{sm}) is performed
around
the large volume limit. Can one also define a perturbation
theory around the limit of very small substringy sizes of the
background? Or even more puzzling, is there a well defined weak-coupling limit
of the T-duality invariant models (\ref{ta}) or (\ref{dpa})?

These questions have been analyzed by several authors from different
viewpoints. From the $\beta-$functions standpoint,
the background
field method was adapted to the doubled coordinates,
expanding them around a classical solution as $X^M=
X^M_{cl} +\Pi^M$.
Since the fluctuation $\Pi^M$
does not in general transform as a vector, in order to have a covariant
expansion,
the expansion parameter is defined
as the tangent vector to the geodesic from $X^M_{cl}$ to $X^M_{cl} + \Pi^M$
whose length
is equal to that of the geodesic.
Since this is a contravariant vector,
the expansion is then organized in terms of covariant objects.

The crucial point to elucidate
in order to consistently apply this method to the double sigma models is how to
define geodesics
in double geometries.
The simplest options starting from the action (\ref{dpa}) are
to consider geodesics of ${\cal G}_{MN}$ or  geodesics
of ${\cal H}_{MN}$, with the
resulting expansions involving covariant derivatives and tensors with respect to
the chosen
metric. The background field method was first applied to the sigma model of Hull's
doubled formalism in \cite{WorldsheetDFT} using geodesics of ${\cal H}_{MN}$.
This was then generalized  in \cite{avra} for the sigma model
(\ref{dpa}) where the expansion was also performed  using geodesics of ${\cal G}_{MN}$
and general expressions for the Weyl and Lorentz anomaly
terms were found.

UV finiteness and worldsheet Weyl invariance
 at one loop were shown in \cite{WorldsheetDFT} to
require the vanishing of the generalized Ricci tensor
 when DFT is restricted to a
fibered background of
the type that the doubled formalism describes. In \cite{copland}, the vanishing $\beta^{\cal H}-$function equation
from the sigma model
(\ref{ta}), with ${\cal H}_{MN}$  arbitrarily
depending on any of
the doubled coordinates,
subject to the strong constraint, was  found to
match the equation
of motion for the generalized metric obtained from the DFT action
(\ref{ActionDFTgenmet}). Hence,
conformal invariance of the double chiral sigma model (\ref{ta}) under the strong constraint, corresponds to the
generalized Ricci flatness equation, and this  implies that DFT is the spacetime
effective field theory of the double worldsheet action. A preliminary
similar result was also found in \cite{copland} for the generalized dilaton.

Although imposing the strong constraint means the theory is no longer
truly doubled, the appearance of the generalized Ricci tensor
in this context is non-trivial
and seems evidence not only of an effective double geometry, but also
of a string theory origin of DFT.  Nevertheless,
 it would be interesting to investigate if these
conclusions still hold beyond the strong constraint, in a truly double space.
Indeed, as extensively discussed in the preceding sections,
 DFT with the strong constraint
 is equivalent to the standard field theory description of the massless
modes of the string.
Actually, the strong constraint implies one
can  perform an $O(D,D)$ rotation so that the fields only depend on
$x^i$ and, since all $O(D,D)$ indices in the action are contracted properly,
its form is preserved under such a rotation.

Furthermore, we have seen that in standard string theory,
 perturbation theory makes sense in the
background field expansion of  the  action (\ref{sm})
if $\alpha '^{1/2}<<R_c$ and in this regime
one can also neglect the massive string states. In order to analyze the validity
of perturbation theory in the double space,
since
$O(n,n)$ duality is a symmetry of string theory on an $n-$torus with
a constant $b_{ij}$ background,
which survives in the effective field theory when it is dimensionally reduced
on $T^n$,
it is convenient to
recall the mass spectrum
of closed strings in these backgrounds.
In terms of the quantized canonical momenta $p_i=\frac{n_i}R$
and winding numbers $\tilde p^i$, the
mass in  $d=D-n$ dimensions is given by
\bea
m^2&=&\frac 1{2\alpha '^2}g_{ij}(v_L^iv_L^j+v_R^iv_R^j)+\frac 2{\alpha '}
(N+\tilde N-2)\, ,\nn\\
v^i_{L,R}&=&\alpha 'g^{ij} \left (\frac{n_j}{R}+b_{jk}\tilde p^kR\right ) \pm
\tilde p^i R\, ,
\eea
where $N, \tilde N$ are the number operators for the left and right moving
oscillators, respectively,
of all the coordinates: compact and non-compact (we
are assuming, for simplicity,
 the same radius $R$ for all compact dimensions). DFT
deals with massless states of the $D-$dimensional theory, i.e. having
 $N=\tilde N =1$, and then it includes all the momentum and winding modes
of the lower dimensional theory, which are massive.
Since it truncates the massive levels (of the decompactified theory), one wonders whether this corresponds to a consistent truncation.
Then, a better understanding of this issue seems necessary in order to
strengthen the
link between string theory and DFT.
By the same token, given that a T-duality symmetric description
treats the compactification scale and its inverse on an equal footing,
it seems important to clarify what is the rank of parameters for which the
coupling constants
are small and the theory is weakly coupled, so that the
perturbative expansion can be trusted.

Another way to tackle these issues is through  the computation of
scattering amplitudes
describing the interactions of winding and momentum states in geometric and
non-geometric backgrounds. A first step in this direction
was taken in
\cite{blum},
where scattering amplitudes of closed string tachyons in an $R-$flux background
were computed
and a very interesting non-associative behavior of the spacetime coordinates
was found. We shall review this work in the next Section, but here
we point out that an effective field theory analysis of
these kinds of scattering amplitudes, which would give an alternative approach
to this question,  is not yet available.

In the absence of a better comprehension, it is important to note that
the background-field equations in a particular duality frame are the same
as for the usual string.
A priori this does not
have to be the case since, as we have seen,
the string winding modes  could in
principle correct the usual  $\beta$-functions. Moreover, given that
T-duality is corrected by worldsheet
instantons
and  the doubled space contains the naive T-dual,
corrections to
the double
geometry could arise.
It is then reasonable to expect that once
 double geometry is understood, one will be able to
elucidate
these questions.
In this sense, a higher loop calculation of the $\beta-$functionals
would be important since the full generalized Riemann tensor
is expected to appear
in case the analogy with  ordinary string theory goes through.
As a matter of fact, as discussed in Section \ref{sec: Double Geometry},
although
a duality covariant generalized Riemann tensor has been  constructed
whose contractions give the generalized Ricci tensor and scalar,
it cannot be completely determined from the
physical fields of DFT as in ordinary
Riemannian
geometry. Better understanding the link between string theory and DFT
might also help to uncover
 the geometry of the double space.

\newpage

\section{Other developments and applications} \label{sec:Further developments}

DFT has proven to be a powerful tool to explore string theoretical features
beyond
the supergravity limit and Riemannian geometry. In the past few years there has
been a great deal of progress on these
issues, growing largely out of the systematic application of symmetries and dualities. We certainly do not have a complete
understanding of DFT, but
an increasing number of promising directions have opened following the original
works and several encouraging  ideas have been put forward. We cannot discuss
all of them  in detail
here but, besides the topics covered above and by way of  conclusion,
we would like to comment on  some
 recent developments and open problems.

\subsection {Non-commutative/non-associative structures in closed string theory
}
In the presence of a constant two-form field,
the coordinates of the end-points of open strings attached to a D-brane
become non-commutative \cite{noncomop}. Moreover, in the background of a
non-trivial $H-$flux, the coordinates
are not only
non-commutative but also non-associative \cite{herbst}.
This behavior is revealed by scattering amplitudes of open string states
and usually interpreted as a consequence of
the structure of interactions in
open string theory, which involve Riemann surfaces with boundaries.
To compute scattering amplitudes,
the vertex operators creating open string states must be inserted
on the boundaries, and
then, a
background that is sensitive to ordering  might distinguish
the insertion points.

In contrast, the sum over world-sheets defining
interactions of closed strings,
contains  Riemann surfaces with no boundaries, in which the vertex
operators are inserted in the bulk.
Therefore, one would not expect to have non-commutative coordinates
in closed string theory because
no unambiguous notion of ordering can be defined in scattering amplitudes.
 However, it has been argued
that in presence of non-geometric fluxes, the coordinates of closed
strings can
become non-commutative or even non-associative \cite{bp,noncommutclo,blum}.

Actually, non-geometric fluxes twist the Poisson structure of the phase-space of
closed strings and
 the non-vanishing
equal time commutator of closed string coordinates in  a $Q-$flux background
has been
conjectured to be given  by
\be
\lim_{\sigma\to
\sigma '} \left [x^i(\tau, \sigma), x^j(\tau, \sigma ')\right ] =
\oint_{C_k}Q^{ij}{}_k dx^k\, ,\label{noncom}
\ee
where $C_k$ is a cycle around which the closed string  wraps, while
non-associativity has been argued to arise in an $R-$flux background in which
\be
\lim_{\sigma ',\sigma ''\rightarrow \sigma} \left (\left [x^i(\tau,
\sigma),\left [x^j(\tau, \sigma '),x^k(\tau, \sigma '')\right ]\right ]+cyclic
\right )=
R^{ijk}\, .
\ee

In particular,
non-commutativity has been studied in the three dimensional background with
 $Q-$flux that is dual to
the flat three-torus with $H-$flux
discussed in Sections 5.2 and 5.3.
Recall that
one can use Buscher's rules (\ref{Buscher})
to map the  flat three-torus with $H-$flux  to a twisted torus with zero
$H-$flux in which the twist
is related to a  geometric flux  $\omega$. A further T-duality
 then yields the non-geometric $Q-$flux background
in which
the metric and two-form  are locally but
not globally well-defined.
In the simple case in which $C_k$
is a circle and the $Q-$flux is constant:  $Q^{ij}{}_k=Q\epsilon^{ij}{}_k$, the
commutator (\ref{noncom})
becomes
\be
\lim_{\sigma\to
\sigma '} \left [x^i(\tau, \sigma), x^j(\tau, \sigma ')\right ]=
2\pi Q\epsilon^{ij}{}_k\tilde p^k\, ,\label{noncomc}
\ee
and
then we see that non-commutativity is a non-local effect related to winding.

Non-associativity of the string coordinates was first observed
 in the theory of
closed strings moving on the three-sphere $S^3$ in the presence of an
$H-$flux background
\cite{bp}.  This theory is described by the exactly solvable $SU(2)_k$
WZW model, and then a conformal field theory computation can be performed.
A non-vanishing
equal-time, equal-position cyclic double commutator of the spacetime
 coordinates,
independent of the world-sheet coordinates, was found.
More recently, a non-trivial cyclic three
product was also found in \cite{blum} from the  scattering
amplitudes of closed string tachyon vertex operators
in an  $R-$flux background. The three tachyon correlator gets
a non-trivial phase in  $R-$space depending on the
operators ordering,
 before enforcing momentum conservation. The non-vanishing cyclic
three-bracket of the
coordinates appears then to be
consistent with the structure of two-dimensional conformal field theory.

The non-associative geometry probed
by closed strings in flat non-geometric $R-$flux backgrounds has been also
studied in \cite{Mylonas:2012pg} from a different perspective. Starting from a
Courant sigma-model on an open membrane, regarded as a topological
sector of closed string dynamics in $R-$space,  the authors derive a twisted Poisson sigma-model on
the boundary of the membrane.
For constant $R-$flux, they obtain closed formulas for the corresponding
non-associative star product and its associator.

Recall that starting from a geometrical background and performing three
T-dualities in these three-dimensional backgrounds, one runs out of
isometric directions.  In particular, in the $R-$flux background,
 the notion of locality is completely lost in  the conventional space.
In DFT instead, the resulting
background  depends on a dual coordinate, and
these global and local issues
 can be avoided.
Thus, by naturally incorporating
all the T-dual backgrounds in a covariant picture through a double space,
DFT
provides a convenient framework for analyzing
non-commutativity/non-associativity. Actually, as discussed in \cite{noncommutclo},
in the doubled phase
space
T-duality would  exchange commutators among the conventional space-time coordinates with others
among the dual ones.
 If coordinates commute in the first setting while the duals do
not, the situation gets exchanged after T-duality.

\subsection{Large gauge transformations in DFT}

While all the results of this review are based on the infinitesimal generalized
diffeomorphisms (\ref{gendiffs}), finite gauge transformations were considered
by O. Hohm and B. Zwiebach in \cite{Hohm:2012gk} and J-H. Park in \cite{Parkdiffeos} (see also \cite{yetgauged}) under the imposition of the
strong constraint. They are defined through exponentiations
of the generalized Lie derivatives, and are interpreted as generalized
coordinate transformations in the doubled space. In \cite{Hohm:2012gk}, a
formula for large gauge transformations was proposed and tested, which is
written in terms of derivatives of the
coordinate maps. Successive
generalized coordinate transformations give a generalized coordinate
transformation that
differs from the direct composition of the original two: it is constructed using
the C-bracket. Interestingly, although these transformations form a group when
acting on fields,
they do not associate when acting on coordinates, and then one wonders whether
this can be related to the works in \cite{bp}.

By now, it is not completely known how to construct a non-trivial patching of
local regions of the doubled manifold
leading to non-geometric  configurations. As we reviewed, the notion of a
T-fold
is based on the idea that field configurations on overlaps can be glued with the
use of T-duality
transformations. In order to address questions of this type in double field
theory we need
a clear picture of the finite gauge transformations. This is a very interesting
line of research.

\subsection{New perspectives on $\alpha '$ corrections}

The effective supergravity action is nicely covariantized under the T-duality group and generalized diffeomorphisms. One can then wonder if a similar
covariantization occurs for the $\alpha'$ corrections to the action. This question was posed in \cite{GeometryZwiebach}, where a first step in this direction was given. Specifically, within a generalized metric formulation, it was shown that the Riemann-squared scalar $R_{ijkl} R^{ijkl}$, familiar in $\alpha'$ corrections to the low-energy effective action of string theory, is not obtained (after the proper implementation of the strong constraint in the supergravity frame $\tilde \partial^i = 0$) from any covariant expression built out of the generalized metric and generalized dilaton (and setting $b_{ij} = \phi = 0$), and quartic in generalized derivatives. For the sake of concreteness, let us be more specific. This obstruction appears due to a problematic contribution in the expansion, taking the form $g^{np}g^{iq} g^{kl}g^{mt}g^{rs} \partial_k g_{mr} \partial_i g_{ns} \partial_q \partial_l g_{pt} $. It was shown that there is no possible covariant combination giving rise to a term like this, and the
origin of this problem can be traced back to the $O(D,D)$ structure of the generalized metric.

To understand the significance of this result, suppose one had succeeded in constructing such a covariant combination related to $R_{ijkl} R^{ijkl}$. Then, one could have written a general four-derivative action from linear combinations of the squares of the generalized curvatures. Being constructed from covariant objects, any of them would be invariant. As argued in \cite{GeometryZwiebach}, this would be unexpected because the field redefinitions $g_{ij}\to g_{ij} + \alpha' (a_1 R_{ij} + a_2 g_{ij} R)$ that respect diffeomorphism invariance, map $\alpha'$-corrected actions into each other, and alter the coefficients of Ricci-squared and R-squared terms. After these field redefinitions, the T-duality transformation of $g_{ij}$ would be $\alpha'$-corrected, in conflict with the original assumption.

Although things are not as easy as one would have liked them to be, a better understanding of $\alpha'$ corrections in DFT would help to understand the contributions of the two-form and  dilaton to the corrections, as commented in \cite{Garousi:2013zca}, and also possibly help to find new patterns, based on duality arguments. Also, a better understanding of this problem might shed light on the mysterious un-physical components of the generalized Riemann tensor, and viceversa.
In any case, $\alpha'$-corrections to supergravity in the context of DFT seem to be a very promising line of research, where plenty of things remain to be done and learnt. Soon after this Review appeared, progress in this direction was done in \cite{Hohm:2013jaa} and \cite{Godazgar:2013bja}.

\subsection{Geometry for non-geometry}

As we have seen, T-duality  appears to imply that the
geometrical structure underlying string theory goes
beyond the usual framework of differential
geometry and suggests an extension
of the standard diffeomorphism group of General Relativity.
A  new geometrical
framework to describe the non-geometric structures was developed   in
\cite{Andriot:2012an,Blumenhagen:2013aia}.
The idea
  of these works is to provide a general formulation
 to study non-geometric backgrounds
in conventional higher dimensional space-time, in a formalism that facilitates the treatment of global issues that are problematic in standard supergravity.

When the  generalized metric has the form
(\ref{gennmet}), it is said to be in the {\it geometric frame}.  A general
$O(D,D)$ transformation mixes the usual metric and two-form fields
in a complicated way and a (T-duality inspired) field redefinition is convenient to re-parameterize the generalized metric such that the description of non-geometric backgrounds becomes more natural (this can be named the {\it non-geometric frame}).
The field redefinition makes a dual metric and a bivector $\beta^{ij}$ enter the game, and these now become the fields of the geometric action for non-geometric fluxes.
Performing these field redefinitions  in the supergravity action (\ref{sugraaction}), makes
the non-geometric fluxes appear, in such a way that the new actions
are well-defined in terms of the new fields. DFT provides a natural framework to interpolate between these two frames, in which geometric and non-geometric backgrounds are better described.

 The new actions can be interpreted as coming from the
differential geometry of  Lie algebroids.
These are generalizations of Lie algebras where the structure
constants can be space-time dependent. Lie algebroids give a natural generalization
 of the familiar  concepts
of standard
Riemannian geometry, such as covariant derivatives, torsion and curvatures.
A detailed account of
the relation between these conventional
objects and those
appearing for Lie algebroids
 is presented in \cite{Blumenhagen:2013aia}.

These are very nice results that specialize in the geometry and dynamics of non-geometric backgrounds.

\subsection{Beyond supergravity: DFT without strong constraint}

As we explained, in order for the generalized Lie derivative to generate closed
transformations, the fields and gauge parameters of the theory must be
constrained, i.e. DFT is a restricted theory (\ref{closure}). One possibility to solve the
constraints is to impose an even stronger restriction: the strong constraint
(\ref{strongconstraint}). This possibility is the most explored one, and allows
for a generic form of fields and gauge parameters, but with a strong restriction
in their coordinate dependence: they can only depend on a (un-doubled) slice of
the double space. This enables a direct relation to supergravity, and puts DFT
in a safe and controlled place. There are however other solutions
\cite{Aldazabal:2011nj,Geissbuhler:2011mx,GDFT,SSMtheory,Musaev,Extended
geometry,GeometryBerman} in which the {\it shape} of the fields is restricted,
but not the coordinate dependence, which can then be truly double. As we
extensively reviewed, in this situation the fields adopt the form of a
Scherk-Schwarz reduction ansatz, and this facilitates to make contact with
gauged supergravities in lower dimensions. The double coordinate dependence here
is encoded in the gaugings, which cover the corners of the configuration space
that are not reached from standard supergravity compactifications.

These doubled solutions correspond to the first attempts of consistently going
beyond supergravity in DFT. Whether these extensions live within string theory is
a question that remains unanswered. This seems most likely to be the case,
because these extensions are precisely governed by the symmetries of string
theory. In any case, DFT provides a (stringy-based) scenario in which
supergravity is only a particular limit, and many explorations beyond this limit
still have to be done.

\subsection{(Exotic) brane orbits  in DFT}
In the open string sector, T-duality exchanges Dirichlet and Neumann boundary
conditions, and then relates D-branes with different dimensionalities. This
situation was nicely depicted in the double torus in \cite{Doublebranes}. After
evaluating the one-loop beta function for the boundary gauge coupling, the
effective field theory for the double D-branes was  obtained, and is described
by a T-duality covariant DBI action of double fields.

In the NS-NS sector, the NS5-brane and KK5-monopole were also considered in the
double torus \cite{NS5/KK5}. Both configurations are related by T-duality, and
the orbit is known to continue. By applying a further T-duality, one obtains the
$5^2_2$-brane (see \cite{deBoer:2010ud,Rotating string} for detailed discussions) which looks like a T-fold, and is a special
case of a $Q$-brane \cite{Hassler:2013wsa}. DFT allows to T-dualize further in
order to obtain an $R$-brane. The picture is analogous to that of duality orbits on
non-geometric fluxes. The {\it exotic} $Q$ and $R$-branes are nicely
accommodated in DFT, and the frameworks of \cite{Andriot:2012an} and
\cite{Blumenhagen:2013aia} suitably describe their underlying geometry. Being
sources of non-geometric fluxes, they exhibit an interesting
non-associative/non-commutative behavior \cite{Hassler:2013wsa}.

The NS5 and KK5 source Bianchi identities on their worldvolumes
\cite{Villadoro:2007tb}, and their exotic T-duals are likely to source the
corresponding T-dual Bianchi identities for non-geometric fluxes
\cite{Exploring}. Interestingly, these Bianchi identities are naturally
identified with the consistency constraints of DFT (\ref{Zetas}).

Brane orbits have been extensively discussed in \cite{Bergshoeff:2012jb} and
\cite{deBoer:2010ud}, and we refer to those papers for a general discussion on
the topic. There are still plenty of unanswered questions, for example regarding
the existence (and validity) of bound states of geometric and non-geometric
branes described by configurations that violate the strong constraint. This
exciting area of research is just beginning, and DFT seems to be a suitable
framework for exploration.

\subsection{New possibilities for upliftings, moduli fixing and dS vacua}

Only a subset of all the possible deformations (gaugings or fluxes) in gauged
supergravities in four-dimensions can be reached from standard compactifications
of $D = 10,11$ supergravity, as we explained. The rest of them (the
non-geometric orbits) on the other hand don't admit supergravity uplifts, and
then one has to appeal to duality arguments in order to make sense of them from
a lower dimensional perspective. DFT (and the more general U-duality covariant
frameworks) provide a suitable scenario to uplift non-geometric orbits in an
extended geometrical sense \cite{Dibitetto:2012rk}. As we explained,
non-geometric fluxes seem to be  necessary ingredients in purely flux-based
moduli stabilization surveys \cite{dS}.  The same happens in dS vacua
explorations. Although there is beautiful recent progress in the quest for
classical (meta)stable dS vacua with non-geometric fluxes \cite{dS}, their
uplift to extended geometry (in particular DFT) or the $10$-dimensional
geometric actions for non-geometric fluxes
\cite{Andriot:2012an,Blumenhagen:2013aia} is still an open question.

Once again, DFT seems to provide a suitable framework to uplift the gauged
supergravities with non-geometric fluxes that give rise to desired
phenomenological features. Progress in this direction was achieved in some
particular gauged supergravities \cite{Dibitetto:2012rk} through consistent
relaxations of the strong constraint.

Also in this direction, the extended geometry of \cite{Extended geometry} might
shed light on the uplifts of the new $SO(8)$ maximal supergravities
\cite{Dall'Agata:2012bb}, which seem to find obstructions in their uplift to $D
= 11$ supergravity \cite{deWit:2013ija}.

\section{Acknowledgments}
{We are very grateful to W. Baron, G. Dibitetto, J. J. Fernandez-Melgarejo, D.
Geissbuhler, M. Grana, V. Penas, D. Roest, A. Rosabal for collaboration in some
works covered by this review. We warmly thank D. Berman, N. Copland, G. Dibitetto, M. Grana,
O. Hohm, J-H. Park, V. Penas, D. Roest, A. Rosabal and B. Zwiebach for enlightening comments and corrections to the review. We are also indebted with E. Andres, W. Baron, O. Bedoya, M. Galante and  S. Iguri
 for helping us to improve the presentation.
CN is specially grateful to V. Rivelles for the invitation to
write this review.
This work was partially supported
by CONICET, UBA and
EPLANET.}

\newpage
\label{sec:References}


\begin{thebibliography}{98}


 \bibitem{Siegel:1993th}
  W.~Siegel,
  ``Superspace duality in low-energy superstrings,''
  Phys.\ Rev.\ D {\bf 48} (1993) 2826
  [hep-th/9305073].

  W.~Siegel,
  ``Two vierbein formalism for string inspired axionic gravity,''
  Phys.\ Rev.\ D {\bf 47} (1993) 5453
  [hep-th/9302036].


\bibitem{Hull:2009mi}
  C.~Hull and B.~Zwiebach,
  ``Double Field Theory,''
  JHEP {\bf 0909} (2009) 099
  [arXiv:0904.4664 [hep-th]].



\bibitem{books}

  J.~Polchinski,
  ``String theory. Vol. 1: An introduction to the bosonic string,
 Vol. 2: Superstring theory and beyond,''
  Cambridge, UK: Univ. Pr. (1998)

  M.~B.~Green, J.~H.~Schwarz and E.~Witten,
  ``Superstring Theory. Vol. 1: Introduction,
Vol. 2: Loop Amplitudes, Anomalies And Phenomenology,''
  Cambridge, Uk: Univ. Pr. ( 1987). ( Cambridge Monographs On Mathematical
Physics)

 K.~Becker, M.~Becker and J.~H.~Schwarz,
  ``String theory and M-theory: A modern introduction,''
  Cambridge, UK: Cambridge Univ. Pr. (2007)

 E.~Kiritsis,
  ``String theory in a nutshell,''
  Princeton University Press, 2007

  B.~Zwiebach,
  ``A first course in string theory,''
  Cambridge, UK: Univ. Pr. (2009)



  L.~E.~Ibanez and A.~M.~Uranga,
  ``String theory and particle physics: An introduction to string
phenomenology,''
  Cambridge, UK: Univ. Pr. (2012)





\bibitem{Giveon:1994fu}
  A.~Giveon, M.~Porrati and E.~Rabinovici,
  ``Target space duality in string theory,''
  Phys.\ Rept.\  {\bf 244} (1994) 77
  [hep-th/9401139].


\bibitem{Alvarez:1994dn}
  E.~Alvarez, L.~Alvarez-Gaume and Y.~Lozano,
  ``An Introduction to T duality in string theory,''
  Nucl.\ Phys.\ Proc.\ Suppl.\  {\bf 41} (1995) 1
  [hep-th/9410237].

  \bibitem{Grana:2005jc}
  M.~Grana,
  ``Flux compactifications in string theory: A Comprehensive review,''
  Phys.\ Rept.\  {\bf 423} (2006) 91
  [hep-th/0509003].

  M.~R.~Douglas and S.~Kachru,
  ``Flux compactification,''
  Rev.\ Mod.\ Phys.\  {\bf 79} (2007) 733
  [hep-th/0610102].

   R.~Blumenhagen, B.~Kors, D.~Lust and S.~Stieberger,
  ``Four-dimensional String Compactifications with D-Branes, Orientifolds and
Fluxes,''
  Phys.\ Rept.\  {\bf 445} (2007) 1
  [hep-th/0610327].

\bibitem{Wecht:2007wu}
  B.~Wecht,
  ``Lectures on Nongeometric Flux Compactifications,''
  Class.\ Quant.\ Grav.\  {\bf 24} (2007) S773
  [arXiv:0708.3984 [hep-th]].

 D.~Andriot,
  ``Non-geometric fluxes versus (non)-geometry,''
  arXiv:1303.0251 [hep-th].

 A.~Chatzistavrakidis and L.~Jonke,
  ``Generalized fluxes in matrix compactifications,''
  arXiv:1305.1864 [hep-th].

\bibitem{Samtleben:2008pe}
  H.~Samtleben,
  ``Lectures on Gauged Supergravity and Flux Compactifications,''
  Class.\ Quant.\ Grav.\  {\bf 25} (2008) 214002
  [arXiv:0808.4076 [hep-th]].

   D.~Roest,
  ``M-theory and gauged supergravities,''
  Fortsch.\ Phys.\  {\bf 53} (2005) 119
  [hep-th/0408175].

\bibitem{Zwiebach:2011rg}
O.~Hohm,
  ``T-duality versus Gauge Symmetry,''
  Prog.\ Theor.\ Phys.\ Suppl.\  {\bf 188} (2011) 116
  [arXiv:1101.3484 [hep-th]].

  B.~Zwiebach,
  ``Double Field Theory, T-Duality, and Courant Brackets,''
  Lect.\ Notes Phys.\  {\bf 851} (2012) 265
  [arXiv:1109.1782 [hep-th]].

\bibitem{lecturesGG}
 N.~Hitchin,
  ``Lectures on generalized geometry,''
  arXiv:1008.0973 [math.DG].

   P.~Koerber,
  ``Lectures on Generalized Complex Geometry for Physicists,''
  Fortsch.\ Phys.\  {\bf 59} (2011) 169
  [arXiv:1006.1536 [hep-th]].

\bibitem{RevBerman}

  D.~S.~Berman and D.~C.~Thompson,
  ``Duality Symmetric String and M-Theory,''
  arXiv:1306.2643 [hep-th].



\bibitem{duff}

M.~J.~Duff,
  ``Duality Rotations In String Theory,''
  Nucl.\ Phys.\ B {\bf 335} (1990) 610.

   M.~J.~Duff and J.~X.~Lu,
  ``Duality Rotations In Membrane Theory,''
  Nucl.\ Phys.\ B {\bf 347} (1990) 394.

\bibitem{tseytlin}
A.~A.~Tseytlin,
  ``Duality symmetric string theory and the cosmological constant problem,''
  Phys.\ Rev.\ Lett.\  {\bf 66} (1991) 545.


   A.~A.~Tseytlin,
  ``Duality symmetric closed string theory and interacting chiral scalars,''
  Nucl.\ Phys.\ B {\bf 350} (1991) 395.

 A.~A.~Tseytlin,
  ``Duality Symmetric Formulation Of String World Sheet Dynamics,''
  Phys.\ Lett.\ B {\bf 242} (1990) 163.

\bibitem{doublegeom}

  C.~M.~Hull,
  ``A Geometry for non-geometric string backgrounds,''
  JHEP {\bf 0510} (2005) 065
  [arXiv:hep-th/0406102].

   C. M. Hull,
  ``Doubled Geometry and T-Folds,''
  JHEP {\bf 0707} (2007) 080
  [hep-th/0605149].

    A.~Dabholkar and C.~Hull,
  ``Generalised T-duality and non-geometric backgrounds,''
  JHEP {\bf 0605} (2006) 009
  [hep-th/0512005].

    C.~M.~Hull and R.~A.~Reid-Edwards,
  ``Gauge symmetry, T-duality and doubled geometry,''
  JHEP {\bf 0808} (2008) 043
  [arXiv:0711.4818 [hep-th]].

    C.~M.~Hull and R.~A.~Reid-Edwards,
  ``Non-geometric backgrounds, doubled geometry and generalised T-duality,''
  JHEP {\bf 0909} (2009) 014
  [arXiv:0902.4032 [hep-th]].

   C.~M.~Hull,
  ``Global aspects of T-duality, gauged sigma models and T-folds,''
  JHEP {\bf 0710} (2007) 057
  [hep-th/0604178].

\bibitem{Kugo:1992md}
  T.~Kugo and B.~Zwiebach,
  ``Target space duality as a symmetry of string field theory,''
  Prog.\ Theor.\ Phys.\  {\bf 87} (1992) 801
  [hep-th/9201040].

  B.~Zwiebach,
  ``Closed string field theory: Quantum action and the B-V master equation,''
  Nucl.\ Phys.\ B {\bf 390} (1993) 33
  [hep-th/9206084].



\bibitem{Backgroundindependent}


  O.~Hohm, C.~Hull and B.~Zwiebach,
  ``Background independent action for double field theory,''
  JHEP {\bf 1007} (2010) 016
  [arXiv:1003.5027 [hep-th]].

  \bibitem{Generalizedmetric}

  O.~Hohm, C.~Hull and B.~Zwiebach,
  ``Generalized metric formulation of double field theory,''
  JHEP {\bf 1008}, 008 (2010)
  [arXiv:1006.4823 [hep-th]].

\bibitem{framelikegeom}

O.~Hohm and S.~K.~Kwak,
  ``Frame-like Geometry of Double Field Theory,''
  J.\ Phys.\ A {\bf 44} (2011) 085404
  [arXiv:1011.4101 [hep-th]].

\bibitem{Hitchin}
 N.~Hitchin,
  ``Generalized Calabi-Yau manifolds,''
  Quart.\ J.\ Math.\ Oxford Ser.\  {\bf 54} (2003) 281
  [math/0209099 [math-dg]].

  M.~Gualtieri,
  ``Generalized complex geometry,''
  math/0401221 [math-dg].

\bibitem{GGafter}

M.~Grana, R.~Minasian, M.~Petrini and A.~Tomasiello,
  ``Supersymmetric backgrounds from generalized Calabi-Yau manifolds,''
  JHEP {\bf 0408} (2004) 046
  [hep-th/0406137].

   M.~Grana, R.~Minasian, M.~Petrini and A.~Tomasiello,
  ``Generalized structures of N=1 vacua,''
  JHEP {\bf 0511} (2005) 020
  [hep-th/0505212].

   M.~Grana, J.~Louis and D.~Waldram,
  ``Hitchin functionals in N=2 supergravity,''
  JHEP {\bf 0601} (2006) 008
  [hep-th/0505264].

   M.~Grana, R.~Minasian, M.~Petrini and A.~Tomasiello,
  ``A Scan for new N=1 vacua on twisted tori,''
  JHEP {\bf 0705} (2007) 031
  [hep-th/0609124].

  M.~Grana, J.~Louis and D.~Waldram,
  ``SU(3) x SU(3) compactification and mirror duals of magnetic fluxes,''
  JHEP {\bf 0704} (2007) 101
  [hep-th/0612237].

   M.~Grana, R.~Minasian, M.~Petrini and D.~Waldram,
  ``T-duality, Generalized Geometry and Non-Geometric Backgrounds,''
  JHEP {\bf 0904} (2009) 075
  [arXiv:0807.4527 [hep-th]].

 L.~Martucci and P.~Smyth,
  ``Supersymmetric D-branes and calibrations on general N=1 backgrounds,''
  JHEP {\bf 0511} (2005) 048
  [hep-th/0507099].

   L.~Martucci,
  ``D-branes on general N=1 backgrounds: Superpotentials and D-terms,''
  JHEP {\bf 0606} (2006) 033
  [hep-th/0602129].

   P.~Koerber and L.~Martucci,
  ``Deformations of calibrated D-branes in flux generalized complex manifolds,''
  JHEP {\bf 0612} (2006) 062
  [hep-th/0610044].

  P.~Koerber and L.~Martucci,
  ``From ten to four and back again: How to generalize the geometry,''
  JHEP {\bf 0708} (2007) 059
  [arXiv:0707.1038 [hep-th]].

   I.~Benmachiche and T.~W.~Grimm,
  ``Generalized N=1 orientifold compactifications and the Hitchin functionals,''
  Nucl.\ Phys.\ B {\bf 748} (2006) 200
  [hep-th/0602241].

    T.~W.~Grimm and J.~Louis,
  ``The Effective action of type IIA Calabi-Yau orientifolds,''
  Nucl.\ Phys.\ B {\bf 718} (2005) 153
  [hep-th/0412277].

    T.~W.~Grimm and J.~Louis,
  ``The Effective action of N = 1 Calabi-Yau orientifolds,''
  Nucl.\ Phys.\ B {\bf 699} (2004) 387
  [hep-th/0403067].



\bibitem{heteroticHohm}

 O.~Hohm and S.~K.~Kwak,
  ``Double Field Theory Formulation of Heterotic Strings,''
  JHEP {\bf 1106} (2011) 096
  [arXiv:1103.2136 [hep-th]].

\bibitem{Andriot:2011iw}
  D.~Andriot,
  ``Heterotic string from a higher dimensional perspective,''
  Nucl.\ Phys.\ B {\bf 855} (2012) 222
  [arXiv:1102.1434 [hep-th]].

\bibitem{TypeIIZwiebach}
O.~Hohm, S.~K.~Kwak and B.~Zwiebach,
  ``Unification of Type II Strings and T-duality,''
  Phys.\ Rev.\ Lett.\  {\bf 107} (2011) 171603
  [arXiv:1106.5452 [hep-th]].

  O.~Hohm, S.~K.~Kwak and B.~Zwiebach,
  ``Double Field Theory of Type II Strings,''
  JHEP {\bf 1109} (2011) 013
  [arXiv:1107.0008 [hep-th]].

\bibitem{TypeIIWaldram}

 A.~Coimbra, C.~Strickland-Constable and D.~Waldram,
  ``Supergravity as Generalised Geometry I: Type II Theories,''
  JHEP {\bf 1111} (2011) 091
  [arXiv:1107.1733 [hep-th]].

  A.~Coimbra, C.~Strickland-Constable and D.~Waldram,
  ``Generalised Geometry and type II Supergravity,''
  Fortsch.\ Phys.\  {\bf 60} (2012) 982
  [arXiv:1202.3170 [hep-th]].

\bibitem{TypeIIPark}

 I.~Jeon, K.~Lee and J.~-H.~Park,
  ``Ramond-Ramond Cohomology and O(D,D) T-duality,''
  JHEP {\bf 1209} (2012) 079
  [arXiv:1206.3478 [hep-th]].


\bibitem{MassiveTypeII}
  O.~Hohm and S.~K.~Kwak,
  ``Massive Type II in Double Field Theory,''
  JHEP {\bf 1111} (2011) 086
  [arXiv:1108.4937 [hep-th]].

\bibitem{SDFT}
   O.~Hohm and S.~K.~Kwak,
  ``N=1 Supersymmetric Double Field Theory,''
  JHEP {\bf 1203} (2012) 080
  [arXiv:1111.7293 [hep-th]].

  I.~Jeon, K.~Lee and J.~-H.~Park,
  ``Supersymmetric Double Field Theory: Stringy Reformulation of Supergravity,''
  Phys.\ Rev.\ D {\bf 85} (2012) 081501
   [Erratum-ibid.\ D {\bf 86} (2012) 089903]
  [arXiv:1112.0069 [hep-th]].

 I.~Jeon, K.~Lee and J.~-H.~Park,
  ``Incorporation of fermions into double field theory,''
  JHEP {\bf 1111} (2011) 025
  [arXiv:1109.2035 [hep-th]].

\bibitem{GeometryPark}

I.~Jeon, K.~Lee and J.~-H.~Park,
  ``Differential geometry with a projection: Application to double field
theory,''
  JHEP {\bf 1104} (2011) 014
  [arXiv:1011.1324 [hep-th]].

  I.~Jeon, K.~Lee and J.~-H.~Park,
  ``Stringy differential geometry, beyond Riemann,''
  Phys.\ Rev.\ D {\bf 84} (2011) 044022
  [arXiv:1105.6294 [hep-th]].

\bibitem{GeometryZwiebach}
O.~Hohm and B.~Zwiebach,
  ``On the Riemann Tensor in Double Field Theory,''
  JHEP {\bf 1205} (2012) 126
  [arXiv:1112.5296 [hep-th]].

 O.~Hohm and B.~Zwiebach,
  ``Towards an invariant geometry of double field theory,''
  arXiv:1212.1736 [hep-th].

\bibitem{GeometryBerman}

 D.~S.~Berman, C.~D.~A.~Blair, E.~Malek and M.~J.~Perry,
  ``The $O_{D,D}$ Geometry of String Theory,''
  arXiv:1303.6727 [hep-th].

\bibitem{TypeIIParkII}

  I.~Jeon, K.~Lee, J.~-H.~Park and Y.~Suh,
  ``Stringy Unification of Type IIA and IIB Supergravities under N=2 D=10
Supersymmetric Double Field Theory,''
  arXiv:1210.5078 [hep-th].

\bibitem{EOMsKwak}
  S.~K.~Kwak,
  ``Invariances and Equations of Motion in Double Field Theory,''
  JHEP {\bf 1010} (2010) 047
  [arXiv:1008.2746 [hep-th]].

 \bibitem{Hull:2009zb}
  C.~Hull and B.~Zwiebach,
  ``The Gauge algebra of double field theory and Courant brackets,''
  JHEP {\bf 0909} (2009) 090
  [arXiv:0908.1792 [hep-th]].

 \bibitem{GDFT}
  M.~Grana and D.~Marques,
  ``Gauged Double Field Theory,''
  JHEP {\bf 1204} (2012) 020
  [arXiv:1201.2924 [hep-th]].

\bibitem{bt} D. S. Berman and N. B. Copland, ``The string partition function in Hull's doubled
formalism,'' Phys. Lett. {\bf B 649} (2007) 325 [arXiv:hep-th/0701080].

\bibitem{WorldsheetDFT}

D.~S.~Berman, N.~B.~Copland and D.~C.~Thompson,
  ``Background Field Equations for the Duality Symmetric String,''
  Nucl.\ Phys.\ B {\bf 791} (2008) 175
  [arXiv:0708.2267 [hep-th]].

 D.~S.~Berman and D.~C.~Thompson,
  ``Duality Symmetric Strings, Dilatons and O(d,d) Effective Actions,''
  Phys.\ Lett.\ B {\bf 662} (2008) 279
  [arXiv:0712.1121 [hep-th]].

  N.~B.~Copland,
  ``Connecting T-duality invariant theories,''
  Nucl.\ Phys.\ B {\bf 854} (2012) 575
  [arXiv:1106.1888 [hep-th]].


   D.~C.~Thompson,
  ``T-duality Invariant Approaches to String Theory,''
  arXiv:1012.4393 [hep-th].


\bibitem{copland}
 N.~B.~Copland,
  ``A Double Sigma Model for Double Field Theory,''
  JHEP {\bf 1204} (2012) 044
  [arXiv:1111.1828 [hep-th]].




  \bibitem{HullU}
  C.~M.~Hull,
  ``Generalised Geometry for M-Theory,''
  JHEP {\bf 0707} (2007) 079
  [hep-th/0701203].

  \bibitem{PachecoU}
  P.~P.~Pacheco and D.~Waldram,
  ``M-theory, exceptional generalised geometry and superpotentials,''
  JHEP {\bf 0809} (2008) 123
  [arXiv:0804.1362 [hep-th]].

  \bibitem{BermanPerry}
   D.~S.~Berman and M.~J.~Perry,
  ``Generalized Geometry and M theory,''
  JHEP {\bf 1106} (2011) 074
  [arXiv:1008.1763 [hep-th]].




  \bibitem{E11programme}
  P.~C.~West,
  ``E(11) and M theory,''
  Class.\ Quant.\ Grav.\  {\bf 18} (2001) 4443
  [hep-th/0104081].

   F.~Riccioni and P.~C.~West,
  ``The E(11) origin of all maximal supergravities,''
  JHEP {\bf 0707} (2007) 063
  [arXiv:0705.0752 [hep-th]].

   F.~Riccioni and P.~C.~West,
  ``E(11)-extended spacetime and gauged supergravities,''
  JHEP {\bf 0802} (2008) 039
  [arXiv:0712.1795 [hep-th]].

    F.~Riccioni, D.~Steele and P.~West,
  ``The E(11) origin of all maximal supergravities: The Hierarchy of
field-strengths,''
  JHEP {\bf 0909} (2009) 095
  [arXiv:0906.1177 [hep-th]].


 P.~West,
  ``Generalised geometry, eleven dimensions and E11,''
  arXiv:1111.1642 [hep-th].

\bibitem{Koepsell:2000xg}
  K.~Koepsell, H.~Nicolai and H.~Samtleben,
  ``An Exceptional geometry for D = 11 supergravity?,''
  Class.\ Quant.\ Grav.\  {\bf 17} (2000) 3689
  [hep-th/0006034].

\bibitem{Hillmann:2009ci}
  C.~Hillmann,
  ``Generalized E(7(7)) coset dynamics and D=11 supergravity,''
  JHEP {\bf 0903} (2009) 135
  [arXiv:0901.1581 [hep-th]].



\bibitem{Uduality1}

  A.~Coimbra, C.~Strickland-Constable and D.~Waldram,
  ``Supergravity as Generalised Geometry II: $E_{d(d)} \times \mathbb{R}^+$ and
M theory,''
  arXiv:1212.1586 [hep-th].

  A.~Coimbra, C.~Strickland-Constable and D.~Waldram,
  ``$E_{d(d)} \times \mathbb{R}^+$ Generalised Geometry, Connections and M
theory,''
  arXiv:1112.3989 [hep-th].


\bibitem{Uduality2}
   D.~S.~Berman, M.~Cederwall, A.~Kleinschmidt and D.~C.~Thompson,
  ``The gauge structure of generalised diffeomorphisms,''
  JHEP {\bf 1301} (2013) 064
  [arXiv:1208.5884 [hep-th]].

\bibitem{Udualities3}

   D.~S.~Berman, H.~Godazgar, M.~J.~Perry and P.~West,
  ``Duality Invariant Actions and Generalised Geometry,''
  JHEP {\bf 1202} (2012) 108
  [arXiv:1111.0459 [hep-th]].

D.~ S.~Berman, H. Godazgar and M. J. Perry,
``SO(5,5) duality in M-theory and generalized geometry,'' Phys.Lett. {\bf B700} (2011) 65-67
 [arXiv:1103.5733 [hep-th]]

 \bibitem{localsymm}
 D.~S.~Berman, H.~Godazgar, M.~Godazgar and M.~J.~Perry,
  ``The Local symmetries of M-theory and their formulation in generalised
geometry,''
  JHEP {\bf 1201} (2012) 012
  [arXiv:1110.3930 [hep-th]].

\bibitem{Thompson:2011uw}
  D.~C.~Thompson,
  ``Duality Invariance: From M-theory to Double Field Theory,''
  JHEP {\bf 1108} (2011) 125
  [arXiv:1106.4036 [hep-th]].

\bibitem{SL5Uduality}
 J.~-H.~Park and Y.~Suh,
  ``U-geometry : SL(5),''
  arXiv:1302.1652 [hep-th].

\bibitem{Cederwall}
   M.~Cederwall, J.~Edlund and A.~Karlsson,
  ``Exceptional geometry and tensor fields,''
  arXiv:1302.6736 [hep-th].

 M.~Cederwall,
  ``Non-gravitational exceptional supermultiplets,''
  arXiv:1302.6737 [hep-th].

\bibitem{E8}
 H.~Godazgar, M.~Godazgar and M.~J.~Perry,
  ``E8 duality and dual gravity,''
  arXiv:1303.2035 [hep-th].

\bibitem{Udualfluxes}
  G.~Aldazabal, E.~Andres, P.~G.~Camara and M.~Grana,
  ``U-dual fluxes and Generalized Geometry,''
  JHEP {\bf 1011} (2010) 083
  [arXiv:1007.5509 [hep-th]].

\bibitem{SSMtheory}
D.~S.~Berman, E.~T.~Musaev, D.~C.~Thompson and D.~C.~Thompson,
  ``Duality Invariant M-theory: Gauged supergravities and Scherk-Schwarz
reductions,''
  JHEP {\bf 1210} (2012) 174
  [arXiv:1208.0020 [hep-th]].

\bibitem{Musaev}
 E.~T.~Musaev,
  ``Gauged supergravities in 5 and 6 dimensions from generalised Scherk-Schwarz
reductions,''
  arXiv:1301.0467 [hep-th].

\bibitem{Extended geometry}

 G.~Aldazabal, M.~Grana, D.~Marques and J.~A.~Rosabal,
  ``Extended geometry and gauged maximal supergravity,''
  arXiv:1302.5419 [hep-th].

\bibitem{Dasgupta:1999ss}
  K.~Dasgupta, G.~Rajesh and S.~Sethi,
  ``M theory, orientifolds and G - flux,''
  JHEP {\bf 9908} (1999) 023
  [hep-th/9908088].

\bibitem{Kachru:2002sk}
  S.~Kachru, M.~B.~Schulz, P.~K.~Tripathy and S.~P.~Trivedi,
  ``New supersymmetric string compactifications,''
  JHEP {\bf 0303} (2003) 061
  [hep-th/0211182].

\bibitem{Hellerman:2002ax}
  S.~Hellerman, J.~McGreevy and B.~Williams,
  ``Geometric constructions of nongeometric string theories,''
  JHEP {\bf 0401} (2004) 024
  [hep-th/0208174].


\bibitem{Dabholkar:2002sy}
  A.~Dabholkar and C.~Hull,
  ``Duality twists, orbifolds, and fluxes,''
  JHEP {\bf 0309} (2003) 054
  [hep-th/0210209].

  \bibitem{stw}
J.~Shelton, W.~Taylor and B.~Wecht, ``Nongeometric flux
compactifications,'' JHEP 0510 (2005) 085, hep-th/0508133.

\bibitem{Othernongeom}
  A.~Flournoy, B.~Wecht and B.~Williams,
  ``Constructing nongeometric vacua in string theory,''
  Nucl.\ Phys.\ B {\bf 706} (2005) 127
  [hep-th/0404217].

  A.~Flournoy and B.~Williams,
  ``Nongeometry, duality twists, and the worldsheet,''
  JHEP {\bf 0601} (2006) 166
  [hep-th/0511126].

  A.~Lawrence, M.~B.~Schulz and B.~Wecht,
  ``D-branes in nongeometric backgrounds,''
  JHEP {\bf 0607} (2006) 038
  [hep-th/0602025].

  W.~Schulgin and J.~Troost,
  ``Backreacted T-folds and non-geometric regions in configuration space,''
  JHEP {\bf 0812} (2008) 098
  [arXiv:0808.1345 [hep-th]].

  F.~Marchesano and W.~Schulgin,
  ``Non-geometric fluxes as supergravity backgrounds,''
  Phys.\ Rev.\ D {\bf 76} (2007) 041901
  [arXiv:0704.3272 [hep-th]].

 K.~Becker and S.~Sethi,
  ``Torsional Heterotic Geometries,''
  Nucl.\ Phys.\ B {\bf 820} (2009) 1
  [arXiv:0903.3769 [hep-th]].

  J.~McOrist, D.~R.~Morrison and S.~Sethi,
  ``Geometries, Non-Geometries, and Fluxes,''
  Adv.\ Theor.\ Math.\ Phys.\  {\bf 14} (2010)
  [arXiv:1004.5447 [hep-th]].



\bibitem{acfi}
  G.~Aldazabal, P.~G.~Camara, A.~Font and L.~E.~Ibanez,
  ``More dual fluxes and moduli fixing,''
  JHEP {\bf 0605}, 070 (2006)
  [arXiv:hep-th/0602089].

\bibitem{doubletorus}
G.~Dall'Agata, N.~Prezas, H.~Samtleben and M.~Trigiante,
  ``Gauged Supergravities from Twisted Doubled Tori and Non-Geometric String
Backgrounds,''
  Nucl.\ Phys.\ B {\bf 799} (2008) 80
  [arXiv:0712.1026 [hep-th]].

\bibitem{HackettJones:2006bp}
  E.~Hackett-Jones and G.~Moutsopoulos,
  ``Quantum mechanics of the doubled torus,''
  JHEP {\bf 0610} (2006) 062
  [hep-th/0605114].

\bibitem{Andriot:2012an}
  D.~Andriot, O.~Hohm, M.~Larfors, D.~Lust and P.~Patalong,
  ``Non-Geometric Fluxes in Supergravity and Double Field Theory,''
  Fortsch.\ Phys.\  {\bf 60} (2012) 1150
  [arXiv:1204.1979 [hep-th]].

  D.~Andriot, O.~Hohm, M.~Larfors, D.~Lust and P.~Patalong,
  ``A geometric action for non-geometric fluxes,''
  Phys.\ Rev.\ Lett.\  {\bf 108} (2012) 261602
  [arXiv:1202.3060 [hep-th]].

  D.~Andriot, M.~Larfors, D.~Lust and P.~Patalong,
  ``A ten-dimensional action for non-geometric fluxes,''
  JHEP {\bf 1109} (2011) 134
  [arXiv:1106.4015 [hep-th]].

  D.~Andriot and A.~Betz,
  ``beta-supergravity: a ten-dimensional theory with non-geometric fluxes, and its geometric framework,''
  arXiv:1306.4381 [hep-th].

  \bibitem{Aldazabal:2011nj}
  G.~Aldazabal, W.~Baron, D.~Marques and C.~Nunez,
  ``The effective action of Double Field Theory,''
  JHEP {\bf 1111} (2011) 052
   [Erratum-ibid.\  {\bf 1111} (2011) 109]
  [arXiv:1109.0290 [hep-th]].

  \bibitem{Geissbuhler:2011mx}
  D.~Geissbuhler,
  ``Double Field Theory and N=4 Gauged Supergravity,''
  JHEP {\bf 1111} (2011) 116
  [arXiv:1109.4280 [hep-th]].


\bibitem{Dibitetto:2012rk}
  G.~Dibitetto, J.~J.~Fernandez-Melgarejo, D.~Marques and D.~Roest,
  ``Duality orbits of non-geometric fluxes,''
  Fortsch.\ Phys.\  {\bf 60} (2012) 1123
  [arXiv:1203.6562 [hep-th]].

  \bibitem{Exploring}
  D.~Geissbuhler, D.~Marques, C.~Nunez and V.~Penas,
  ``Exploring Double Field Theory,''
  arXiv:1304.1472 [hep-th].


  \bibitem{Blumenhagen:2013aia}


  R.~Blumenhagen, A.~Deser, E.~Plauschinn, F.~Rennecke and C.~Schmid,
  ``The Intriguing Structure of Non-geometric Frames in String Theory,''
  arXiv:1304.2784 [hep-th].

  R.~Blumenhagen, A.~Deser, E.~Plauschinn and F.~Rennecke,
  ``Non-geometric strings, symplectic gravity and differential geometry of Lie
algebroids,''
  JHEP {\bf 1302} (2013) 122
  [arXiv:1211.0030 [hep-th]].

  R.~Blumenhagen, A.~Deser, E.~Plauschinn and F.~Rennecke,
  ``A bi-invariant Einstein-Hilbert action for the non-geometric string,''
  Phys.\ Lett.\ B {\bf 720} (2013) 215
  [arXiv:1210.1591 [hep-th]].

\bibitem{Berman:2013cli}
  D.~S.~Berman and K.~Lee,
  ``Supersymmetry for Gauged Double Field Theory and Generalised Scherk-Schwarz Reductions,''
  arXiv:1305.2747 [hep-th].

\bibitem{oxidation}
 R.~Blumenhagen, X.~Gao, D.~Herschmann and P.~Shukla,
  ``Dimensional Oxidation of Non-geometric Fluxes in Type II Orientifolds,''
  arXiv:1306.2761 [hep-th].

 \bibitem{Hohm:2013nja}
  O.~Hohm and H.~Samtleben,
  ``Gauge theory of Kaluza-Klein and winding modes,''
  arXiv:1307.0039 [hep-th].

\bibitem{Condeescu:2013yma}
  C.~Condeescu, I.~Florakis, C.~Kounnas and D.~Lust,
  ``Gauged supergravities and non-geometric Q/R-fluxes from asymmetric orbifold CFT's,''
  arXiv:1307.0999 [hep-th].

\bibitem{avra}

 G.~Dall'Agata and N.~Prezas,
  ``Worldsheet theories for non-geometric string backgrounds,''
  JHEP {\bf 0808} (2008) 088
  [arXiv:0806.2003 [hep-th]].

S.~D.~Avramis, J.~-P.~Derendinger and N.~Prezas,
  ``Conformal chiral boson models on twisted doubled tori and non-geometric
string vacua,''
  Nucl.\ Phys.\ B {\bf 827} (2010) 281
  [arXiv:0910.0431 [hep-th]].

\bibitem{Halmagyi:2008dr}
  N.~Halmagyi,
  ``Non-geometric String Backgrounds and Worldsheet Algebras,''
  JHEP {\bf 0807} (2008) 137
  [arXiv:0805.4571 [hep-th]].

   N.~Halmagyi,
  ``Non-geometric Backgrounds and the First Order String Sigma Model,''
  arXiv:0906.2891 [hep-th].

  \bibitem{Mylonas:2012pg}
  D.~Mylonas, P.~Schupp and R.~J.~Szabo,
  ``Membrane Sigma-Models and Quantization of Non-Geometric Flux Backgrounds,''
  JHEP {\bf 1209} (2012) 012
  [arXiv:1207.0926 [hep-th]].




  \bibitem{OtherDFT}
  I.~Jeon, K.~Lee and J.~-H.~Park,
  ``Double field formulation of Yang-Mills theory,''
  Phys.\ Lett.\ B {\bf 701} (2011) 260
  [arXiv:1102.0419 [hep-th]].

  O.~Hohm,
  ``On factorizations in perturbative quantum gravity,''
  JHEP {\bf 1104} (2011) 103
  [arXiv:1103.0032 [hep-th]].

  I.~Vaisman,
  ``On the geometry of double field theory,''
  J.\ Math.\ Phys.\  {\bf 53} (2012) 033509
  [arXiv:1203.0836 [math.DG]].

  I.~Vaisman,
  ``Towards a double field theory on para-Hermitian manifolds,''
  arXiv:1209.0152 [math.DG].


  E.~Malek,
  ``U-duality in three and four dimensions,''
  arXiv:1205.6403 [hep-th].

   E.~Malek,
  ``Timelike U-dualities in Generalised Geometry,''
  arXiv:1301.0543 [hep-th].

  A.~Chatzistavrakidis and L.~Jonke,
  ``Matrix theory origins of non-geometric fluxes,''
  JHEP {\bf 1302} (2013) 040
  [arXiv:1207.6412 [hep-th]].

M.~P.~Garcia del Moral,
  ``Dualities as symmetries of the Supermembrane Theory,''
  arXiv:1211.6265 [hep-th].

  M.~P.~Garcia del Moral, J.~Pena and A.~Restuccia,
  ``T-duality Invariance of the Supermembrane,''
  arXiv:1211.2434 [hep-th].

J.~Maharana,
  ``The Worldsheet Perspective of T-duality Symmetry in String Theory,''
  Invited Review article for Int. Journal of Mod. Phys. A 2013
  [arXiv:1302.1719 [hep-th]].

J.~Maharana,
  ``Duality Symmetry of String Theory: A Worldsheet Perspective,''
  Phys.\ Lett.\ B {\bf 695} (2011) 370
  [arXiv:1010.1727 [hep-th]].

J.~Maharana,
  ``T-duality of NSR superstring: The worldsheet perspective,''
  Int.\ J.\ Mod.\ Phys.\ A {\bf 27} (2012) 1250140
  [arXiv:1203.3357 [hep-th]].

  M.~Aldi and R.~Heluani,
  ``Dilogarithms, OPE and twisted T-duality,''
  arXiv:1105.4280 [math-ph].

M.~B.~Schulz,
  ``T-folds, doubled geometry, and the SU(2) WZW model,''
  JHEP {\bf 1206} (2012) 158
  [arXiv:1106.6291 [hep-th]].

  N.~Kan, K.~Kobayashi and K.~Shiraishi,
  ``Equations of Motion in Double Field Theory: From particles to scale
factors,''
  Phys.\ Rev.\ D {\bf 84} (2011) 124049
  [arXiv:1108.5795 [hep-th]].

D.~S.~Berman, E.~T.~Musaev and M.~J.~Perry,
  ``Boundary Terms in Generalized Geometry and doubled field theory,''
  Phys.\ Lett.\ B {\bf 706} (2011) 228
  [arXiv:1110.3097 [hep-th]].

 I.~Bakhmatov,
 ``Fermionic T-duality and U-duality in type II supergravity,''
  arXiv:1112.1983 [hep-th].

   M.~Hatsuda and T.~Kimura,
  ``Canonical approach to Courant brackets for D-branes,''
  JHEP {\bf 1206} (2012) 034
  [arXiv:1203.5499 [hep-th]].

   M.~Hatsuda and K.~Kamimura,
  ``SL(5) duality from canonical M2-brane,''
  JHEP {\bf 1211} (2012) 001
  [arXiv:1208.1232 [hep-th]].

   N.~B.~Copland, S.~M.~Ko and J.~-H.~Park,
  ``Superconformal Yang-Mills quantum mechanics and Calogero model with
OSp(N|2,R) symmetry,''
  JHEP {\bf 1207} (2012) 076
  [arXiv:1205.3869 [hep-th]].

   T.~Kimura and S.~Sasaki,
  ``Gauged Linear Sigma Model for Exotic Five-brane,''
  arXiv:1304.4061 [hep-th].

  M.~Hatsuda and K.~Kamimura,
  ``M5 algebra and SO(5,5) duality,''
  arXiv:1305.2258 [hep-th].

  H.~Wu and H.~Yang,
  ``Double Field Theory Inspired Cosmology,''
  arXiv:1307.0159 [hep-th].

\bibitem{nonAbelianT}

  G.~Itsios, C.~Nunez, K.~Sfetsos and D.~C.~Thompson,
  ``On Non-Abelian T-Duality and new N=1 backgrounds,''
  Phys.\ Lett.\ B {\bf 721} (2013) 342
  [arXiv:1212.4840 [hep-th]].

  G.~Itsios, C.~Nunez, K.~Sfetsos and D.~C.~Thompson,
  ``Non-Abelian T-duality and the AdS/CFT correspondence:new N=1 backgrounds,''
  arXiv:1301.6755 [hep-th].

  G.~Itsios, Y.~Lozano, E.~.O Colgain and K.~Sfetsos,
  ``Non-Abelian T-duality and consistent truncations in type-II supergravity,''
  JHEP {\bf 1208} (2012) 132
  [arXiv:1205.2274 [hep-th]].

 Y.~Lozano, E.~.O Colgain, K.~Sfetsos and D.~C.~Thompson,
  ``Non-abelian T-duality, Ramond Fields and Coset Geometries,''
  JHEP {\bf 1106} (2011) 106
  [arXiv:1104.5196 [hep-th]].

\bibitem{Giveon:1988tt}
  A.~Giveon, E.~Rabinovici and G.~Veneziano,
  ``Duality in String Background Space,''
  Nucl.\ Phys.\ B {\bf 322} (1989) 167.

\bibitem{Maharana:1992my}
  J.~Maharana and J.~H.~Schwarz,
  ``Noncompact symmetries in string theory,''
  Nucl.\ Phys.\ B {\bf 390} (1993) 3
  [hep-th/9207016].

\bibitem{buscher1}
T. Buscher, ``A Symmetry of the String Background Field Equations,'' Phys. Lett.
{\bf 194B} (1987) 59.

\bibitem{buscher2} T. Buscher, ``Path Integral Derivation of Quantum Duality in
Nonlinear Sigma Models,''
Phys. Lett. {\bf 201B} (1988) 466.

\bibitem{Courant} T. Courant, ``Dirac Manifolds.'' Trans. Amer. Math. Soc.
319: 631-661, 1990.

 D. Roytenberg, ``Courant algebroids, derived brackets and even symplectic supermanifolds'',
Ph.D. Thesis, U.C. Berkeley, arXiv:math/9910078.

\bibitem{Schon}
  J.~Schon and M.~Weidner,
   ``Gauged N=4 supergravities,''
  JHEP {\bf 0605} (2006) 034
  [hep-th/0602024].

\bibitem{Scherk:1979zr}

  J.~Scherk and J.~H.~Schwarz,
  ``Spontaneous Breaking of Supersymmetry Through Dimensional Reduction,''
  Phys.\ Lett.\ B {\bf 82} (1979) 60.

J.~Scherk and J.~H.~Schwarz,
  ``How to Get Masses from Extra Dimensions,''
  Nucl.\ Phys.\ B {\bf 153}, 61 (1979).

  \bibitem{Kaloper:1999yr}
  N.~Kaloper and R.~C.~Myers,
  ``The Odd story of massive supergravity,''
  JHEP {\bf 9905} (1999) 010
  [hep-th/9901045].

\bibitem{Grana:2013ila}
  M.~Grana, R.~Minasian, H.~Triendl and T.~Van Riet,
  ``The moduli problem in Scherk-Schwarz compactifications,''
  arXiv:1305.0785 [hep-th].

  \bibitem{dS}



  U.~H.~Danielsson, G.~Shiu, T.~Van Riet and T.~Wrase,
  ``A note on obstinate tachyons in classical dS solutions,''
  arXiv:1212.5178 [hep-th].

  U.~Danielsson and G.~Dibitetto,
  ``On the distribution of stable de Sitter vacua,''
  JHEP {\bf 1303} (2013) 018
  [arXiv:1212.4984 [hep-th]].

 J.~Blaback, U.~Danielsson and G.~Dibitetto,
  ``Fully stable dS vacua from generalised fluxes,''
  arXiv:1301.7073 [hep-th].

  C.~Damian and O.~Loaiza-Brito,
  ``More stable dS vacua from S-dual non-geometric fluxes,''
  arXiv:1304.0792 [hep-th].

 C.~Damian, O.~Loaiza-Brito, L.~Rey and M.~Sabido,
  ``Slow-Roll Inflation in Non-geometric Flux Compactification,''
  arXiv:1302.0529 [hep-th].

 B.~de Carlos, A.~Guarino and J.~M.~Moreno,
  ``Flux moduli stabilisation, Supergravity algebras and no-go theorems,''
  JHEP {\bf 1001} (2010) 012
  [arXiv:0907.5580 [hep-th]].

   B.~de Carlos, A.~Guarino and J.~M.~Moreno,
  ``Complete classification of Minkowski vacua in generalised flux models,''
  JHEP {\bf 1002} (2010) 076
  [arXiv:0911.2876 [hep-th]].

   J.~Shelton, W.~Taylor and B.~Wecht,
  ``Generalized Flux Vacua,''
  JHEP {\bf 0702} (2007) 095
  [hep-th/0607015].



   A.~Micu, E.~Palti and G.~Tasinato,
  ``Towards Minkowski Vacua in Type II String Compactifications,''
  JHEP {\bf 0703} (2007) 104
  [hep-th/0701173].

   E.~Palti,
  ``Low Energy Supersymmetry from Non-Geometry,''
  JHEP {\bf 0710} (2007) 011
  [arXiv:0707.1595 [hep-th]].

  K.~Becker, M.~Becker, C.~Vafa and J.~Walcher,
  ``Moduli Stabilization in Non-Geometric Backgrounds,''
  Nucl.\ Phys.\ B {\bf 770} (2007) 1
  [hep-th/0611001].

 \bibitem{FreedWitten}

   C.~M.~Hull and R.~A.~Reid-Edwards,
  ``Flux compactifications of M-theory on twisted Tori,''
  JHEP {\bf 0610} (2006) 086
  [hep-th/0603094].



  C.~M.~Hull and R.~A.~Reid-Edwards,
  ``Flux compactifications of string theory on twisted tori,''
  Fortsch.\ Phys.\  {\bf 57} (2009) 862
  [hep-th/0503114].

R.~A.~Reid-Edwards,
  ``Flux compactifications, twisted tori and doubled geometry,''
  JHEP {\bf 0906}, 085 (2009)
  [arXiv:0904.0380 [hep-th]].

   G.~Dall'Agata and S.~Ferrara,
  ``Gauged supergravity algebras from twisted tori compactifications with
fluxes,''
  Nucl.\ Phys.\ B {\bf 717} (2005) 223
  [hep-th/0502066].

  G.~Aldazabal, P.~G.~Camara and J.~A.~Rosabal,
  ``Flux algebra, Bianchi identities and Freed-Witten anomalies in F-theory
compactifications,''
  Nucl.\ Phys.\ B {\bf 814} (2009) 21
  [arXiv:0811.2900 [hep-th]].

   G.~Dall'Agata, G.~Villadoro and F.~Zwirner,
  ``Type-IIA flux compactifications and N=4 gauged supergravities,''
  JHEP {\bf 0908} (2009) 018
  [arXiv:0906.0370 [hep-th]].

  L.~Andrianopoli, M.~A.~Lledo and M.~Trigiante,
  ``The Scherk-Schwarz mechanism as a flux compactification with internal
torsion,''
  JHEP {\bf 0505} (2005) 051
  [hep-th/0502083].

  G.~Villadoro and F.~Zwirner,
  ``N=1 effective potential from dual type-IIA D6/O6 orientifolds with general
fluxes,''
  JHEP {\bf 0506} (2005) 047
  [hep-th/0503169].

  J.~-P.~Derendinger, C.~Kounnas, P.~M.~Petropoulos and F.~Zwirner,
  ``Superpotentials in IIA compactifications with general fluxes,''
  Nucl.\ Phys.\ B {\bf 715} (2005) 211
  [hep-th/0411276].

  D.~A.~Lowe, H.~Nastase and S.~Ramgoolam,
  ``Massive IIA string theory and matrix theory compactification,''
  Nucl.\ Phys.\ B {\bf 667} (2003) 55
  [hep-th/0303173].


  D.~Andriot, R.~Minasian and M.~Petrini,
  ``Flux backgrounds from Twists,''
  JHEP {\bf 0912} (2009) 028
  [arXiv:0903.0633 [hep-th]].

 R.~A.~Reid-Edwards,
  ``Geometric and non-geometric compactifications of IIB supergravity,''
  JHEP {\bf 0812} (2008) 043
  [hep-th/0610263].


  R.~A.~Reid-Edwards and B.~Spanjaard,
  ``N=4 Gauged Supergravity from Duality-Twist Compactifications of String Theory,''
  JHEP {\bf 0812} (2008) 052
  [arXiv:0810.4699 [hep-th]].


   A.~Chatzistavrakidis and L.~Jonke,
  ``Matrix theory compactifications on twisted tori,''
  Phys.\ Rev.\ D {\bf 85} (2012) 106013
  [arXiv:1202.4310 [hep-th]].

 \bibitem{Derendinger:2007xp}
  J.~-P.~Derendinger, P.~M.~Petropoulos and N.~Prezas,
  ``Axionic symmetry gaugings in N=4 supergravities and their higher-dimensional
origin,''
  Nucl.\ Phys.\ B {\bf 785} (2007) 115
  [arXiv:0705.0008 [hep-th]].

  \bibitem{ExceptionalFluxComp}

  G.~Dibitetto, A.~Guarino and D.~Roest,
  ``How to halve maximal supergravity,''
  JHEP {\bf 1106}, 030 (2011)
  [arXiv:1104.3587 [hep-th]].

  G.~Dibitetto, A.~Guarino and D.~Roest,
  ``Exceptional Flux Compactifications,''
  JHEP {\bf 1205} (2012) 056
  [arXiv:1202.0770 [hep-th]].

  G.~Aldazabal, D.~Marques, C.~Nunez and J.~A.~Rosabal,
  ``On Type IIB moduli stabilization and N = 4, 8 supergravities,''
  Nucl.\ Phys.\  B {\bf 849}, 80 (2011)
  [arXiv:1101.5954 [hep-th]].


  \bibitem{Blumenhagen:2012pc}
  R.~Blumenhagen, A.~Deser, E.~Plauschinn and F.~Rennecke,
  ``Bianchi Identities for Non-Geometric Fluxes - From Quasi-Poisson Structures
to Courant Algebroids,''
  arXiv:1205.1522 [hep-th].

 R.~Blumenhagen, A.~Deser, E.~Plauschinn and F.~Rennecke,
  ``Palatini-Lovelock-Cartan Gravity - Bianchi Identities for Stringy Fluxes,''
  Class.\ Quant.\ Grav.\  {\bf 29} (2012) 135004
  [arXiv:1202.4934 [hep-th]].

  \bibitem{Uduality}

 C.~M.~Hull and P.~K.~Townsend,
  ``Unity of superstring dualities,''
  Nucl.\ Phys.\ B {\bf 438} (1995) 109
  [hep-th/9410167].

  \bibitem{Cremmer:1997ct}
  E.~Cremmer, B.~Julia, H.~Lu and C.~N.~Pope,
  ``Dualization of dualities. 1.,''
  Nucl.\ Phys.\ B {\bf 523} (1998) 73
  [hep-th/9710119].

  \bibitem{Obers:1998fb}
  N.~A.~Obers and B.~Pioline,
  ``U duality and M theory,''
  Phys.\ Rept.\  {\bf 318} (1999) 113
  [hep-th/9809039].




 \bibitem{de Wit:2007mt}
  B.~de Wit, H.~Samtleben and M.~Trigiante,
  ``The Maximal D=4 supergravities,''
  JHEP {\bf 0706} (2007) 049
  [arXiv:0705.2101 [hep-th]].
\bibitem{de Wit:1982ig}
  B.~de Wit and H.~Nicolai,
  ``N=8 Supergravity,''
  Nucl.\ Phys.\ B {\bf 208} (1982) 323.
\bibitem{deWit:2002vt}
  B.~de Wit, H.~Samtleben and M.~Trigiante,
  ``On Lagrangians and gaugings of maximal supergravities,''
  Nucl.\ Phys.\ B {\bf 655} (2003) 93
  [hep-th/0212239].

  \bibitem{LeDiffon:2008sh}
  A.~Le Diffon and H.~Samtleben,
  ``Supergravities without an Action: Gauging the Trombone,''
  Nucl.\ Phys.\ B {\bf 811} (2009) 1
  [arXiv:0809.5180 [hep-th]].\\
  A.~Le Diffon, H.~Samtleben and M.~Trigiante,
  ``N=8 Supergravity with Local Scaling Symmetry,''
  JHEP {\bf 1104} (2011) 079
  [arXiv:1103.2785 [hep-th]].

\bibitem{deWit:2008ta}
  B.~de Wit, H.~Nicolai and H.~Samtleben,
  ``Gauged Supergravities, Tensor Hierarchies, and M-Theory,''
  JHEP {\bf 0802} (2008) 044
  [arXiv:0801.1294 [hep-th]].

  B.~de Wit and H.~Samtleben,
  ``The End of the p-form hierarchy,''
  JHEP {\bf 0808} (2008) 015
  [arXiv:0805.4767 [hep-th]].

  J.~Palmkvist,
  ``The tensor hierarchy algebra,''
  arXiv:1305.0018 [hep-th].

\bibitem{Dall'Agata:2012bb}
  G.~Dall'Agata, G.~Inverso and M.~Trigiante,
  ``Evidence for a family of SO(8) gauged supergravity theories,''
  Phys.\ Rev.\ Lett.\  {\bf 109} (2012) 201301
  [arXiv:1209.0760 [hep-th]].

  \bibitem{deWit:2013ija}
  B.~de Wit and H.~Nicolai,
  ``Deformations of gauged SO(8) supergravity and supergravity in eleven
dimensions,''
  arXiv:1302.6219 [hep-th].



\bibitem{Hohm:2013jma}
  O.~Hohm and H.~Samtleben,
  ``U-duality covariant gravity,''
  arXiv:1307.0509 [hep-th].

 O.~Hohm and H.~Samtleben,
  ``Exceptional Form of D=11 Supergravity,''
  arXiv:1308.1673 [hep-th].



\bibitem{siegel1}
W.~Siegel,
  ``Covariantly Second Quantized String,''
  Phys.\ Lett.\ B {\bf 142} (1984) 276.

\bibitem{rv}M.~Rocek and E.~P.~Verlinde,
  ``Duality, quotients, and currents,''
  Nucl.\ Phys.\ B {\bf 373}, 630 (1992)
  [hep-th/9110053].

\bibitem{frj}  B.~E.~Fridling and A.~Jevicki,
  ``Dual Representations and Ultraviolet Divergences in Nonlinear sigma
Models,''
  Phys.\ Lett.\ B {\bf 134} (1984) 70.

\bibitem{sfl}
K. Sfetsos, ``Canonical equivalence of non-isometric sigma-models and
Poisson-Lie'', Nucl. Phys. B 517
(1998) 549 [arXiv:hep-th/9710163].

E.~Alvarez, L.~Alvarez-Gaume and Y.~Lozano,
  ``A Canonical approach to duality transformations,''
  Phys.\ Lett.\ B {\bf 336} (1994) 183
  [hep-th/9406206].

 Y. Lozano, ``NonAbelian duality and canonical transformations,'' Phys. Lett. B
355, 165 (1995)
[arXiv:hep-th/9503045].

Y. Lozano, ``Duality and canonical transformations,'' Mod. Phys. Lett. A 11,
2893 (1996) [arXiv:hepth/
9610024].



\bibitem{fj}
 R.~Floreanini and R.~Jackiw,
  ``Selfdual Fields as Charge Density Solitons,''
  Phys.\ Rev.\ Lett.\  {\bf 59} (1987) 1873.

M.~Henneaux and C.~Teitelboim,
  ``Dynamics Of Chiral (selfdual) P Forms,''
  Phys.\ Lett.\ B {\bf 206} (1988) 650.

\bibitem{schs} J.~H.~Schwarz and A.~Sen,
  ``Duality symmetric actions,''
  Nucl.\ Phys.\ B {\bf 411}, 35 (1994)
  [hep-th/9304154].

\bibitem{rt}
  M.~Rocek and A.~A.~Tseytlin,
  ``Partial breaking of global D = 4 supersymmetry, constrained
superfields, and
three-brane actions,''
  Phys.\ Rev.\ D {\bf 59}, 106001 (1999)
  [hep-th/9811232].

\bibitem{patalong}
  S.~Groot Nibbelink and P.~Patalong,
  ``A Lorentz invariant doubled worldsheet theory,''
  Phys.\ Rev.\ D {\bf 87} (2013) 041902
  [arXiv:1207.6110 [hep-th]].

\bibitem{sfetsos}
K.~Sfetsos, K.~Siampos and D.~C.~Thompson,
  ``Renormalization of Lorentz non-invariant actions and manifest T-duality,''
  Nucl.\ Phys.\ B {\bf 827}, 545 (2010)
  [arXiv:0910.1345 [hep-th]].


\bibitem{klimcik} C.~Klimcik and P.~Severa,
  ``Poisson-Lie T duality and loop groups of Drinfeld doubles,''
  Phys.\ Lett.\ B {\bf 372}, 65 (1996)
  [hep-th/9512040].

\bibitem{re}  R.~A.~Reid-Edwards,
  ``Bi-Algebras, Generalised Geometry and T-Duality,''
  arXiv:1001.2479 [hep-th].



\bibitem{blum}
R.~Blumenhagen, A.~Deser, D.~Lust, E.~Plauschinn and F.~Rennecke,
  ``Non-geometric Fluxes, Asymmetric Strings and Nonassociative Geometry,''
  J.\ Phys.\ A {\bf 44}, 385401 (2011)
  [arXiv:1106.0316 [hep-th]].



\bibitem{noncomop}
 C.~-S.~Chu and P.~-M.~Ho,
  ``Noncommutative open string and D-brane,''
  Nucl.\ Phys.\ B {\bf 550}, 151 (1999)
  [hep-th/9812219].

V.~Schomerus,
  ``D-branes and deformation quantization,''
  JHEP {\bf 9906}, 030 (1999)
  [hep-th/9903205].

 N.~Seiberg and E.~Witten,
  ``String theory and noncommutative geometry,''
  JHEP {\bf 9909}, 032 (1999)
  [hep-th/9908142].

F.~Ardalan, H.~Arfaei and M.~M.~Sheikh-Jabbari,
  ``Mixed branes and M(atrix) theory on noncommutative torus,''
  hep-th/9803067.

\bibitem{herbst}
M. Herbst, A. Kling and M. Kreuzer, ``Star products from open strings
in curved backgrounds,'' JHEP 0109, 014 (2001) [hep-th/0106159].

 M. Herbst, A. Kling and M. Kreuzer, ``Noncommutative tachyon action
and D-brane geometry,'' JHEP 0208, 010 (2002) [hep-th/0203077].

L. Cornalba and R. Schiappa, ``Nonassociative star product deformations
for D-brane world volumes in curved backgrounds,'' Commun. Math.
Phys. 225, 33 (2002) [hep-th/0101219].

\bibitem{bp}  R.~Blumenhagen and E.~Plauschinn,
  ``Nonassociative Gravity in String Theory?,''
  J.\ Phys.\ A {\bf 44}, 015401 (2011)
  [arXiv:1010.1263 [hep-th]].

  R.~Blumenhagen,
  ``Nonassociativity in String Theory,''
  arXiv:1112.4611 [hep-th].

  E.~Plauschinn,
  ``Non-geometric fluxes and non-associative geometry,''
  PoS CORFU {\bf 2011}, 061 (2011)
  [arXiv:1203.6203 [hep-th]].

\bibitem{noncommutclo}


D.~Lust,
  ``T-duality and closed string non-commutative (doubled) geometry,''
  JHEP {\bf 1012}, 084 (2010)
  [arXiv:1010.1361 [hep-th]].

C.~Condeescu, I.~Florakis and D.~Lust,
  ``Asymmetric Orbifolds, Non-Geometric Fluxes and Non-Commutativity in Closed
String Theory,''
  JHEP {\bf 1204}, 121 (2012)
  [arXiv:1202.6366 [hep-th]].

 D.~Lust,
  ``Twisted Poisson Structures and Non-commutative/non-associative Closed String
Geometry,''
  PoS CORFU {\bf 2011}, 086 (2011)
  [arXiv:1205.0100 [hep-th]].

D.~Andriot, M.~Larfors, D.~Lust and P.~Patalong,
  ``(Non-)commutative closed string on T-dual toroidal backgrounds,''
  arXiv:1211.6437 [hep-th].


\bibitem{Hohm:2012gk}
  O.~Hohm and B.~Zwiebach,
  ``Large Gauge Transformations in Double Field Theory,''
  JHEP {\bf 1302}, 075 (2013)
  [arXiv:1207.4198 [hep-th]].

\bibitem{Parkdiffeos}

J.~-H.~Park,
  ``Comments on double field theory and diffeomorphism,''
  arXiv:1304.5946 [hep-th].

\bibitem{yetgauged}

K.~Lee and J.~-H.~Park,
  ``Covariant action for a string in doubled yet gauged spacetime,''
  arXiv:1307.8377 [hep-th].

\bibitem{Garousi:2013zca}
  M.~R.~Garousi,
  ``On Riemann curvature corrections to type II supergravity,''
  arXiv:1303.4034 [hep-th].

  M.~R.~Garousi,
  ``Ricci curvature corrections to type II supergravity,''
  Phys.\ Rev.\ D {\bf 87} (2013) 025006
  [arXiv:1210.4379 [hep-th]].

   M.~R.~Garousi,
  ``T-duality of the Riemann curvature corrections to supergravity,''
  Phys.\ Lett.\ B {\bf 718} (2013) 1481
  [arXiv:1208.4459 [hep-th]].


\bibitem{Hohm:2013jaa}
  O.~Hohm, W.~Siegel and B.~Zwiebach,
  ``Doubled alpha'-Geometry,''
  arXiv:1306.2970 [hep-th].

\bibitem{Godazgar:2013bja}

H.~Godazgar and M.~Godazgar,
  ``Duality completion of higher derivative corrections,''
  arXiv:1306.4918 [hep-th].


\bibitem{Doublebranes}

  C.~Albertsson, T.~Kimura and R.~A.~Reid-Edwards,
  ``D-branes and doubled geometry,''
  JHEP {\bf 0904} (2009) 113
  [arXiv:0806.1783 [hep-th]].

    C.~Albertsson, S.~-H.~Dai, P.~-W.~Kao and F.~-L.~Lin,
  ``Double Field Theory for Double D-branes,''
  JHEP {\bf 1109} (2011) 025
  [arXiv:1107.0876 [hep-th]].

  \bibitem{NS5/KK5}
 S.~Jensen,
  ``The KK-Monopole/NS5-Brane in Doubled Geometry,''
  JHEP {\bf 1107} (2011) 088
  [arXiv:1106.1174 [hep-th]].

 \bibitem{deBoer:2010ud}
  J.~de Boer and M.~Shigemori,
  ``Exotic branes and non-geometric backgrounds,''
  Phys.\ Rev.\ Lett.\  {\bf 104} (2010) 251603
  [arXiv:1004.2521 [hep-th]].

   J.~de Boer and M.~Shigemori,
  ``Exotic Branes in String Theory,''
  arXiv:1209.6056 [hep-th].

\bibitem{Rotating string}

  T.~Kikuchi, T.~Okada and Y.~Sakatani,
  ``Rotating string in doubled geometry with generalized isometries,''
  Phys.\ Rev.\ D {\bf 86} (2012) 046001
  [arXiv:1205.5549 [hep-th]].

  \bibitem{Hassler:2013wsa}
  F.~Hassler and D.~Lust,
  ``Non-commutative/non-associative IIA (IIB) Q- and R-branes and their
intersections,''
  arXiv:1303.1413 [hep-th].


\bibitem{Villadoro:2007tb}
  G.~Villadoro and F.~Zwirner,
  ``Beyond Twisted Tori,''
  Phys.\ Lett.\ B {\bf 652} (2007) 118
  [arXiv:0706.3049 [hep-th]].

  G.~Villadoro and F.~Zwirner,
  ``On general flux backgrounds with localized sources,''
  JHEP {\bf 0711} (2007) 082
  [arXiv:0710.2551 [hep-th]].





\bibitem{Bergshoeff:2012jb}


  E.~A.~Bergshoeff, A.~Marrani and F.~Riccioni,
  ``Brane orbits,''
  Nucl.\ Phys.\ B {\bf 861} (2012) 104
  [arXiv:1201.5819 [hep-th]].


  E.~A.~Bergshoeff and F.~Riccioni,
  ``The D-brane U-scan,''
  arXiv:1109.1725 [hep-th].


  E.~A.~Bergshoeff and F.~Riccioni,
  ``Branes and wrapping rules,''
  Phys.\ Lett.\ B {\bf 704} (2011) 367
  [arXiv:1108.5067 [hep-th]].


  E.~A.~Bergshoeff and F.~Riccioni,
  ``Dual doubled geometry,''
  Phys.\ Lett.\ B {\bf 702} (2011) 281
  [arXiv:1106.0212 [hep-th]].

  E.~A.~Bergshoeff and F.~Riccioni,
  ``String Solitons and T-duality,''
  JHEP {\bf 1105} (2011) 131
  [arXiv:1102.0934 [hep-th]].


\end{thebibliography}
\end{document}